\documentclass[aps,prd,superscriptaddress,nofootinbib,preprintnumbers,floatfix]{revtex4-1}
\usepackage{tikz}
\usepackage{longtable}
\usepackage{epstopdf}
\usepackage{graphicx}
\usepackage{slashed}
\usepackage{dcolumn}
\usepackage{bm}
\usepackage{amssymb,amsthm,amsmath,amstext,amsbsy,amsopn}
\usepackage{bbm}
\usepackage{nicefrac}
\usepackage{mathrsfs}
\usepackage{pstricks}
\usepackage{hyperref}
\usepackage{leftidx}
\usepackage{environ}
\usepackage{mathtools}
\usepackage{xspace}
\usepackage{overpic}
\usepackage{subfigure}
\usepackage{tensor}
\usepackage{gensymb} 
\usepackage{cancel}
\usepackage{float}
\usepackage{color}
\usepackage{calligra}
\usepackage{soul}
\usepackage{changes}
\usepackage{appendix}
\usepackage{comment}
\usepackage{multirow}
\usepackage{physics}

\begin{document}

\title{Elastic lepton–proton two-photon exchange scattering: An exact HB$\chi$PT analysis including hadronic effects at NNLO}
\author{Rakshanda Goswami}
    \email{r.goswami@iitg.ac.in}
	\affiliation{Department of Physics, Indian Institute of Technology Guwahati, Guwahati - 781039, Assam, India.}
    
\author{Pulak Talukdar}
    \email{pulaktalukdar45@gmail.com}
	\affiliation{Department of Physics, S. B. Deorah College, Guwahati - 781007, Assam, India.}
    
\author{Bhoomika Das}
    \email{bhoomika.das@iitg.ac.in}
	\affiliation{Department of Physics, Indian Institute of Technology Guwahati, Guwahati - 781039, Assam, India.}
    
\author{Udit Raha}
    \email{udit.raha@iitg.ac.in}
	\affiliation{Department of Physics, Indian Institute of Technology Guwahati, Guwahati - 781039, Assam, India.}
    
\author{Fred Myhrer}
    \email{myhrer@mailbox.sc.edu}
	\affiliation{Department of Physics and Astronomy, University of South Carolina, Columbia, SC 29208, USA.}
\begin{abstract}
We present an exact analytical evaluation of the two-photon exchange (TPE) correction to the 
elastic lepton–proton differential scattering cross section at low energies within the framework of 
heavy-baryon chiral perturbation theory. Our analysis focuses on the kinematic regime relevant to the 
ongoing MUSE experiment, and we therefore restrict the intermediate states to the dominant elastic 
channel. All loop integrals are evaluated analytically without approximations. Radiative and chiral 
recoil contributions of the proton are included, retaining kinematical and dynamical TPE corrections to 
the cross section through next-to-next-to-leading order [i.e., ${\mathcal O}(\alpha/M^2)$] accuracy in 
the recoil expansion where $M$ is the proton mass. At this chiral order, pion-loop contributions 
demonstrate that structure-dependent TPE effects arise through the proton form factors. Our analytical 
results for the scattering cross section reveal non-vanishing residual proton structure effects of 
${\mathcal O}(\alpha/M^2)$, despite substantial cancellations between TPE box and crossed-box contributions.
Such effects were entirely absent at this accuracy in our earlier analysis based on the 
soft-photon approximation. Although the next-to-leading-order contributions are numerically sizable, the 
next-to-next-to-leading-order TPE corrections are found to be small, thereby indicating that the chiral 
expansion exhibits reasonably good perturbative convergence.
\end{abstract}
\maketitle

\section{Introduction}
\label{sec:intro}
In recent years, the two-photon exchange (TPE) process in elastic (anti-)lepton-proton ($\ell^\pm$-p) 
scatterings has garnered considerable interest in the high-energy physics community. This was primarily 
driven by several unexpected findings in precision experiments over the last two decades, which could be 
ostensibly resolved by including the TPE effects of the radiative corrections to the cross section. These
observations include the discrepancies between polarized and unpolarized measurements of the proton's form
factors, e.g.,
Refs.~\cite{Rosenbluth:1950yq,Akhiezer:1974em,Arnold:1980zj,Gayou:2001qt,Jones:1999rz,Perdrisat:2006hj,Punjabi:2015bba,Puckett:2010,Arrington:2003,Guichon:2003,Blunden:2003sp,Rekalo:2004wa,Blunden:2005ew,Carlson:2007sp,Arrington:2011}, and 
between scattering and atomic spectroscopy measurements of the proton charge-radius puzzle, e.g., Refs.~\cite{Pohl:2010zza,Pohl:2013,Mohr:2012tt,Antognini:1900ns,Bernauer:2014,Carlson:2015,Bernauer:2020ont,Gao:2021sml}. 
It has been realized that the TPE radiative correction to the one-photon exchange (OPE) process plays 
a decisive role in the precise determination of the unpolarized elastic cross section, representing one 
of the dominant sources of systematic uncertainty that can no longer be neglected at the sub-percent 
level accuracy of current experiments. In particular, the interference between the OPE and TPE amplitudes
modifies the elastic scattering cross section at the few-percent level, thereby affecting the extraction 
of fundamental proton properties. The most direct experimental access to the TPE contribution is provided 
by the measurement of the lepton–anti-lepton charge asymmetry ($A_{\ell^\pm}$), defined as the relative 
difference between the elastic cross sections for oppositely charged leptons. Since, the interference 
term is {\it charge-odd}, namely, that it changes sign under lepton charge conjugation, this isolates the 
interference between the OPE (${\mathcal M}_{\gamma}$) and TPE (${\mathcal M}_{\gamma\gamma}$) amplitudes, 
thereby providing direct sensitivity to the real part of the TPE correction {\it via} the asymmetry 
observable, 
$A_{\ell^\pm}\propto {\mathbb R}{\rm e}\left({\mathcal M}^*_{\gamma}\,{\mathcal M}_{\gamma\gamma}\right)$. 
Experimental Collaborations, such as OLYMPUS (DESY)~\cite{OLYMPUS:2016gso}, VEPP-3 
(Novosibirsk)~\cite{Rachek:2014fam}, and CLAS (Jefferson Lab)~\cite{CLAS:2016fvy} have previously measured 
the electron–proton charge asymmetry by alternately scattering electron and positron beams off protons 
under identical conditions. However, the data from these measurements suffer from limited statistics at 
very low-momentum transfers, below $|Q^2|\lesssim 0.1$~GeV${}^2/c^2$, precisely where the systematic 
uncertainties associated with the TPE are the most pronounced. In this context, the MUSE 
Collaboration~\cite{Gilman:2013eiv} is soon poised to publish a simultaneous measurement of the charge 
asymmetry using both electron and muon beams, as well as their antiparticles, at unprecedentedly 
low-momentum transfers, $Q^2 < 0.1$~GeV${}^2/c^2$. 

Since the TPE intermediate hadronic state can involve all possible excitations of the proton, its 
evaluation is theoretically challenging relying on a variety of complementary theoretical approaches have 
been developed over the past decades. The existing methods include hadronic
models~\cite{Blunden:2003sp,Chen:2004tw,Gorchtein:2004ac,Kondratyuk:2007hc,Afanasev:2007ii,Maximon:2000hm,Vanderhaeghen:2000ws,Tomalak:2014dja,Lee:2015jqa,Tomalak:2015aoa,Tomalak:2015hva,Koshchii:2017dzr,Bucoveanu:2018soy,Guo:2022kfo}, dispersion relations~\cite{Lee:2015jqa,Tomalak:2015hva,Guo:2022kfo,Hammer:2003ai,Gorchtein:2006mq,Borisyuk:2008es,Tomalak:2014sva,Lorenz:2014yda,Tomalak:2016vbf,Tomalak:2017npu,Tomalak:2018jak}, and effective field theoretical analysis based on
non-relativistic QED (NRQED)~\cite{Hill:2011wy,Peset:2021iul} and chiral dynamics~\cite{Lorenz:2014yda,Nevado:2007dd,Birse:2012eb,Alarcon:2013cba,Peset:2015zga,Alarcon:2020wjg,Talukdar:2019dko,Talukdar:2020aui,Cao:2021nhm,Choudhary:2023rsz,Goswami:2025zoe}. Furthermore, recent lattice QCD simulations~\cite{Fu:2021dja,Fu:2022fgh},
albeit computational challenges, have shown promising first-principle results for evaluating the TPE 
contribution. At low-energies heavy baryon chiral perturbation theory (HB$\chi$PT) provides a convenient 
analytical tool for studying TPE effects of elastic lepton–proton scatterings (see e.g., 
Refs.~\cite{Birse:2012eb,Talukdar:2019dko,Talukdar:2020aui,Choudhary:2023rsz,Goswami:2025zoe}), due to its 
transparent and systematic diagrammatic formulation. This low-energy EFT offers a systematic, 
model-independent approach by exploiting the approximate chiral symmetry of QCD and its dynamical breakings.
In HB$\chi$PT observables are simultaneously expanded in powers of small external momenta compared to the 
chiral-symmetry breaking scale $\Lambda_\chi\simeq 1$~GeV$/c$, or the inverse powers of the proton mass, 
$M\simeq 938$~MeV. This allows the proton to be treated as a heavy, non-relativistic source where 
perturbative momentum-dependent corrections can be incorporated order-by-order. This allows a controlled 
organization of higher-order {\it radiative} and proton  {\it recoil} effects, see, e.g.,
Refs.~\cite{Gasser:1982ap,Jenkins:1990jv,Bernard:1992qa,Ecker:1994pi,Bernard:1995dp,Fettes:1998ud,Fettes:2000fd,scherer2003}. 

The evaluation of the TPE contributions to lepton-proton radiative corrections within HB$\chi$PT, 
up-to-and-including next-to-leading order (NLO), was first undertaken by Talukdar {\it et al.} in 
Ref.~\cite{Talukdar:2019dko}. Motivated by earlier studies of low- and intermediate-energy TPE effects, e.g.,
Refs.~\cite{Maximon:2000hm,Vanderhaeghen:2000ws,Koshchii:2017dzr,Bucoveanu:2018soy}, Talukdar {\it et al.}
employed HB$\chi$PT for the first time to provide a model-independent evaluation of TPE corrections in 
elastic lepton–proton scattering. Their first TPE analysis utilized SPA~\cite{Tsai:1961zz,Mo:1968cg} to 
simplify the 4-point loop-integrals arising in the evaluation of the box and crossed-box diagrams. 
Subsequently, the TPE contribution in HB$\chi$PT at NLO accuracy was evaluated \underline{exactly} by 
Choudhary {\it et al.} in Ref.~\cite{Choudhary:2023rsz}. Their study demonstrated that the inclusion of the 
full TPE loop kinematic region including configurations with two {\it hard photon} exchange, yields a 
significant contribution to the elastic cross section, which is missed in SPA. In addition, our 
\underline{exact} TPE approach yielded a LO contribution akin to the McKinley and 
Feshbach~\cite{McKinley:1948zz}. In contrast, the SPA approach of Talukdar {\it et al.} yielded a vanishing
LO contribution. Unfortunately, within the standard HB$\chi$PT power-counting scheme the NLO corrections do
not incorporate proton’s internal hadronic structure within the radiative corrections, effectively treating 
the proton as a point-like particle. To account for the important finite-size effects arising from the 
proton’s structure, which in HB$\chi$PT is introduced {\it via} the pion-loop contributions and low-energy
constants (LECs) such as the anomalous magnetic moment, $\kappa_p=1.795$, the TPE must include the dominant 
part of the next-to-next-to-leading order (NNLO) modifications stemming from the $\gamma$pp vertex 
corrections, {\it vis-a-vis} proton’s electromagnetic form factors. In our recent SPA work by Goswami 
{\it et al.}~\cite{Goswami:2025zoe}, we improved the SPA analysis of Talukdar {\it et al.} by 
{\it partly}\footnote{A complete treatment of all NNLO contributions 
requires the evaluation of the full set of two-loop TPE diagrams in which pion exchange occurs in one of the 
loops. In the recent SPA work of Goswami {\it et al.}~\cite{Goswami:2025zoe}, only the dominant ``reducible" 
two-loop topologies were considered, in which the pion-loops factorize within the TPE box and crossed-box 
diagrams, effectively renormalizing and dressing the proton-photon vertices through form factors. At NNLO, 
however, there also exits several “non–form-factor” type (see Ref.~\cite{Goswami:2025zoe}), involving 
non-factorizable pion-loops that do not contribute to vertex renormalization. These genuine two-loop TPE 
contributions are considerably more challenging to compute analytically, and are therefore omitted in the 
present work. Nevertheless, such diagrams are expected to yield contributions that are numerically 
suppressed relative to the typical NNLO corrections.} including these NNLO effects and by 
employing a more rigorous treatment of the loop-integrals, while retaining the SPA used in Talukdar 
{\it et al.}'s original approach in Ref.~\cite{Talukdar:2019dko}. Particularly in the analysis of Goswami 
{\it et al.} it was observed that at the intended ${\mathcal O}(\alpha^3/M^2)$ precision of the radiative 
corrected elastic cross section the NNLO TPE contributions, especially those stemming from the proton’s 
internal structure, either {\it fortuitously} cancel at this order, or manifest themselves only at higher 
recoil order, namely, ${\mathcal O}(\alpha^3/M^3)$. In addition, ${\mathcal O}(\alpha^3/M^2)$ effects also 
arise from proton recoil correction terms present in the NNLO squared amplitudes as well as from the 
kinematically suppressed LO and NLO terms. The latter corrections are dependent on the outgoing lepton 
energy ($E^\prime$) and velocity ($\beta^\prime$), which can also be expressed {\it via} the following 
recoil-expansions: 
\begin{equation}
E^\prime=E+\frac{Q^2}{2M}\, \quad \text{and} \quad
\beta^\prime=\beta+\frac{Q^2(1-\beta^2)}{2ME\beta}-\frac{Q^4(2\beta^4-1-\beta^2)}{8M^2E^2\beta^3}
+\mathcal{O}\left(\frac{1}{M^3}\right)\,,
\label{eq:E_beta_expansions}
\end{equation} 
respectively. Here $E$ and $\beta$ are the incident lepton beam variables. Consequently, the continued use 
of the SPA methodology, particularly in the low-energy MUSE kinematic region, is discouraged, as was already 
emphasized in Ref.~\cite{Goswami:2025zoe}. In the present work, we therefore abandon the SPA framework and 
extend the NLO \underline{exact} TPE analysis of Choudhary {\it et al.}~\cite{Choudhary:2023rsz} to include
the dominant NNLO proton's finite-size contributions along with the other ${\mathcal O}(\alpha^3/M^2)$ terms
mentioned above {\it via} the heavy baryon formalism.  

This paper is organized as follows. In Sec.~2, we briefly outline the essential features of our EFT 
methodology, along with the kinematical setup required for our analytical evaluations. The analytical details
are then presented in Sec.~3, where we extend the NLO TPE evaluations of Choudhary 
{\it et al.}~\cite{Choudhary:2023rsz} by including all possible kinematically suppressed NNLO [i.e., 
${\mathcal O}(\alpha/M^2)$] corrections to the elastic cross section arising from the LO and NLO TPE diagrams
of ${\mathcal O}(e^4)$ and ${\mathcal O}(e^4/M)$, respectively. In Sec.~4,  we evaluate the genuine dynamical 
NNLO TPE corrections to the cross section including the pion-loop or hadronic structure contributions. Here 
we include the additional contribution from all possible NNLO TPE diagrams of ${\mathcal O}(e^4/M^2)$, 
retaining corrections terms up to ${\mathcal O}(\alpha/M^2)$ accuracy. We provide our numerical results in 
Sec.~5 and analyze them in the context of low-energy MUSE kinematics as specified in 
Ref.~\cite{Gilman:2013eiv}. Section~6 contains our summary and conclusions. At the end, we include two 
appendices where we relegate the explicit analytical expressions for several modified loop-integrals that 
appeared in our past works in Refs.~\cite{Choudhary:2023rsz,Goswami:2025zoe}, as well as several additional
new ones appearing for the first time in the present analysis. 

\section{HB$\chi$PT: General formalism and kinematics}   
We briefly discuss the basic EFT methodology of the HB$\chi$PT formalism whose details could be obtained, 
e.g., in 
Refs.~\cite{Gasser:1982ap,Jenkins:1990jv,Bernard:1992qa,Ecker:1994pi,Bernard:1995dp,Fettes:1998ud,Fettes:2000fd,scherer2003}. 
HB$\chi$PT emerges from a non-relativistic expansion of the relativistic pion-nucleon Lagrangian in 
inverse powers of $M$. The original relativistic $\chi$PT Lagrangian, invariant under the action of the
{\it chiral gauge group} ${\mathcal G}=$ SU(2)$_L\otimes$ SU(2)$_R$, is constructed such that the 
corresponding pions emerge as the Goldstone bosons associated with the spontaneous broking of chiral 
symmetry. The pion reside in the coset space ${\mathcal G}/{\mathcal H}$, and are defined {\it via} the
{\it non-linear realization}: $K=\sqrt{LU^\dagger R^\dagger}R\sqrt{U}$, where the elements $L\in$ 
SU(2)$_L$, $R\in$ SU(2)$_R$, and $U$ is the non-linearly realized pion field $\vec{\pi}$. The so-called 
{\it compensator} element $K\in {\mathcal H}=$ SU(2)$_{V=L+R}\subseteq {\mathcal G}$ belongs to the 
unbroken vector (flavor) subgroup $\mathcal H$ of the chiral group $\mathcal G$. Likewise, the 
relativistic iso-spinor nucleon field $\Psi_N=({\rm p} \,\,\, {\rm n})^{\rm T} $ can also be shown to 
transform non-linearly as $\Psi_N\to K(U,L,R)\Psi_N$, 
leading to the following representation of the covariant derivatives in this theory, namely,
\begin{equation}
D_\mu \Psi_N \equiv (\partial_\mu + \Gamma_\mu)\Psi_N\,, \quad  \text{and} \quad
\nabla_\mu U = \partial_\mu U - i({\rm v}_\mu+a_\mu) U + iU ({\rm v}_\mu-a_\mu)\,, 
\end{equation} 
where the {\it chiral connection}
$\Gamma_\mu$ and {\it chiral vielbein} $u_\mu$ enter as 
\begin{equation}
\Gamma_\mu = \frac{1}{2}[u^\dagger,\partial_\mu u] -\frac{i}{2}u^\dagger({\rm v}_\mu+a_\mu)u 
- \frac{i}{2}u({\rm v}_\mu-a_\mu)u^\dagger\,; \quad u=\sqrt{U}\,\,\, \text{and} 
\quad u_\mu= i u^\dagger\nabla_\mu U u^\dagger\,. 
\end{equation}
Here ${\rm v}_\mu$ and $a_\mu$ are the external vector and axial fields, respectively. Furthermore, 
induced by the action of $\mathcal G$, the following transformations of the basic elements of the
effective Lagrangian manifest:
\begin{equation}
D_\mu \to K D_\mu K^\dagger\,,\quad  \Gamma_\mu \to K\Gamma_\mu K^\dagger + K\partial_\mu K^\dagger\,,\quad  
\text{and} \quad u_\mu \to Ku_\mu K^\dagger\,,
\end{equation}
motivating the construction of the most general effective relativistic Lagrangian, allowed by all 
possible low-energy symmetries, such as invariance under P, C, T, cluster decomposition, Lorentz, and 
chiral transformations. 

It is well known that a fully relativistic formulation of $\chi$PT for the pion–nucleon system 
encounters severe practical difficulties in diagrammatic evaluations. In particular, the power-counting
scheme fails to maintain a {\it one-to-one correspondence} between the expansion in pion-loops and the 
chiral expansion, thereby raising concerns about the efficacy of low-energy perturbative convergence. 
However, this issue was shown to be effectively circumvented by adopting a non-relativistic 
expansion~\cite{Gasser:1982ap,Jenkins:1990jv,Bernard:1992qa,Ecker:1994pi,Bernard:1995dp}, wherein one 
projects out from the relativistic nucleon spinor field $\Psi_N$ only the ``large" Dirac components 
$\mathcal N_v \sim {\mathcal O}(M)$, which depend on the four-velocity, we choose $v_\mu = (1, {\bf 0})$ 
in the nucleon rest frame, using the velocity projector ${\mathcal P}^+_v = (1+v\!\!\!/)/2$, i.e., 
\begin{equation}
    \Psi_N(x) =e^{-iMv\cdot x}\left[{\mathcal N}_v(x) + h_v(x)\right]\,, \qquad \text{with}
\end{equation}
\begin{equation}
{\mathcal N}_v(x) = e^{iMv\cdot x}{\mathcal P}^+_v \Psi_N(x)\,, \,\,  \text{and} \quad
    v\!\!\!/{\mathcal N}_v(x) = {\mathcal N}_v(x) \equiv \binom{{\rm p}_v}{{\rm n}_v}\, . 
\end{equation}
Here the effects of the ``small" Dirac components $h_v$ are integrated out in the theory. For an early
review, see Ref.~\cite{Bernard:1995dp} which also contains some early references. A pedagogical 
derivation of HB$\chi$PT can be found in the book given by Ref.~\cite{Scherer2012}. Nevertheless, an 
additional re-parametrization invariance allows us to express the proton's four-momenta in an arbitrary
boosted frame as $P^\mu=Mv^\mu + p^\mu_p$, where $p^\mu_p$ is the small nucleon, {\it residual} 
off-shell four-momentum with $v\cdot p_p \ll M$. The advantage of this formulation lies in the 
decoupling of the nucleon mass in the equation of motion up to corrections suppressed by powers of 
$1/M$, for example, $v\cdot\partial {\mathcal N}_v={\mathcal O}(1/M)$. Consequently the $M$ dependence
of the heavy nucleon propagator is moved to the interaction vertices, which are ordered by a 
$1/M$-{\it recoil expansion}. Thus, up to chiral-order $\nu=2$ [i.e., ${\mathcal O}(1/M^2)$], the proton
propagator is  represented as 
\begin{eqnarray} 
iS^{(p)}_{\rm full}(l) = \frac{i}{v\cdot l+i0} + \frac{i}{2M}\left[1-\frac{l^2}{(v\cdot l +i0)^2}\right] 
+ \frac{i}{4M^2}\left[\frac{(v \cdot l)^3-l^2 (v \cdot l)}{(v\cdot l +i0)^2}\right]
+ {\mathcal O}\left(\frac{1}{M^{3}}\right)\,, 
\label{eq:p_prop}
\end{eqnarray}
where $l^\mu$ denotes the generic residual off-shell proton momentum. The simultaneous chiral and 
nucleon recoil expansion scheme restores the aforementioned one-to-one diagrammatic correspondence. An 
immediate practical benefit of the HB$\chi$PT formalism is that the vital low-energy recoil effects of 
the heavy proton are naturally captured in a controlled manner. The SU(2) HB$\chi$PT pion-nucleon 
Lagrangian is well established in the 
literature~\cite{Ecker:1994pi,Bernard:1995dp,Fettes:1998ud,Fettes:2000fd,Fettes:2000gb,Bernard:1998gv}. 
Given the lengthy sequence of operator structures and the phenomenologically determined LECs that 
parameterize the short ranged and unresolved ultraviolet (UV) physics, the explicit form of the effective
Lagrangian will not be presented here, and instead we refer to the literature, see e.g., 
Ref.~\cite{Bernard:1995dp}. In fact, in our previous TPE analysis reported in Ref.~\cite{Goswami:2025zoe},
we employed the same relevant parts of the effective Lagrangian, to which we hereby refer the reader for 
the detailed expressions. It is important to emphasize that the heavy proton is treated within a 
manifestly non-relativistic framework, whereas the pions, and other light particles, such as leptons 
($\ell^\pm \equiv e^\pm, \mu^\pm$) and  photon are treated relativistically. We symbolically express the 
most general form of the HB$\chi$PT Lagrangian as terms up to NNLO, i.e., with chiral dimension $\nu=2$,
at the intended level of accuracy in this work.  
\begin{eqnarray}
\mathcal{L}_{\ell\pi N\gamma} &=& \mathcal{L}^{\rm QED}_{\ell} + \mathcal{L}^{\rm eff}_{\pi N}\,, \quad \text{where}
\nonumber\\
\mathcal{L}^{\rm eff}_{\pi N} &=& \mathcal{L}^{(2)}_\pi + \sum_{\nu=0}^{2}\mathcal{L}_{\pi N}^{(\nu)}\,.
\end{eqnarray}
In particular, the pion-loop corrections arising at NNLO generate UV divergences, which necessitate 
local counter terms for renormalization, see, e.g., Sec.~\ref{sec:intro}. 

To facilitate a straightforward comparison, we have, for the most part, kept the notation and 
convention close to Ref~\cite{Goswami:2025zoe}. The choice of a laboratory frame (lab-frame) where the
target proton is at rest greatly simplifies our evaluations and kinematics, especially bearing in mind
the low energy MUSE kinematic region of interest~\cite{Gilman:2013eiv} (also see Table I in
Ref.~\cite{Talukdar:2019dko}). In this work we shall consider 
$Q_{\mu} = (p-p^\prime)_{\mu} = (p^\prime_p-p_p)_\mu$ as the four-momentum transfer for the elastic 
scattering process, with $p(p^\prime)$ and $p_p(p^\prime_p)$ as the incoming (outgoing) lepton 
four-momentum and the proton's {\it residual} four-momentum, respectively, where in the lab. frame 
$p_p =(0,{\bf 0})$. Furthermore, we wish to emphasize that terms like $v\cdot Q$ and $v\cdot p^\prime_p$
are $\mathcal{O}(1/M)$, i.e., $v\cdot p^\prime_p=v\cdot Q=E-E^\prime=-Q^2/2M$.   

In summary, a complete and \underline{exact} analytical study of the TPE contribution up to NLO accuracy
was already performed in Ref.~\cite{Choudhary:2023rsz}. The present work aims to extend that analysis by 
incorporating the NNLO corrections to assess the convergence behavior of HB$\chi$PT and thereby obtain a 
more robust estimate of the higher order systematic uncertainties. In the following two sections we 
identify the relevant sources of these NNLO contributions. 

\section{Kinematical NNLO recoil corrections from LO and NLO TPE diagrams} 
\label{sec:three} 
Here we shall extend the analysis of Choudhary {\it et al.}~\cite{Choudhary:2023rsz} of the fractional TPE 
corrections $\pm\delta^{\rm (LO+NLO)}_{\gamma\gamma}$ to the elastic (anti)lepton-proton ($\ell^\mp$-p) 
differential cross section, namely,
\begin{eqnarray}
\left[\frac{{\rm d}\sigma_{el}(Q^2)}{{\rm d}\Omega^\prime_l}\right]^{(\ell^\mp)}_{\rm LO+NLO} 
&=& \left[\frac{{\rm d}\sigma_{el}(Q^2)}{{\rm d}\Omega^\prime_l}\right]_0
\left[1\pm \delta^{\rm (LO+NLO)}_{\gamma\gamma}(Q^2)\right]\,, \quad \text{where}
\nonumber\\
\delta^{\rm (LO+NLO)}_{\gamma\gamma}(Q^2) 
&=& \frac{2{\mathcal R}e\!\!\sum\limits_{spins}\left[{\mathcal M}_{\gamma}^{(0)*}
\left({\mathcal M}^{\rm (LO)}_{\gamma\gamma}+{\mathcal M}^{\rm (NLO)}_{\gamma\gamma}\right)
+{\mathcal M}_{\gamma}^{(1)*}{\mathcal M}^{\rm (LO)}_{\gamma\gamma}\right]}{\sum\limits_{spins} 
\left|{\mathcal M}_{\gamma}^{(0)}\right|^2} \, .
\label{eq:delta_TPE_LO}
\end{eqnarray}
These terms were evaluated analytically up-to-and-including  NLO [i.e., $\mathcal{O}(\alpha/M)$] where 
the LO OPE term~\cite{Talukdar:2020aui} is
\begin{eqnarray}
\left[\frac{{\rm d}\sigma_{el}(Q^2)}{{\rm d}\Omega^\prime_l}\right]_0 
= \frac{1}{64\pi^2M^2}\left(\frac{E^\prime\beta^\prime}{E\beta}\right)\frac{1}{4}\sum\limits_{spins} 
\left|{\mathcal M}_{\gamma}^{(0)}\right|^2
=\frac{\alpha^2 E^\prime\beta^\prime}{Q^2E\beta}\left(1-\frac{Q^2}{4M^2}\right)\left[\frac{Q^2+4EE^\prime}{Q^2}\right]\,.\quad\,
\label{eq:diff_LO}
\end{eqnarray}
In Eq.~\eqref{eq:delta_TPE_LO} the expressions ${\mathcal M}_{\gamma}^{(0,1)}$ denote the (LO, NLO) OPE
amplitudes whereas ${\mathcal M}^{\rm (LO,NLO)}_{\gamma\gamma}$ represent the sum of the (LO, NLO) TPE 
amplitudes shown in Fig.~\ref{fig:LO_NLO_TPE} for lepton-proton ($\ell^-$-p) scattering. Their 
expressions are as follows:
\begin{eqnarray}
\label{eq:M0}
{\mathcal M}^{(0)}_\gamma &=& 
-\,\frac{e^2}{Q^2} \big[\bar{u}(p^\prime)\gamma^\mu u(p) \big] \left[\chi^{\dagger}(p^\prime_p) v_\mu \chi(p_p)\right]\,, 
\\
\label{eq:M1}
\mathcal{M}^{(1)}_\gamma \!&=&\! -\,\frac{e^2}{2 M Q^2}[{\bar u}(p^\prime)\gamma^\mu\,u(p)]\,\bigg[\chi^\dagger(p_p^\prime)
\Big\{(p_p + p_p^\prime)_\mu - v_\mu v \cdot (p_p + p_p^\prime) 
\nonumber\\
&&\hspace{5.2cm} + (2+\kappa_s+\kappa_v)[S_\mu,S\cdot Q]\Big\}\, \chi(p_p)\bigg]\,, \quad \text{where}
\\
{\mathcal M}^{\rm (LO)}_{\gamma\gamma} &=& {\mathcal M}^{(a)}_{\rm box} + {\mathcal M}^{(b)}_{\rm xbox}\,, \quad \text{and} \quad
{\mathcal M}^{\rm (NLO)}_{\gamma\gamma} = {\mathcal M}^{(c)}_{\rm box} + \cdots + {\mathcal M}^{(h)}_{\rm xbox} 
+ {\mathcal M}^{(i)}_{\rm seagull}\,\,.
\end{eqnarray} 
The general frame-independent integral representations of these individual TPE amplitudes are provided in 
Ref.~\cite{Choudhary:2023rsz,Goswami:2025zoe}. Specializing to the lab-frame and performing the shift
$k\rightarrow -k+Q$ for each loop-integration four-momentum variable $k$ in the NLO crossed-box amplitudes 
${\mathcal{M}^{(b)}_{\rm xbox}}$, ${\mathcal{M}^{(d)}_{\rm xbox}}$, ${\mathcal{M}^{(f)}_{\rm xbox}}$, and 
${\mathcal{M}^{(h)}_{\rm xbox}}$, and with further partial cancellation of the $v_\mu v_\nu$ terms among 
the amplitudes $\mathcal{M}^{(c)}_{\rm box}$, $\mathcal{M}^{(f)}_{\rm box}$, $\mathcal{M}^{(g)}_{\rm box}$, 
$\mathcal{M}^{(h)}_{\rm xbox}$ and $\mathcal{M}^{(i)}_{\rm seagull}$, we arrive at the following 
\underline{modified} integral representations of the LO and NLO amplitudes~\cite{Goswami:2025zoe}:
%
\begin{figure*}[tbp]
\centering
\includegraphics[scale=0.45]{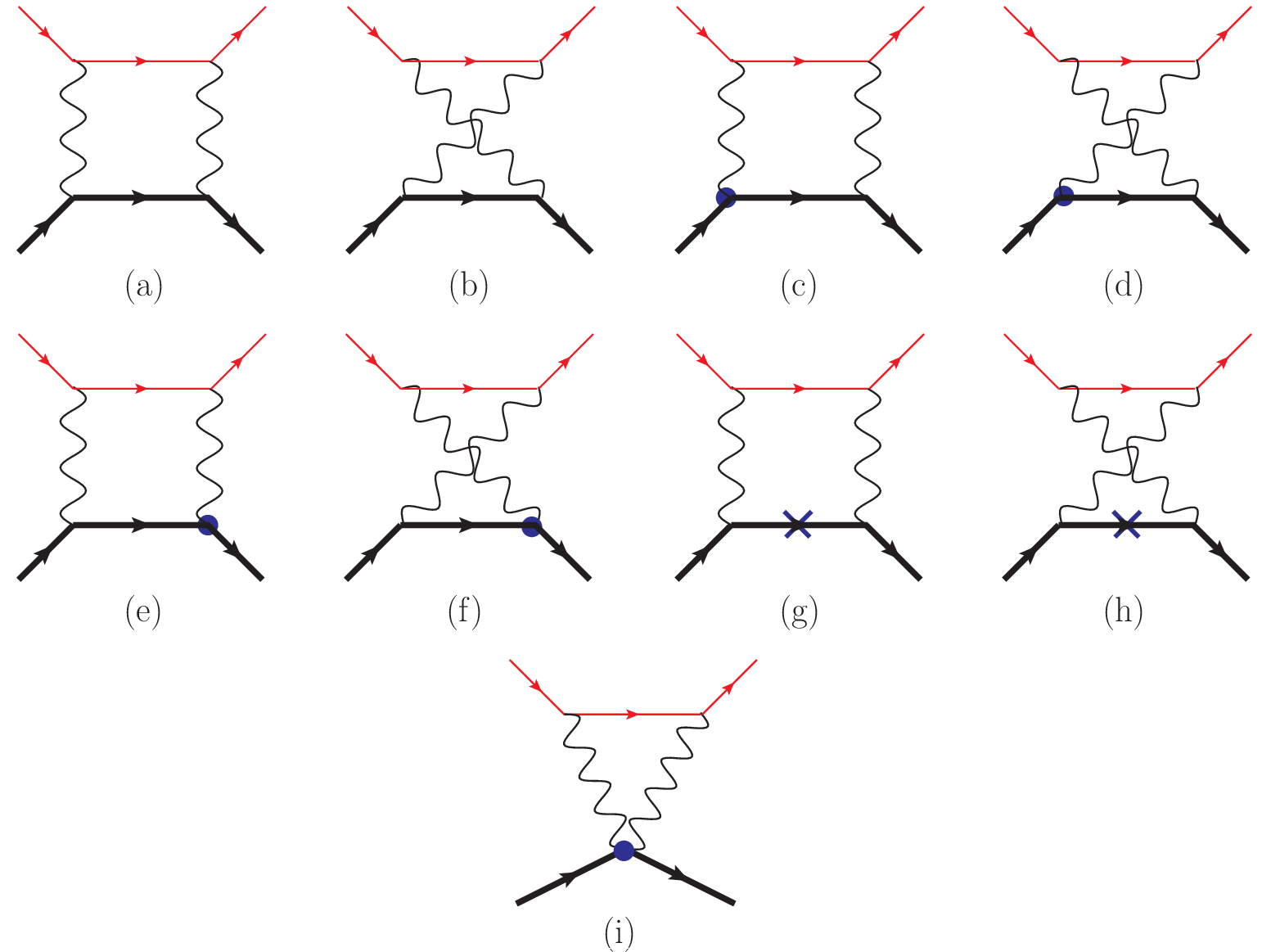}
\caption{The LO [i.e., $\mathcal{O}(\alpha^2)$] and NLO [i.e., $\mathcal{O}(\alpha^2/M)$] TPE diagrams 
         contributing to the $\mathcal{O}(\alpha^3/M)$ lepton-proton elastic differential cross-section,
         are shown. The thick, thin, and wiggly lines denote the proton, lepton, and photon propagators.
         The blobs and crosses denote the insertions of the NLO proton-photon vertices and 
         ${\mathcal O}(1/M)$ proton propagators, respectively (figure reproduced from 
         Ref.~\cite{Goswami:2025zoe}). }
\label{fig:LO_NLO_TPE}
\end{figure*} 
%
\begin{eqnarray}
    {\mathcal{M}^{(a)}_{\rm box}} &=& e^4 \int \frac{{\rm d}^4 k}{(2\pi)^4 i}
    \frac{\left[\Bar{u}(p^\prime)\gamma^{\mu}(\slashed{p}-\slashed{k}+m_l)\gamma^{\nu}u(p)\right]
    \left[\chi^{\dagger}(p^\prime_p)v_{\mu}v_{\nu}\chi(p_p)\right]}{(k^2+i0)\, [(Q-k)^2+i0]\,(k^2-2k\cdot p+i0)\,
    (v\cdot k +i0)} \,,
\\
    {\mathcal{M}^{(b)}_{\rm xbox}} &=& e^4 \int \frac{{\rm d}^4 k}{(2\pi)^4 i}
    \frac{\left[\Bar{u}(p^\prime)\gamma^{\mu}(\slashed{p}+\slashed{k}-\slashed{Q}+m_l)\gamma^{\nu}u(p)\right]
    \left[\chi^{\dagger}(p^\prime_p)v_{\mu}v_{\nu}\chi(p_p)\right]}{(k^2+i0)\, [(Q-k)^2+i0]\,(k^2+2k\cdot p^\prime+i0)\, 
    (v\cdot k +i0)} \,,
\\
\nonumber\\
    {\widetilde{\mathcal{M}}^{(c)}_{\rm box}} &=& \frac{e^4}{2 M} \int \frac{{\rm d}^4 k}{(2\pi)^4 i}
    \frac{[\Bar{u}(p^\prime)\gamma^{\mu}(\slashed{p}-\slashed{k}+m_l)\gamma^{\nu}u(p)]
    \Big[\chi^{\dagger}(p^\prime_p)v_{\mu} k_{\nu}\chi(p_p)\Big]}{(k^2+i0)\, [(Q-k)^2+i0]\, (k^2-2k\cdot p+i0)\, (v\cdot k +i0)} \,,
\\
    {\widetilde{\mathcal{M}}^{(d)}_{\rm xbox}} &=& \frac{e^4}{2M} \int \frac{{\rm d}^4 k}{(2\pi)^4 i}
    \frac{[\Bar{u}(p^\prime)\gamma^{\mu}(\slashed{p}+\slashed{k}-\slashed{Q}+m_l)\gamma^{\nu}u(p)]
    \Big[\chi^{\dagger}(p^\prime_p)v_{\nu} k_{\mu}\chi(p_p)\Big]}{(k^2+i0)\, [(Q-k)^2+i0]\, (k^2+2k\cdot p^\prime+i0)\, (v\cdot k+i0)}\,
\end{eqnarray}    
\begin{eqnarray}
    {\widetilde{\mathcal{M}}^{(e)}_{\rm box}} &=& \frac{e^4}{2M} \int \frac{{\rm d}^4 k}{(2\pi)^4 i}
    \frac{[\Bar{u}(p^\prime)\gamma^{\mu}(\slashed{p}-\slashed{k}+m_l)\gamma^{\nu}u(p)]
    \Big[\chi^{\dagger}(p^\prime_p)v_{\nu}(Q+k)_{\mu}\chi(p_p)\Big]}{(k^2+i0)\, [(Q-k)^2+i0]\, (k^2-2k\cdot p+i0)\, (v\cdot k +i0)} \,,
\\
    {\widetilde{\mathcal{M}}^{(f)}_{\rm xbox}} &=& \frac{e^4}{2M} \int \frac{{\rm d}^4 k}{(2\pi)^4 i}
    \frac{[\Bar{u}(p^\prime)\gamma^{\mu}(\slashed{p}+\slashed{k}-\slashed{Q}+m_l)\gamma^{\nu}u(p)]
    \Big[\chi^{\dagger}(p^\prime_p)v_{\mu} (k+Q)_{\nu}\chi(p_p)\Big]}{(k^2+i0)\, [(Q-k)^2+i0]\, (k^2+2k\cdot p^\prime+i0)\, (v\cdot k +i0)} \,,
\\
\nonumber\\
    {\widetilde{\mathcal{M}}^{(g)}_{\rm box}} &=& \frac{e^4}{2M} \int \frac{{\rm d}^4 k}{(2\pi)^4 i}
    \frac{[\Bar{u}(p^\prime)\gamma^{\mu}(\slashed{p}-\slashed{k}+m_l)\gamma^{\nu}u(p)]
    \Big[\chi^{\dagger}(p^\prime_p)v_{\mu} v_{\nu} \chi(p_p)\Big]}{(k^2+i0)\, [(Q-k)^2+i0]\, (k^2-2k\cdot p+i0)} 
    \left(-\frac{k^2}{(v\cdot k +i0)^2}\right) \,,  \qquad\,\,
\\
    {\widetilde{\mathcal{M}}^{(h)}_{\rm xbox}} &=& \frac{e^4}{2M}\! \int \!\frac{{\rm d}^4 k}{(2\pi)^4 i}
    \frac{[\Bar{u}(p^\prime)\gamma^{\mu}(\slashed{p}+\slashed{k}-\slashed{Q}+m_l)\gamma^{\nu}u(p)]
    \Big[\chi^{\dagger}(p^\prime_p)v_{\mu} v_{\nu} \chi(p_p)\Big] }{(k^2+i0)\, [(Q-k)^2+i0]\, (k^2+2k\cdot p^\prime+i0)}\!\!
    \left(-\frac{k^2}{(v\cdot k +i0)^2}\right), \quad\,\quad \,\,
\\
\nonumber\\
\text{and} && {\widetilde{\mathcal{M}}^{(i)}_{\rm seagull}} 
= -\frac{e^4}{M} \int \frac{{\rm d}^4 k}{(2\pi)^4 i}
\frac{[\Bar{u}(p^\prime)\gamma^{\mu}(\slashed{p}-\slashed{k}+m_l)\gamma_{\mu}u(p)]
\Big[\chi^{\dagger}(p^\prime_p)\chi(p_p)\Big]}{(k^2+i0)\, [(Q-k)^2+i0]\, (k^2-2k\cdot p+i0) } \,,
\end{eqnarray}
where $m_l$ denotes the lepton mass in the above integrals. It is noteworthy that the contributions 
arising from the interference of amplitudes in Eq.~\eqref{eq:delta_TPE_LO}, namely, 
${\mathcal M}_{\gamma}^{(1)*}{\mathcal M}^{\rm (LO)}_{\gamma\gamma}$, which generate the 
${\mathcal O}(\alpha/M^2)$ fractional correction terms to the elastic cross section, 
$\delta^{(a_1)}_{\rm box}$ and $\delta^{(b_1)}_{\rm box}$ [see Eqs.~\eqref{a1-TPE} and \eqref{b1-TPE}] , 
are an order beyond the intended NLO accuracy of our previous exact TPE work in 
Ref.~\cite{Choudhary:2023rsz}.

Now we analytically re-evaluate the the nine TPE diagrams (a) - (i), in order to determine their additional
${\mathcal O}(\alpha/M^2)$ fractional corrections contributing to the elastic differential cross
section at our target accuracy of ${\mathcal O}(\alpha^3/M^2)$. We follow the same reduction strategy used 
in Ref.~\cite{Choudhary:2023rsz}, employing successive applications of integration by parts (IBP) identities
and the method of partial fractions to simplify the corresponding Feynman amplitudes. As in 
Ref.~\cite{Choudhary:2023rsz}, we first present the unexpanded expressions for the fractional contributions
arising from the LO and NLO TPE diagrams, accurate to ${\mathcal O}(\alpha/M^2)$]:
\begin{eqnarray}
\delta^{\rm (LO+NLO)}_{\gamma\gamma}(Q^2) &=& \left[\delta^{(a)}_{\rm box}(Q^2) + \delta^{(b)}_{\rm xbox}(Q^2)\right]_{\rm LO} 
+ \left[\delta^{(a_1)}_{\rm box}(Q^2) + \delta^{(b_1)}_{\rm xbox}(Q^2) + \delta^{(c)}_{\rm box}(Q^2) + \delta^{(d)}_{\rm xbox}(Q^2) \right.
\nonumber\\ 
&& \left. +\, \delta^{(e)}_{\rm box}(Q^2)  + \delta^{(f)}_{\rm xbox}(Q^2) + \delta^{(g)}_{\rm box}(Q^2) + \delta^{(h)}_{\rm xbox}(Q^2) 
 + \delta^{\rm (seagull)}_{\gamma\gamma}(Q^2)\right]_{\rm NLO}\,, 
\end{eqnarray}
where
\begin{eqnarray}
\label{a-TPE}
\delta^{(a)}_{\rm box}(Q^2) &=& {2{\mathbb R}e\sum\limits_{spins}
\bigg[\mathcal{M}^{(0)*}_{\gamma} {\mathcal M}^{(a)}_{\rm box}\bigg]}\bigg/
{\sum\limits_{spins}\left|\mathcal{M}^{(0)}_\gamma\right|^2}
\nonumber\\
&=& -\,4 \pi\alpha \left[\frac{Q^2}{Q^2+4E E^\prime}\right]  
{\mathbb R}e\left\{\int \frac{{\rm d}^4 k}{{(2 \pi ) }^4 i}
\frac{{\rm Tr}\left[(\slashed{p}+m_l)\,\not{\!v}\,(\slashed{p}^{\prime}+m_l)\,\slashed{v}\, 
(\slashed{p}-\slashed{k}+m_l)\,\slashed{v}\right]}{(k^2+i0)\,
[(Q-k)^2+i0]\,(k^2-2 k\cdot p+i0)\,(v\cdot k+i0)}\right\}
\nonumber\\
&=& -\,8 \pi \alpha \left[\frac{Q^2}{Q^2+4 E E^\prime}\right] 
{\mathbb R}e \bigg\{E^\prime I^-(p,0|0,1,1,1) +EI^-(p,0|1,0,1,1)  -(Q^2+8E E^\prime)
\nonumber\\
&& \hspace{3.8cm} \times\, I^-(p,0|1,1,1,0) +E(Q^2+8E E^\prime)I^-(p,0|1,1,1,1) 
\nonumber\\
&& \hspace{3.8cm} -\, (E+E^\prime)I^-(p,0|1,1,0,1) \bigg\}\,,
\\
\label{b-TPE}
\delta^{(b)}_{\rm xbox}(Q^2) &=& {2{\mathbb R}e \sum\limits_{spins}
\bigg[\mathcal{M}^{(0)*}_{\gamma} {\mathcal M}^{(b)}_{\rm xbox}\bigg]}\bigg/
{\sum\limits_{spins}\left|\mathcal{M}^{(0)}_\gamma\right|^2}
\nonumber\\
&=& -\,4 \pi\alpha \left[\frac{Q^2}{Q^2+4E E^\prime}\right]  
{\mathbb R}e\left\{\int \frac{{\rm d}^4 k}{{(2 \pi)}^4 i}
\frac{{\rm Tr}\left[(\slashed{p}+m_l)\,\slashed{v}\,(\slashed{p}^{\, \prime}+m_l)\,\slashed{v}\, 
(\slashed{p}+\slashed{k}-\slashed{Q}+m_l)\,\slashed{v}\right]}{(k^2+i0)\,
[(Q-k)^2+i0]\,(k^2 + 2k\cdot p^\prime+i0)\,(v\cdot k+i0)}\right\}\,,
\nonumber\\
&=& -\,8 \pi \alpha \left[\frac{Q^2}{Q^2+4 E E^\prime}\right] 
{\mathbb R}e \bigg\{EI^+(p^\prime,0|0,1,1,1)  +E^\prime I^+(p^\prime,0|1,0,1,1) +(Q^2+8E E^\prime)
\nonumber\\
&& \hspace{3.8cm} \times\, I^+(p^\prime,0|1,1,1,0)  +E^\prime (Q^2+8E E^\prime) I^+(p^\prime,0|1,1,1,1) 
\nonumber\\
&& \hspace{3.8cm} -\, (E+E^\prime)I^+(p^\prime,0|1,1,0,1)\bigg\}\,,
\\
\label{a1-TPE}
\delta^{(a_1)}_{\rm box}(Q^2) &=& {2{\mathbb R}e \sum\limits_{spins}
\bigg[\mathcal{M}^{(1)*}_{\gamma} {\mathcal M}^{(a)}_{\rm box}\bigg]}\bigg/
{\sum\limits_{spins}\left|\mathcal{M}^{(0)}_\gamma\right|^2}
\nonumber\\
&=& -\,\frac{2\pi\alpha}{ M} \left[\frac{Q^2}{Q^2+4E E^\prime}\right] \left(Q_{\alpha}+v_{\alpha} \frac{Q^2}{2M}\right)
\nonumber\\
&& \times\, {\mathbb R}e \left\{\int \frac{{\rm d}^4 k}{{(2 \pi)}^4 i}
\frac{{\rm Tr}\left[(\slashed{p}+m_l)\,\gamma^{\alpha}\,(\slashed{p}^{\prime}+m_l)\,\slashed{v}\, 
(\slashed{p}-\slashed{k}+m_l)\,\slashed{v}\,\right]}{(k^2+i0)\,
[(Q-k)^2+i0]\,(k^2-2 k\cdot p+i0)\,(v\cdot k+i0)}\right\}
= \frac{Q^2}{4M^2} \delta^{(a)}_{\rm box}(Q^2)\,,
\\
\nonumber\\
\label{b1-TPE}
\delta^{(b_1)}_{\rm xbox}(Q^2) &=& {2{\mathbb R}e \sum\limits_{spins}
\bigg[\mathcal{M}^{(1)*}_{\gamma} {\mathcal M}^{(b)}_{\rm xbox}\bigg]}\bigg/
{\sum\limits_{spins}\left|\mathcal{M}^{(0)}_\gamma\right|^2}
\nonumber\\
&=& -\,\frac{ 2\pi\alpha }{M} \left[\frac{Q^2}{Q^2+4E E^\prime}\right] \left(Q_{\alpha}+v_{\alpha} \frac{Q^2}{2M}\right) 
\nonumber\\
&& \times\, {\mathbb R}e \left\{\int \frac{{\rm d}^4 k}{{(2 \pi)}^4 i}
\frac{{\rm Tr}\left[(\slashed{p}+m_l)\,\gamma^{\alpha}\,(\slashed{p}^{\, \prime}+m_l)\,\slashed{v}\, 
(\slashed{p}+\slashed{k}-\slashed{Q}+m_l)\,\slashed{v}\right]}{(k^2+i0)\,
[(Q-k)^2+i0]\,(k^2 + 2k\cdot p^\prime+i0)\,(v\cdot k+i0)}\right\}
= \frac{Q^2}{4M^2} \delta^{(b)}_{\rm xbox}(Q^2)\,,
\\
\nonumber\\
\label{c-TPE}
\delta^{(c)}_{\rm box}(Q^2) &=& {2{\mathbb R}e \sum\limits_{spins}
\bigg[\mathcal{M}^{(0)*}_{\gamma} \widetilde{\mathcal M}^{(c)}_{\rm box}\bigg]}\bigg/
{\sum\limits_{spins}\left|\mathcal{M}^{(0)}_\gamma\right|^2} 
\nonumber \\
&=& -\,\frac{2 \pi\alpha}{M}\left[\frac{Q^2}{Q^2+4EE^\prime}\right]  
{\mathbb R}e \left\{\int \frac{{\rm d}^4 k}{{(2 \pi)}^4 i}
\frac{{\rm Tr}\left[(\slashed{p}+m_l)\,\slashed{v}\,(\slashed{p}^{\, \prime}+m_l)\,\slashed{v}\, 
(\slashed{p}-\slashed{k}+m_l)\, \slashed{k}\right]}{(k^2+i0)\,
\left[(Q-k)^2+i0\right]\,(k^2-2 k\cdot p+i0)\,(v\cdot k+i0)}\right\}
\nonumber \\
&=& \frac{4 \pi \alpha }{M} Q^2\, {\mathbb R}e \left\{I^-(p,0|1,1,0,1)\right\}\,,
\end{eqnarray}
\begin{eqnarray}
\label{d-TPE}
\delta^{(d)}_{\rm xbox}(Q^2) \!&=&\! {2{\mathbb R}e \sum\limits_{spins}
\bigg[\mathcal{M}^{(0)*}_{\gamma} \widetilde{\mathcal M}^{(d)}_{\rm xbox}\bigg]}\bigg/
{\sum\limits_{spins}\left|\mathcal{M}^{(0)}_\gamma\right|^2}
\nonumber \\
&=& -\, \frac{2 \pi\alpha}{M}\left[\frac{Q^2}{Q^2+4EE^\prime}\right]  
{\mathbb R}e \left\{\int \frac{{\rm d}^4 k}{{(2 \pi)}^4 i}\frac{{\rm Tr}\left[(\slashed{p}+m_l)\,\slashed{v}\,
(\slashed{p}^{\, \prime}+m_l)\,\slashed{k}\,(\slashed{p}+\slashed{k}-\slashed{Q}+m_l)\,\slashed{v}\right]}{(k^2+i0)\,
\left[(Q-k)^2+i0\right]\,(k^2+2 k\cdot p^{\prime}+i0)\,(v\cdot k+i0)}\right\}
\nonumber\\
&=& -\,\frac{4\pi\alpha}{M} Q^2\, {\mathbb R}e \left\{I^-(p,0|1,1,0,1)\right\}\,,
\\
\nonumber\\
\label{e-TPE}
\delta^{(e)}_{\rm box}(Q^2) &=& {2{\mathbb R}e \sum\limits_{spins}
\bigg[\mathcal{M}^{(0)*}_{\gamma} \widetilde{\mathcal M}^{(e)}_{\rm box}\bigg]}\bigg/
{\sum\limits_{spins}\left|\mathcal{M}^{(0)}_\gamma\right|^2}
\nonumber\\
&=&\! -\, \frac{2 \pi\alpha}{M}\left[\frac{Q^2}{Q^2+4EE^\prime}\right]  
{\mathbb R}e \left\{\int \frac{{\rm d}^4 k}{{(2 \pi)}^4 i}\frac{{\rm Tr}\left[(\slashed{p}+m_l)\,\slashed{v}\,
(\slashed{p}^{\, \prime}+m_l)\,(\slashed{k} +\slashed{Q})\, (\slashed{p}-\slashed{k}+m_l)\, \slashed{v}\right]}{(k^2+i0)\,
\left[(Q-k)^2+i0\right]\,(k^2-2 k\cdot p+i0)\,(v\cdot k+i0)}\right\}
\nonumber\\
&=& \frac{4\pi\alpha}{M}\left[\frac{Q^2}{Q^2+4 E E^\prime}\right] {\mathbb R}e \bigg\{4E^2 Q^2 I^-(p,0|1,1,1,1) 
-4 E^2 I^-(p,0|1,0,1,1)-(Q^2-4E^2) 
\nonumber\\
&& \hspace{3.6cm} \times\, I^-(p,0|1,1,0,1)  +2(Q^2+2 E E^\prime) I^-(p,0|0,1,1,1) 
\nonumber\\
&& \hspace{3.6cm} -\,4E Q^2 I^-(p,0|1,1,1,0) \bigg\}\,,
\\
\nonumber\\
\label{f-TPE}
 \delta^{(f)}_{\rm xbox}(Q^2) &=& {2{\mathbb R}e \sum\limits_{spins}
\bigg[\mathcal{M}^{(0)*}_{\gamma} \widetilde{\mathcal M}^{(f)}_{\rm xbox}\bigg]}\bigg/
{\sum\limits_{spins}\left|\mathcal{M}^{(0)}_\gamma\right|^2}
\nonumber\\
&=& -\,\frac{2 \pi\alpha}{M}\left[\frac{Q^2}{Q^2+4EE^\prime}\right]  
{\mathbb R}e \left\{\int \frac{{\rm d}^4 k}{{(2 \pi)}^4 i}\frac{{\rm Tr}\left[(\slashed{p}+m_l)\,\slashed{v}\,
(\slashed{p}^{\, \prime}+m_l)\,\slashed{v}\, (\slashed{p}+\slashed{k}-\slashed{Q}+m_l)\, 
(\slashed{k}+\slashed{Q})\right]}{(k^2+i0)\,\left[(Q-k)^2+i0\right]\,(k^2+2 k\cdot p^\prime+i0)\,(v\cdot k+i0)}\right\}   
\nonumber\\
&=& -\,\frac{4\pi\alpha}{M}\left[\frac{Q^2}{Q^2+4 E E^\prime}\right] {\mathbb R}e 
\bigg\{4{E^\prime}^2 Q^2 I^+(p^\prime,0|1,1,1,1)  -4 {E^\prime}^2 I^+(p^\prime,0|1,0,1,1) -(Q^2-4{E^\prime}^2) 
\nonumber\\
&& \hspace{3.9cm}  \times\,I^+(p^\prime,0|1,1,0,1)+ 2(Q^2+2 EE^\prime)I^+(p^\prime,0|0,1,1,1) 
\nonumber\\
&& \hspace{3.9cm} +\, 4E^\prime Q^2 I^+(p^\prime,0|1,1,1,0)\bigg\}\,, 
\\
\nonumber\\
\label{g-TPE}
\delta^{(g)}_{\rm box}(Q^2) &=& {2{\mathbb R}e \sum\limits_{spins}
\bigg[\mathcal{M}^{(0)*}_{\gamma} \widetilde{\mathcal M}^{(g)}_{\rm box}\bigg]}\bigg/
{\sum\limits_{spins}\left|\mathcal{M}^{(0)}_\gamma\right|^2}
\nonumber\\
&=& \frac{2\pi\alpha}{M}\left[\frac{Q^2}{Q^2+4EE^\prime}\right] 
{\mathbb R}e \left\{\int \frac{{\rm d}^4 k}{{(2 \pi)}^4 i}\frac{{\rm Tr}[(\slashed{p}+m_l)\slashed{v}\,
(\slashed{p}^{\,\prime}+m_l)\,\slashed{v}\, (\slashed{p}-\slashed{k}+m_l)\, 
\slashed{v}]}{\left[(Q-k)^2+i0\right]\,(k^2-2 k\cdot p+i0)\,(v\cdot k+i0)^2}\right\}
\nonumber\\
&=& -\,\frac{4 \pi \alpha }{M}\left[\frac{Q^2}{Q^2+4 E E^\prime}\right] {\mathbb R}e 
\bigg\{(Q^2+8 E E^\prime)I^-(p,0|0,1,1,1) -\, \left[EQ^2+8E^2E^\prime+2m^2_l(E+E^\prime)\right] 
\nonumber\\
&& \hspace{3.9cm} \times\, I^{-}(p,0|0,1,1,2) -2(E^\prime p + E p^\prime)\cdot T^{-}_1(p,0|0,1,1,2) \bigg\} \,, \qquad \text{and}
\\
\nonumber\\
\label{h-TPE}
\delta^{(h)}_{\rm xbox}(Q^2) &=& {2{\mathbb R}e \sum\limits_{spins}
\bigg[\mathcal{M}^{(0)}_{\gamma} \widetilde{\mathcal M}^{(h)}_{\rm xbox}\bigg]}\bigg/
{\sum\limits_{spins}\left|\mathcal{M}^{(0)}_\gamma\right|^2}
\nonumber\\
&=& \frac{2 \pi\alpha}{M}\left[\frac{Q^2}{Q^2+4EE^\prime}\right]  
{\mathbb R}e \left\{\int \frac{{\rm d}^4 k}{{(2 \pi)}^4 i}\frac{{\rm Tr}[(\slashed{p}+m_l)\slashed{v}\,
(\slashed{p}^{\, \prime}+m_l)\,\slashed{v}\, (\slashed{p}+\slashed{k}-\slashed{Q}+m_l)\slashed{v}]}
{\left[(Q-k)^2+i0\right]\,(k^2+2 k\cdot p^{\prime}+i0)\,(v\cdot k+i0)^2}\right\}
\nonumber\\
&=& \frac{4 \pi \alpha }{M}\left[\frac{Q^2}{Q^2+4 E E^\prime}\right] {\mathbb R}e 
\bigg\{(Q^2+8 E E^\prime)I^+(p^\prime,0|0,1,1,1) -\left[EQ^2-2E^\prime Q^2-8EE^{\prime 2}\right. 
\nonumber\\
&& \hspace{3.55cm} \left. -\,2m^2_l(E+E^\prime)\right]I^{+}(p^\prime,0| 0,1,1,2) 
-2(E^\prime p + Ep^\prime)\cdot T^{+}_1(p^\prime,0|0,1,1,2) \bigg\}. \qquad\,
\end{eqnarray}
We note that the contribution $\delta^{\rm (seagul)}_{\gamma\gamma}$ from the seagull diagram (i) has already 
been evaluated exactly up to ${\mathcal O}(\alpha/M^2)$ in our previous work~\cite{Goswami:2025zoe}, and hence,
not repeated here. There are ten 3-point scalar irreducible loop-integrals (master-integrals) which appear in
the above expressions, which are generically defined {\it via} the following 4-point scalar reducible 
loop-integrals, see Ref.~\cite{Choudhary:2023rsz}:
\begin{eqnarray}
I^-(p,\omega|n_1,n_2,n_3,n_4) &=& \frac{1}{i} \int \frac{{\rm d}^4k}{(2\pi)^4}
\frac{1}{(k^2+i0)^{n_1} [(k-Q)^2+i0]^{n_2}(k^2-2k\cdot p+i0)^{n_3}(v\cdot k+\omega +i0)^{n_4}}\,,\qquad \text{and}
\nonumber\\
I^+(p^\prime ,\omega|n_1,n_2,n_3,n_4) &=& \frac{1}{i} 
\int \frac{{\rm d}^4k}{(2\pi)^4}
\frac{1}{(k^2+i0)^{n_1} [(k-Q)^2+i0]^{n_2}(k^2+2k\cdot p^\prime+i0)^{n_3}(v\cdot k+\omega +i0)^{n_4}}\,,
\end{eqnarray}
where the indices $n_{1,2,3,4} \in {\mathbb Z}$, and $\omega$ are arbitrary real-valued parameter. In our
case $\omega$ can either be taken as zero, or any other finite quantity, e.g., $\omega\to v\cdot Q=-Q^2/2M$, 
$\omega\to v\cdot p=E$, $\omega\to -v\cdot p^\prime=-E^\prime$, etc., depending on the specific requirement. 
In Ref.~\cite{Choudhary:2023rsz}, the 3-point loop-integrals were evaluated exactly, however, retaining 
terms up to ${\mathcal O}(1/M)$ only. Subsequently, In our more recent SPA-based TPE 
analysis~\cite{Goswami:2025zoe}, several of these results were extended to include contributions up to 
${\mathcal O}(1/M^2)$, consistent with the inclusion of the dynamical NNLO contributions arising from the 
$\nu = 2$ chiral-order Lagrangian. In this work, we systematically account for all possible kinematically 
suppressed corrections of ${\mathcal O}(1/M^2)$. These arise not only from the relevant loop-integrals 
appearing in the LO and NLO fractional contributions to the cross section, Eqs.~\eqref{a-TPE} - 
\eqref{h-TPE}, but also from phase-space kinematics involving the outgoing lepton energy $E^\prime$  An 
example is provided by the recoil expansion of the kinematical pre-factor $Q^2/(Q^2+4E E^\prime)$ that 
enters the TPE corrections derived from the LO cross section, Eq.~\eqref{eq:diff_LO} [also see 
Eq.~\eqref{eq:E_beta_expansions}], namely,  
\begin{eqnarray}
\frac{Q^2}{Q^2+4 E E^\prime} &=& \left[\frac{Q^2}{Q^2+4 E^2}\right] 
- \frac{1}{M} \left[\frac{2 E Q^4}{(Q^2+4 E^2)^2}\right] + \frac{1}{M^2} \left[\frac{4 E^2 Q^6}{(Q^2+4 E^2)^3}\right] 
+ {\mathcal O}\left(\frac{1}{M^3}\right)\,.
\end{eqnarray} 

Regarding the 4-point scaler loop-integrals $I^-(p,0|1,1,1,1)$ and $I^+(p^\prime,0|1,1,1,1)$ appearing in 
Eqs.~\eqref{a-TPE} - \eqref{h-TPE}, their direct evaluation was found to be technically challenging. 
Instead, Ref.~\cite{Choudhary:2023rsz} employs an astute analytical trick, wherein these intricate 
reducible loop-integrals are evaluated by reducing them to a combination of simpler 3-point master-integrals
using a judicious application of partial fraction decomposition and IBP identities to obtain the following 
results:
\begin{eqnarray}
\label{eq:I(1111)}
I^{-} (p,0|1,1,1,1)  &=& 
\frac{1}{Q^2} \left[I^{-} (p,0|1,0,1,1)+ I^{-}(p,0|0,1,1,1) 
-2Z^-\left(\Delta,i\sqrt{-Q^2}/2,m_l,E\right)\right], \qquad \text{and}
\nonumber\\
I^{+} (p^\prime,0|1,1,1,1)  
&=& \frac{1}{Q^2} \left[I^{+}(p^\prime,0|1,0,1,1) + I^{+}(p^\prime,0|0,1,1,1) 
-2Z^+\left(\Delta^\prime,i\sqrt{-Q^2}/2,m_l,-E^\prime\right)\right]\,,
\end{eqnarray}
where
\begin{eqnarray}
Z^-(\Delta,i\sqrt{-Q^2}/2,m_l,E) &=&
\frac{1}{i}\int \frac{{\rm d}^4 k}{(2\pi)^4} 
\frac{1}{\left[(k+\Delta)^2-\frac{1}{4}Q^2+i0\right]\,(k^2-m^{2}_l+i0)\,\left(v\cdot k+E+i0\right)}\,, \quad\, \text{and}
\nonumber\\ 
Z^+(\Delta^\prime,i\sqrt{-Q^2}/2,m_l,-E^\prime) &=& 
\frac{1}{i}\int \frac{{\rm d}^4 k}{(2\pi)^4} 
\frac{1}{\left[(k+\Delta^\prime)^2-\frac{1}{4}Q^2+i0\right]\,(k^2-m^{2}_l+i0)\,\left(v\cdot k-E^\prime+i0 \right)}\,,
\end{eqnarray}
are the two additional 3-point master-integrals expressed in term of the four-vectors $\Delta_\mu$
and $\Delta^\prime_\mu$, defined as $\Delta_\mu=-\Delta^\prime_\mu=(p-Q/2)_\mu$. The \underline{exact} 
analytical expressions for these two master-integrals, truncated at order $1/M^2$, are provided in 
Appendix~A. In doing so we extend to next order the original ${\mathcal O}(1/M)$ results for the $Z^\pm$ 
integrals obtained in our previous \underline{exact} TPE analysis~\cite{Choudhary:2023rsz}.

Regarding the fractional contributions $\delta^{(g)}_{\rm box}$ and $\delta^{(h)}_{\rm xbox}$ arising 
from the NLO TPE diagrams (g) and (h), these terms additionally involve two 3-point tensor integrals, 
defined as
\begin{eqnarray}
T^{-\mu}_1(p,0|0,1,1,2) &=& \frac{1}{i} \int \frac{{\rm d}^4k}{(2\pi)^4}
\frac{(k-p)^\mu}{[(k-Q)^2+i0]\, (k^2-2k\cdot p+i0)\, (v\cdot k +i0)^{2}}\,, \qquad \text{and} 
\nonumber\\
T^{+\mu}_1(p^\prime,0|0,1,1,2) &=& \frac{1}{i} \int \frac{{\rm d}^4k}{(2\pi)^4}
\frac{(k+p^\prime)^\mu}{[(k-Q)^2+i0]\, (k^2+2k\cdot p^\prime+i0)\,(v\cdot k +i0)^{2}}\,.
\end{eqnarray}
At the targeted ${\mathcal O}(1/M^2)$ accuracy, it suffices to evaluate these loop-integrals up to 
${\mathcal O}(1/M)$, since the expressions for $\delta^{(g),(h)}_{\rm box}$ carry an overall $1/M$ 
pre-factor [see Eqs.~\eqref{g-TPE} and \eqref{h-TPE}]. The required ${\mathcal O}(1/M)$ expressions for
these integrals were already derived analytically in Ref.~\cite{Choudhary:2023rsz} [cf. Eqs.~(B30) and 
(B31) of Appendix~B therein], which we directly adopt in this work. 

As noted by Choudhary {\it et al.}~\cite{Choudhary:2023rsz}, the LO TPE box (a) and crossed-box (b) 
diagrams are the only ones whose fractional contributions to the elastic cross section, 
$\delta_{\rm box}^{(a)}$, Eq.~\eqref{a-TPE}, and $\delta_{\rm xbox}^{(b)}$, Eq.~\eqref{b-TPE}, contain 
infrared (IR) divergences. These divergences originate from the IR-singular behavior of the 
loop-integrals $I^{-}(p,0|1,0,1,1)$ and $I^{+}(p^\prime,0|1,0,1,1)$. However, it was shown that when the
two contributions are combined in the sum $\delta^{(ab)}_{\gamma\gamma}$, the LO [i.e., 
${\mathcal O}(\alpha M^0)$] IR-divergent terms cancel, leaving only a residual, kinematically 
suppressed ${\mathcal O}(\alpha/M)$ divergence consistent with NLO contributions. These NLO IR-divergent
terms are exactly canceled by corresponding divergences arising from the inclusion of the charge-odd 
lepton-proton interference bremsstrahlung contributions, as explicitly demonstrated in our earlier work
in Ref.~\cite{Talukdar:2020aui} and subsequently in Ref.~\cite{Das:2025jfh}. {\it En passant}, it is 
worth noting that the expressions $\delta_{\rm box}^{(e)}$, Eq.~\eqref{e-TPE}, and 
$\delta_{\rm xbox}^{(f)}$, Eq.~\eqref{f-TPE}, corresponding to the NLO TPE box (e) and crossed-box (f) 
diagrams, also contain terms involving the same IR-divergent integrals $I^{-}(p,0|1,0,1,1)$ and 
$I^{+}(p^\prime,0|1,0,1,1)$. In this case, however, both get canceled by the accompanying 4-point 
loop-functions $I^{-}(p,0|1,1,1,1)$ and $I^{+}(p^\prime,0|1,1,1,1)$, which themselves are expressed in 
terms of the same IR-divergent integrals [see Eqs.~\eqref{eq:I(1111)}]. 

In what follows we will display the analytical expressions for the finite parts of the TPE contributions 
arising from LO and NLO pairs of box and crossed-box amplitudes (a) - (g), as well as the seagull 
amplitude (i)  [cf. Fig.~\ref{fig:LO_NLO_TPE}], accounting for all kinematically suppressed terms up to 
${\mathcal O}(\alpha/M^2)$ in each case. Particularly in our analytical results at this recoil order, we 
find it legitimate to replace all appearances of $E^\prime$ and $\beta^\prime$ by $E$ and $\beta$, 
respectively, since no further recoil order terms are needed. It is noteworthy that diagrams (a) – (d) 
exhibit substantial cancellations between the box and crossed-box contributions, driven by the underlying 
analytic structure of the corresponding $\pm$ type of loop-integrals in each case. In fact the total 
contribution from the (c) and (d) diagrams vanishes altogether, as was also found in 
Ref.~\cite{Choudhary:2023rsz}, i.e., 
$\delta^{(cd)}_{\gamma\gamma}(Q^2) = \delta_{\rm box}^{(c)}(Q^2) +  \delta_{\rm xbox}^{(d)}(Q^2) = 0\,$. 
This behavior, however, do not pertain to diagrams (e) – (h), which predominantly contribute 
constructively leading to sizable enhancements of the cross section. The finite results are enumerated 
below:\\

\noindent$\bullet$ First, we express the total finite contribution from the LO TPE box (a) and 
crossed-box (b) diagrams as\footnote{Henceforth, an overbar on the fractional TPE contributions 
$\delta_{\gamma\gamma}$ will denote only the finite part, with the IR-divergent terms proportional to
$\left[\frac{1}{\epsilon}-\gamma_E+\ln\bigg(\frac{4\pi \mu^2}{m_l^2}\bigg) \right]$ subtracted according
to the standard $\overline{\rm MS}$ scheme.}:
\begin{eqnarray}
\overline{\delta^{(ab)}_{\gamma\gamma}}(Q^2) &=&  \delta_{\rm box}^{(a)}(Q^2) + \delta_{\rm xbox}^{(b)}(Q^2) 
- \delta^{\rm (box)}_{\rm IR}(Q^2)
\nonumber\\
&\equiv&  \delta^{(0)}_{\gamma\gamma}(Q^2) + \overline{\delta^{(ab;1/M)}_{\gamma\gamma}}(Q^2) 
+ \overline{\delta^{(ab;1/M^2)}_{\gamma\gamma}}(Q^2) + {\mathcal O}\left(\frac{\alpha}{M^3}\right)\,,  \qquad \text{where}
\end{eqnarray}
\begin{eqnarray}
\delta^{\rm (box)}_{\rm IR}(Q^2) &=& 
\frac{\alpha}{\pi\beta}\left[\frac{1}{\epsilon}-\gamma_E+\ln\bigg(\frac{4\pi \mu^2}{m_l^2}\bigg) \right]
\left\{\ln \sqrt{\frac{1+\beta}{1-\beta}}-\frac{\beta}{\beta^\prime}
\ln \sqrt{\frac{1+\beta^\prime}{1-\beta^\prime}}\,  \right\}
\nonumber\\
&=& -\,\frac{\alpha Q^2}{2 \pi M E \beta^2}
\left[\frac{1}{\epsilon}-\gamma_E+\ln\bigg(\frac{4\pi \mu^2}{m_l^2}\bigg) \right] 
\left\{1+\bigg(\beta-\frac{1}{\beta}\bigg) \ln \sqrt{\frac{1+\beta}{1-\beta}}\, \right\} 
\nonumber\\
&& +\,\frac{3\alpha Q^4}{8\pi M^2 E^2 \beta^4}
\left[\frac{1}{\epsilon}-\gamma_E+\ln\bigg(\frac{4\pi \mu^2}{m_l^2}\bigg) \right]
\left\{1-\frac{2}{3}\beta^2 +\bigg(\beta-\frac{1}{\beta}\bigg) \ln \sqrt{\frac{1+\beta}{1-\beta}}\,\right\}  
+ \mathcal{O}\left(\frac{\alpha}{M^3}\right)\,.
\label{eq:delta_IR}
\end{eqnarray}
Here, we have isolated IR divergence using dimensional regularization (DR) by analytically continuing 
the integrals to $D$-dimensional space-time with pole, $\epsilon = (4-D)/2<0$. $\gamma_E=0.577216...$ 
is the Euler-Mascheroni constant, and $\mu$ is an arbitrary subtraction scale. As mentioned, the 
genuine LO [i.e.,  ${\mathcal O}(\alpha M^0)$] terms exhibit significant cancellations between 
$\delta_{\rm box}^{(a)}$ and $\delta_{\rm xbox}^{(b)}$, leaving a residual term solely arising from 
the 3-point master-integral $I(Q|1,1,0,1)\equiv I^{-}(p,0|1,1,0,1)=I^{+}(p^\prime,0|1,1,0,1)$. Unlike 
the SPA approach, as demonstrated in Ref.~\cite{Choudhary:2023rsz}, these terms yield a non-vanishing 
LO TPE contribution akin to the well-known McKinley and Feshbach~\cite{McKinley:1948zz} result:
\begin{eqnarray}
\delta^{(0)}_{\gamma\gamma}(Q^2) = 
32\pi\alpha E \left[\frac{Q^2}{Q^2+4E^2}\right] {\mathbb R}e \left[ I^{(0)}(Q|1,1,0,1) \right] 
= \pi \alpha \frac{\sqrt{{-Q^2}}}{2E}\left[\frac{1}{1+\frac{Q^2}{4 E^2}}\right]\, ,
\label{eq:delta-ab_LO}
\end{eqnarray}
where the loop-function, $I^{(0)}(Q|1,1,0,1)\sim {\mathcal O}(M^0)$, denotes the leading recoil order
component of the loop-integral $I(Q|1,1,0,1)$. Its analytical expression up to ${\mathcal O}(1/M^2)$ 
in the recoil expansion is evaluated in Eq.~\eqref{eq:1101} of Appendix~A. In our previous 
\underline{exact} TPE work of Ref.~\cite{Choudhary:2023rsz} only the ${\mathcal O}(\alpha/M)$ 
contribution to the cross section stemming from the LO TPE diagrams was derived. Here, we reproduce the
same result for completeness: 

\begin{eqnarray}
\overline{\delta^{(ab;1/M)}_{\gamma\gamma}}(Q^2) 
&=& -\, 16\pi\alpha E\, {\mathbb R}e \Bigg\{\overline{\delta^{(1/M)}I^{+}}(p^\prime,0|1,0,1,1) 
+\delta^{(1/M)}I^{-}(p,0|0,1,1,1) + \delta^{(1/M)}I^{+}(p^\prime,0|0,1,1,1)
\nonumber\\
&&\hspace{2.1cm} -\left[\frac{Q^2+8 E^2}{Q^2+4E^2}\right] \bigg(\delta^{(1/M)} Z^{-}(\Delta,i\sqrt{-Q^2}/2,m_l,E) 
\nonumber\\
&&\hspace{4.7cm} +\,\delta^{(1/M)} Z^{+}(\Delta^\prime,i\sqrt{-Q^2}/2,m_l,-E^\prime)\bigg) 
\nonumber\\
&&\hspace{2.1cm}- \left[\frac{2 Q^2}{Q^2+4E^2} \right]\delta^{(1/M)} I(Q|1,1,0,1) \Bigg\}
\nonumber\\
&& +\, 8\pi\alpha\frac{Q^2}{M} {\mathbb R}e \Bigg\{\overline{I^{(0)}}(p,0|1,0,1,1) 
+ \left[\frac{4E^2}{Q^2+4E^2}\right] I^{(0)}(p,0|0,1,1,1) +\left[\frac{Q^2(Q^2-4E^2)}{\left(Q^2+4E^2\right)^2}\right]
\nonumber\\
&&\hspace{2.1cm}  \times\, I^{(0)}(Q|1,1,0,1) -\left[\frac{Q^2+8 E^2}{Q^2+4E^2}\right]Z^{(0)}(\Delta,i\sqrt{-Q^2}/2,m_l,E) \Bigg\} \, . 
\label{eq:delta-ab_NLO}    
\end{eqnarray}
The kinematically suppressed ${\mathcal O}(\alpha/M^2)$ component represents our \underline{new}  
extension of the above IR-finite part, and is given by the following expression:
\begin{eqnarray}   
\label{eq:delta-ab_NNLO} 
\overline{\delta^{(ab;1/M^2)}_{\gamma\gamma}}\!(Q^2) 
&=&\! -\,16\pi\alpha E\, {\mathbb R}e\Bigg\{ 
\overline{\delta^{(1/M^2)}I^{+}}(p^\prime,0|1,0,1,1) +\delta^{(1/M^2)}I^-(p,0|0,1,1,1) +\delta^{(1/M^2)}I^+(p^\prime,0|0,1,1,1) 
\nonumber\\
&&\! \hspace{2.1cm} -\,\left[\frac{2Q^2}{Q^2+4 E^2}\right]\delta^{(1/M^2)}I(Q|1,1,0,1) -\left[\frac{Q^2+8E^2}{Q^2+4 E^2}\right]
\nonumber\\
&&\! \hspace{2.1cm} \times\, \bigg(\delta^{(1/M^2)}Z^-(\Delta,i\sqrt{-Q^2}/2,m_l,E)
+\delta^{(1/M^2)}Z^+(\Delta^\prime,i\sqrt{-Q^2}/2,m_l,-E^\prime)\bigg) \Bigg\}
\nonumber\\
\hspace{2.1cm}&&\! -\, \frac{4\pi\alpha Q^2}{M(Q^2+4E^2)}\, {\mathbb R}e\Bigg\{
2(Q^2+4E^2)\overline{\delta^{(1/M)}I^+}(p^\prime,0|1,0,1,1)  +Q^2\delta^{(1/M)}I^-(p,0|0,1,1,1) 
\nonumber\\
&&\! \hspace{3.2cm} +\,(Q^2+8E^2)\delta^{(1/M)}I^+(p^\prime,0|0,1,1,1) -2\left[\frac{Q^2(Q^2-4 E^2)}{Q^2+4 E^2}\right]
\nonumber\\
&&\! \hspace{3.2cm}  \times\,\delta^{(1/M)}I(Q|1,1,0,1) 
\!-\! \left[\frac{8Q^2E^2}{Q^2+4E^2}\right]\!\delta^{(1/M)}\!Z^-(\Delta,i\sqrt{-Q^2}/2,m_l,E) 
\nonumber\\
&&\! \hspace{3.2cm} -\,2  \left[\frac{Q^4+16Q^2E^2 +32 E^4}{Q^2 +4E^2}\right] 
\delta^{(1/M)} Z^+(\Delta^\prime,i\sqrt{-Q^2}/2,m_l,-E^\prime)\bigg\} 
\nonumber\\
&&\!  -\, \frac{16\pi\alpha Q^6 E}{M^2(Q^2+4E^2)^2}\, {\mathbb R}e\Bigg\{ 
\left[\frac{Q^2-4 E^2}{Q^2+4 E^2}\right] I^{(0)}(Q|1,1,0,1) -I^{(0)}(p,0|0,1,1,1) 
\nonumber\\
&&\! \hspace{3.5cm} +\, Z^{(0)}(\Delta^\prime,i\sqrt{-Q^2}/2,m_l,-E^\prime) 
\Bigg\}
\end{eqnarray}
All the 3-point master-integrals appearing in the above expressions were originally evaluated 
analytically up to ${\mathcal O}(1/M)$ in the work of Choudhary {\it et al.}~\cite{Choudhary:2023rsz}. 
As for the master-integrals, $I^{-}(p,0|0,1,1,1),\,\,\,I^{+}(p^\prime,0|0,1,1,1),\,\,\,
I^{-}(p,0|1,0,1,1)$\\
$I^{+}(p^\prime,0|1,0,1,1),\,\,\,\,I^{-}(p,0|1,1,1,0)$, and $I^{+}(p^\prime,0|1,1,1,0)$, their 
${\mathcal O}(1/M^2)$ extensions were subsequently obtained in the SPA-based TPE work by Goswami 
{\it et al.}~\cite{Goswami:2025zoe}. Whereas, the ${\mathcal O}(1/M^2)$ extensions to the 3-point 
master-integrals $Z^-(\Delta,i\sqrt{-Q^2}/2,m_l,E)$, 
$Z^+(\Delta^\prime,i\sqrt{-Q^2}/2,m_l,-E^\prime)$, $I^-(p,0|1,1,0,1)$, and $I^+(p^\prime,0|1,1,0,1)$, 
are evaluated for the first time in this work and presented in Appendix~A. Note that the functions 
$I^{(0)}$ and $Z^{(0)}$ denote the LO [i.e., ${\mathcal O}(M^0)$] components of the corresponding $I^\pm$
and $Z^\pm$ integrals, respectively (up to an overall sign); see Ref.~\cite{Choudhary:2023rsz} for 
details. In particular, $\overline{I^{(0)}},\,\,\overline{\delta^{(1/M)}I^{+}}(p^\prime,0|1,0,1,1)$, and 
$\overline{\delta^{(1/M^2)}I^{+}}(p^\prime,0|1,0,1,1)$ in the above expressions denote the IR-finite 
parts of the corresponding IR-divergent master-integral $I^{+}(p^\prime,0|1,0,1,1)$, truncated at 
${\mathcal O}(1/M^2)$ (also see Appendix~A of Ref.~\cite{Goswami:2025zoe}), namely,
\begin{eqnarray}
I^+(p^\prime,0|1,0,1,1) &=& \frac{1}{i} 
\int \frac{{\rm d}^4k}{(2\pi)^4}\frac{1}{(k^2+i0)\,(k^2+2k\cdot p^\prime+i0)\,(v\cdot k+i0)}
\nonumber\\
&=& \frac{1}{(4\pi)^2\beta^\prime E^\prime}\Bigg[\left\{\frac{1}{\epsilon}-\gamma_E
+\ln{\left(\frac{4\pi\mu^2}{m_l^2}\right)}\right\}\ln{\sqrt{\frac{1+\beta^\prime}{1-\beta^\prime}}}
-i\pi\left\{\frac{1}{\epsilon}-\gamma_E+\ln{\left(\frac{4\pi \mu^2}{m_l^2}\right)}\right\}\Bigg]
\nonumber\\
&& -\,\overline{I^{(0)}}(p,0|1,0,1,1) +\overline{\delta^{(1/M)}I^+}(p^\prime,0|1,0,1,1) 
+\overline{\delta^{(1/M^2)}I^+}(p^\prime,0|1,0,1,1) 
+ \mathcal{O}\left(\frac{1}{M^3}\right)\,, \qquad 
\end{eqnarray}
where the IR-finite parts, after expressing the outgoing lepton energy $E^\prime$ and velocity 
$\beta^\prime$ in terms of the corresponding incident energy $E$ and velocity $\beta$, are given by
\begin{eqnarray}
\overline{I^{(0)}}(p,0|1,0,1,1) 
&=& \frac{1}{(4\pi)^2\beta E}\Bigg[{\rm Li}_2\left(\frac{2\beta}{1+\beta}\right)
+\ln^2{\sqrt{\frac{1+\beta}{1-\beta}}}\,\,\Bigg]\,,
\\
\overline{\delta^{(1/M)}I^+}(p^\prime,0|1,0,1,1) 
&=& \frac{Q^2}{2(4\pi)^2 M E^2 \beta^3}\Bigg[{\rm Li}_2\left(\frac{2\beta}{1+\beta}\right)
+\ln^2{\sqrt{\frac{1+\beta}{1-\beta}}} -2\ln{\sqrt{\frac{1+\beta}{1-\beta}}}\,\,\Bigg]\,, \,\, \text{and}
\\
\overline{\delta^{(1/M^2)}I^+}(p^\prime,0|1,0,1,1) 
&=& -\,\frac{Q^4}{8(4\pi)^2 M^2 E^3 \beta^5}\Bigg[2\beta +(3-\beta^2)\,{\rm Li}_2\left(\frac{2\beta}{1+\beta}\right) 
+(3-\beta^2)\ln^2{\sqrt{\frac{1+\beta}{1-\beta}}} 
\nonumber\\
&&\hspace{3.2cm} +\,2(\beta^2-4)\ln{\sqrt{\frac{1+\beta}{1-\beta}}}\,\, \Bigg]\,.
\end{eqnarray}
The function $\rm{Li}_2$ in the above equations represents the standard di-logarithm or Spence function 
defined as
\begin{eqnarray}
{\rm Li}_2(z) =- \int_0^z {\rm d}t\, \frac{\ln(1-t)}{t}\,,\quad  \forall z\in {\mathbb C}\,.
\label{eq:Li2}
\end{eqnarray}

\noindent$\bullet$ 
Second, the finite part of the contribution arising from the interference of the NLO OPE amplitude 
$M^{(1)}_\gamma$ with the LO box (a) and crossed-box (b) TPE amplitudes, after subtracting the residual 
${\mathcal O}(\alpha/M^3)$ IR divergence, is given by
\begin{eqnarray}
\overline{\delta^{(a_1b_1)}_{\gamma\gamma}}(Q^2) =  \delta_{\rm box}^{(a_1)}(Q^2) +  \delta_{\rm xbox}^{(b_1)}(Q^2) 
- \frac{Q^2}{4M^2}\delta^{\rm (box)}_{\rm IR}(Q^2) =  \overline{\delta^{(a_1b_1;1/M^2)}_{\gamma\gamma}}(Q^2)
+ {\mathcal O}\left(\frac{\alpha}{M^3}\right)\,,
\end{eqnarray}
where
\begin{eqnarray}
\overline{\delta^{(a_1b_1;1/M^2)}_{\gamma\gamma}}(Q^2) &=&  \frac{Q^2}{4M^2} \delta^{(0)}_{\gamma\gamma}(Q^2) 
\nonumber\\
&=& -\,\frac{\pi\alpha (-Q^2)^{3/2}}{8M^2 E}\left[\frac{1}{1+\frac{Q^2}{4 E^2}}\right]\,. 
\label{eq;3.42} 
\end{eqnarray}
In other words, since the leading TPE IR divergence scales as 
$\delta^{\rm (box)}_{\rm IR}\sim{\mathcal O}(\alpha/M)$ [see Eq.~\eqref{eq:delta_IR}], no new 
$\mathcal{O}(\alpha/M^2)$ divergence is generated by the contribution $\delta^{(a_1b_1)}_{\gamma\gamma}$,
other than those terms originating from the $1/M$-recoil expansion of $\delta^{\rm (box)}_{\rm IR}$ itself
originating from the contribution $\overline{\delta^{(ab)}_{\gamma\gamma}}$ already considered. \\   

\noindent$\bullet$ 
Third, the largest corrections originate from the NLO TPE box (e) and crossed-box (f) diagrams, 
whose contributions add constructively and may be expressed as
\begin{eqnarray}
\delta^{(ef)}_{\gamma\gamma}(Q^2) &=&  \delta_{\rm box}^{(e)}(Q^2) +  \delta_{\rm xbox}^{(f)}(Q^2) 
\nonumber\\
&\equiv&  \delta^{(ef;1/M)}_{\gamma\gamma}(Q^2) + \delta^{(ef;1/M^2)}_{\gamma\gamma}(Q^2) 
+ {\mathcal O}\left(\frac{\alpha}{M^3}\right)\,,  
\end{eqnarray}
where the ${\mathcal O}(\alpha/M)$ component $\delta^{(ef;1/M)}_{\gamma\gamma}$, previously derived 
in Ref.~\cite{Choudhary:2023rsz}, is reproduced below for completeness. Whereas, the kinematically 
suppressed ${\mathcal O}(\alpha/M^2)$ component $\delta^{(ef;1/M^2)}_{\gamma\gamma}$ constitutes
the necessary extension of that result in the present work. Thus, we have
\begin{eqnarray}
\label{eq:delta-ef_NLO}
\delta^{(ef;1/M)}_{\gamma\gamma}(Q^2)
&=& \frac{16\pi\alpha Q^2}{M (Q^2+4E^2)}\, {\mathbb R}e \Bigg\{(Q^2+4E^2) I^{(0)}(p,0|0,1,1,1) -2Q^2 EI(Q|1,1,1,0) 
\nonumber \\
&&\hspace{3.2cm}  -\, 4E^2 Z^{(0)}(\Delta,i\sqrt{-Q^2}/2,m_l,E) \Bigg\} \,,
\end{eqnarray}
and
\begin{eqnarray}
\delta^{(ef;1/M^2)}_{\gamma\gamma}(Q^2) 
&=&  \frac{8\pi\alpha Q^2}{M(Q^2+4E^2)}\, {\mathbb R}e \Bigg\{
(Q^2+4E^2)\bigg( \delta^{(1/M)}I^-(p,0|0,1,1,1)- \delta^{(1/M)}I^+(p^\prime,0|0,1,1,1)\bigg)
\nonumber\\
&&\hspace{2.9cm} -\,4E^2 \bigg(\delta^{(1/M)}Z^-(\Delta,i\sqrt{-Q^2}/2,m_l,E) 
\nonumber\\
&&\hspace{4.1cm} -\,\delta^{(1/M)}Z^+(\Delta^\prime,i\sqrt{-Q^2}/2,m_l,-E^\prime)\bigg) \Bigg\}
\nonumber\\
&& -\,\frac{8\pi \alpha Q^4}{M^2 (Q^2+4E^2)^2}\, {\mathbb R}e \bigg\{
2(Q^2+4E^2)E I^{(0)}(Q|1,1,0,1) +Q^2(Q^2-4E^2) I^{(0)}(p,0|1,1,1,0)
\nonumber\\
&&\hspace{3.5cm} +\, 4Q^2 E Z^{(0)}(\Delta,i\sqrt{-Q^2}/2,m_l,E) \bigg) \Bigg\}\,.
\label{eq:delta-ef_NNLO}
\end{eqnarray} 

\noindent$\bullet$ 
Fourth, the total contribution from the NLO TPE box (g) and crossed-box (h) 
diagrams can be expressed as follows:
\begin{eqnarray}
\delta^{(gh)}_{\gamma\gamma}(Q^2) &=&  \delta_{\rm box}^{(g)}(Q^2) +\delta_{\rm xbox}^{(h)}(Q^2) 
\nonumber\\
&=& \delta^{(gh;1/M)}_{\gamma\gamma}(Q^2) + \delta^{(gh;1/M^2)}_{\gamma\gamma}(Q^2) 
+ {\mathcal O}\left(\frac{\alpha}{M^3}\right)\,,  
\end{eqnarray}
where the ${\mathcal O}(\alpha/M)$ component $\delta^{(gh;1/M)}_{\gamma\gamma}$, previously 
derived in Ref.~\cite{Choudhary:2023rsz}, is reproduced below for completeness, namely,
\begin{eqnarray}
\label{eq:delta-gh_NLO}
\delta^{(gh);1/M}_{\gamma\gamma}(Q^2) 
&=& -\,\frac{4\pi\alpha Q^2}{M(Q^2+4E^2)}\, {\mathbb R}e
\Bigg\{ 2\left[Q^2\left(1+\frac{1}{\beta^2}\right)+8E^2\right] I^{(0)}(p,0|0,1,1,1) 
\nonumber \\
&&\hspace{3.5cm} +\,\left[\frac{Q^2+4E^2\beta^2}{E\beta^2}\right]\left(I(Q|0,1,0,2) - I^{(0)}(p,0|0,0,1,2)\right) \Bigg\}
\nonumber\\
&& -\,\frac{2\pi\alpha Q^4}{M(Q^2+4E^2)} \left[Q^2 \left(1+\frac{2}{\beta^2}\right)+8E^2\right] 
{\mathbb R}e \left[I^{(0)}(p,0|0,1,1,2)\right]
\nonumber\\
&& -\,\frac{\alpha Q^2}{\pi M E\beta^3}\left[\ln\sqrt{\frac{1+\beta}{1-\beta}}-\beta\right]  \, . 
\end{eqnarray}
Here, the function $I^{(0)}(p,0|0,0,1,2)$ denotes the ${\mathcal O}(M^0)$ component of 2-point
master-integrals $I^-(p,0|0,0,1,2)$ and $I^+(p^\prime,0|0,0,1,2)$, as evaluated in 
Ref.~\cite{Choudhary:2023rsz}. The 2-point master-integral $I(Q|0,1,0,2)$, which is also 
computed in the same reference, contains the same UV-divergent pieces as 
$I^{(0)}(p,0|0,0,1,2)$. These divergences cancel identically when the two contributions are 
combined in the above expression, yielding a finite result. The term 
$\delta^{(gh;1/M^2)}_{\gamma\gamma}$ denotes the ${\mathcal O}(\alpha/M^2)$ extension of the 
above results obtained in the present work, which, for brevity, is presented in the unexpanded 
form in terms of the relevant loop-functions as
\begin{eqnarray}
\label{eq:delta-gh_NNLO}
\delta^{(gh;1/M^2)}_{\gamma\gamma}(Q^2) 
&=&\frac{4\pi\alpha Q^2}{M(Q^2+4E^2)}\, {\mathbb R}e \Bigg\{
2E(Q^2+4 E^2)\bigg(\delta^{(1/M)}I^-(p,0|0,1,1,2)+\delta^{(1/M)}I^+(p^\prime,0|0,1,1,2)\bigg)
\nonumber\\
&&\hspace{2.9cm} +\, \left[Q^2\left(1+\frac{1}{\beta^2}\right)+8 E^2\right] 
\bigg(\delta^{(1/M)}I^+(p^\prime,0|0,1,1,1)
\nonumber\\
&&\hspace{7.1cm} -\,\delta^{(1/M)}I^-(p,0|0,1,1,1)\bigg)
\nonumber\\
&&\hspace{2.9cm} +\,\left(2E+\frac{Q^2}{2E\beta^2}\right) 
\delta^{(1/M)}I^+(p^\prime,0|0,0,1,2) \Bigg\}
\nonumber\\
&+& \frac{\pi\alpha Q^4}{M^2E^2 \beta^4 (Q^2+4 E^2)^2} \Bigg\{
2 \beta^2 E^2 \bigg(32 \beta^2 E^4+(4 E^2 +16E^2 \beta^2 )Q^2+Q^4(1+2\beta^2)\bigg)
\nonumber\\
&&\hspace{3.6cm} \times\,\delta^{(M^0)}I^+(p^\prime,0|0,1,1,2) 
+4 E Q^2 (1-\beta^2)\bigg(Q^2  +4E^2 (1+\beta^2)\bigg)
\nonumber\\
&&\hspace{3.6cm} \times\, I^{(0)}(p,0|0,1,1,1)  +2 E^2 \beta^2 (1-\beta^2) (Q^2-4E^2)
\nonumber\\
&&\hspace{3.6cm}  \times\,\left(I(Q|0,1,0,2) -I^{(0)}(p,0|0,0,1,2)\right)\Bigg\}
\nonumber\\
&+& \frac{2 \pi\alpha Q^8 }{M^2 E \beta^4 (Q^2+ 4E^2)^2} 
\bigg(2 E^2  (2-\beta^4 ) + Q^2 (1-\beta^2)\bigg) {\mathbb R}e\left[I^{(0)}(p,0|0,1,1,2)\right]\,.
\end{eqnarray}
The 3-point master-integrals $I^-(p,0|0,1,1,2)$ and $I^+(p^\prime,0|0,1,1,2)$ needed 
in the above results, contain ``super-leading-order" contributions proportional to the function 
$I^{(0)}(p,0|0,1,1,2)$ that scale as ${\mathcal O}(M)$, as we first demonstrated in 
Ref.~\cite{Choudhary:2023rsz}. The same functions also appears in our recent SPA-based TPE 
analysis~\cite{Goswami:2025zoe}, where a slightly modified notation is introduced to denote the
various $1/M$-recoil order components, which we adopt here, namely,
\begin{eqnarray}
I^-(p,0|0,1,1,2) &=& \frac{1}{i} 
\int \frac{{\rm d}^4k}{(2\pi)^4}\frac{1}{[(k-Q)^2+i0]\,(k^2-2k\cdot p+i0)\,(v\cdot k+i0)^2}
\nonumber\\
&\equiv& M I^{(0)}(p,0|0,1,1,2) +\delta^{(1/M)}I^-(p,0|0,1,1,2)
+ \mathcal{O}\left(\frac{1}{M^2}\right)\,,
\label{eq:I-0112}
\end{eqnarray}
where
\begin{eqnarray}
I^{(0)}(p,0|0,1,1,2) &=& -\,\frac{4}{(4 \pi)^2 Q^2 E \beta}\ln{\sqrt{\frac{1+\beta}{1-\beta}}}\,, \quad \text{and}
\\
\delta^{(1/M)}I^-(p,0|0,1,1,2) &=& -\,\frac{3 Q^2}{(4\pi)^2 M E^3 \beta^4 }\,.
\end{eqnarray} 
Likewise, the expression of the loop-integral $I^+(p^\prime,0|0,1,1,2)\sim {\mathcal O}(M)$ is given as 
\begin{eqnarray}
I^+(p^\prime,0|0,1,1,2) &=& \frac{1}{i} 
\int \frac{{\rm d}^4k}{(2\pi)^4}\frac{1}{[(k-Q)^2+i0]\,(k^2+2k\cdot p^\prime+i0)\,(v\cdot k+i0)^2}
\nonumber\\
&\equiv&\!\!-MI^{(0)}(p,0|0,1,1,2) +\delta^{(M^0)}I^+(p^\prime,0|0,1,1,2)+\delta^{(1/M)}I^+(p^\prime,0|0,1,1,2) 
+ \mathcal{O}\left(\frac{1}{M^2}\right)\,,
\label{eq:I+0112}
\end{eqnarray}
where
\begin{eqnarray}
\delta^{(M^0)}I^+(p^\prime,0|0,1,1,2) 
&=& -\,\frac{2}{(4\pi)^2 E^2 \beta^3}\left[\ln{\sqrt{\frac{1+\beta}{1-\beta}}-\beta}\right]\,, \quad \text{and}
\\
\delta^{(1/M)}I^+(p^\prime,0|0,1,1,2) &=& \frac{Q^2}{2(4\pi)^2 M E^3\beta^5}
\left(-3\beta +(3-\beta^2)\ln\sqrt{{\frac{1+\beta}{1-\beta}}}\,\,\right)\,.
\end{eqnarray}

\noindent$\bullet$ 
Fifth, the total contribution from the NLO TPE seagull diagram (i) is evaluated exactly in our earlier 
SPA-based analysis~\cite{Goswami:2025zoe}:   
\begin{eqnarray}
\delta^{\rm (seagull)}_{\gamma \gamma}(Q^2) \equiv \delta_{\gamma \gamma}^{({\rm seagull};1/M)}(Q^2) 
+\delta_{\gamma \gamma}^{({\rm seagull};1/M^2)}(Q^2) + {\mathcal O}\left(\frac{\alpha}{M^3}\right)\,,
\end{eqnarray}
where the ${\mathcal O}(\alpha/M)$ and the kinematically suppressed ${\mathcal O}(\alpha/M^2)$ 
components are reproduced below:
\begin{eqnarray}
\label{eq:delta-seagull_NLO}
\delta_{\gamma \gamma}^{({\rm seagull};1/M)}(Q^2)
&=&\frac{16\alpha m^2_l E }{\pi M (Q^2+4E^2) \nu_l^2 }
\nonumber\\
&&\times\,\Bigg[\left(\frac{1+\nu_l^2}{2\nu_l}\right) \bigg\{\frac{\pi^2}{3}
+\ln^2{\sqrt{\frac{\nu_l+1}{\nu_l-1}}+{\rm Li}_2\left(\frac{\nu_l-1}{\nu_l+1}\right)}\bigg\}
-\ln\sqrt{-\frac{Q^2}{m_l^2}}\,\,\Bigg]\,,\quad\,\,
\end{eqnarray}
and 
\begin{eqnarray}
\label{eq:delta-seagull_NNLO}
\delta_{\gamma \gamma}^{({\rm seagull};1/M^2)}(Q^2)
&=& -\,\frac{32\alpha m^2_l Q^2 E^2}{\pi M^2 (Q^2+4E^2)^2 \nu^2_l }
\nonumber\\
&&\times\,\Bigg[\left(\frac{1+\nu_l^2}{2\nu_l}\right)\bigg\{\frac{\pi^2}{3}
+\ln^2{\sqrt{\frac{\nu_l+1}{\nu_l-1}}+{\rm Li}_2\left(\frac{\nu_l-1}{\nu_l+1}\right)}\bigg\}
-\ln\sqrt{-\frac{Q^2}{m_l^2}}\,\,\Bigg]
\nonumber\\
&& +\,\frac{\alpha Q^4}{\pi M^2 (Q^2+4E^2) \nu_l^2}
\nonumber\\
&&\times\, \Bigg[\frac{2m^2_l}{Q^2\nu_l}\bigg\{\frac{\pi^2}{3}
+\ln^2{\sqrt{\frac{\nu_l+1}{\nu_l-1}}+{\rm Li}_2\left(\frac{\nu_l-1}{\nu_l+1}\right)}\bigg\}
+2\ln\sqrt{-\frac{Q^2}{m_l^2}}\,\,\Bigg]\,,
\end{eqnarray}
respectively, and where $\nu_l=\sqrt{1-4m^2_l/Q^2}$. The latter correction is also included in the
total ${\mathcal O}(\alpha/M^2)$ TPE contribution $\delta^{\rm (LO+NLO;2)}_{\gamma\gamma}$, as 
considered below. The seagull contribution, which depends primarily on the square of the lepton mass, 
$m_l$, is exceedingly small in the case of electron-proton scattering when compared with the other 
non-vanishing contributions from the NLO diagram.\\

\begin{figure*}[tbp]
\begin{center}
\includegraphics[width=0.48\linewidth]{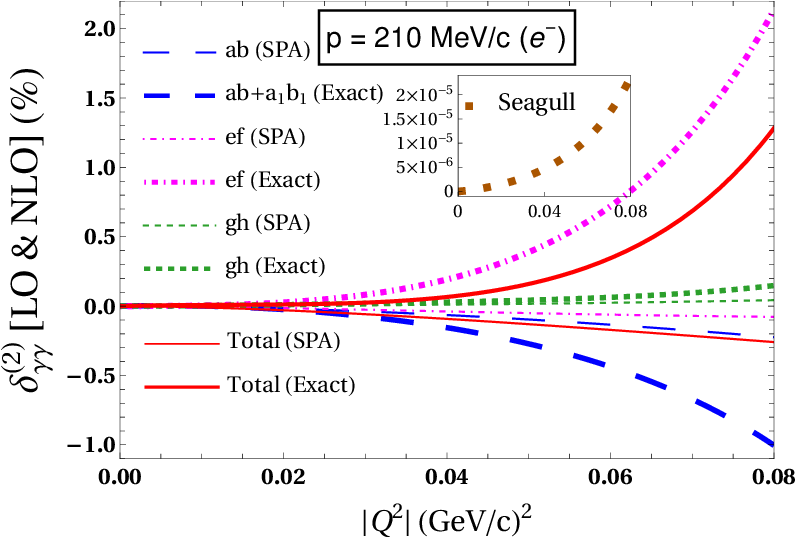}~\quad~\includegraphics[width=0.48\linewidth]{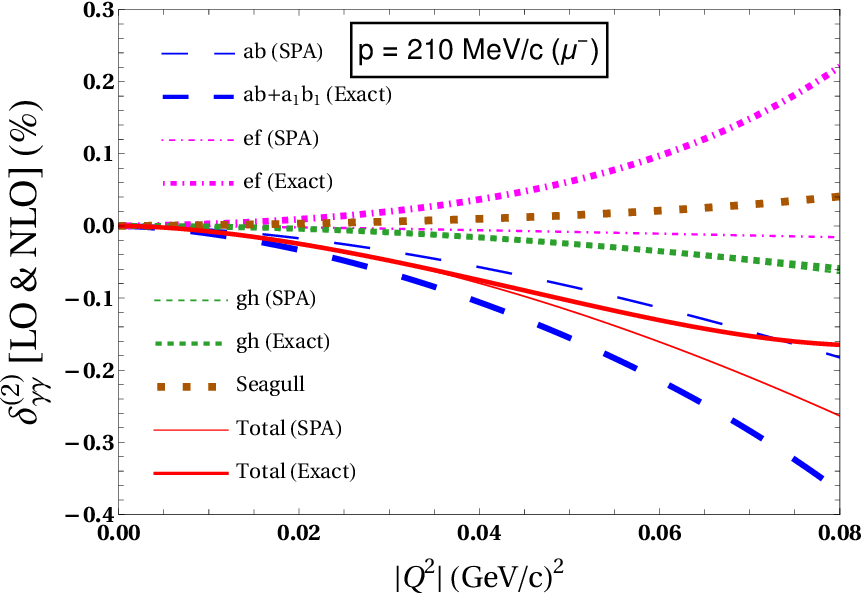}

\vspace{0.5cm}
    
\includegraphics[width=0.48\linewidth]{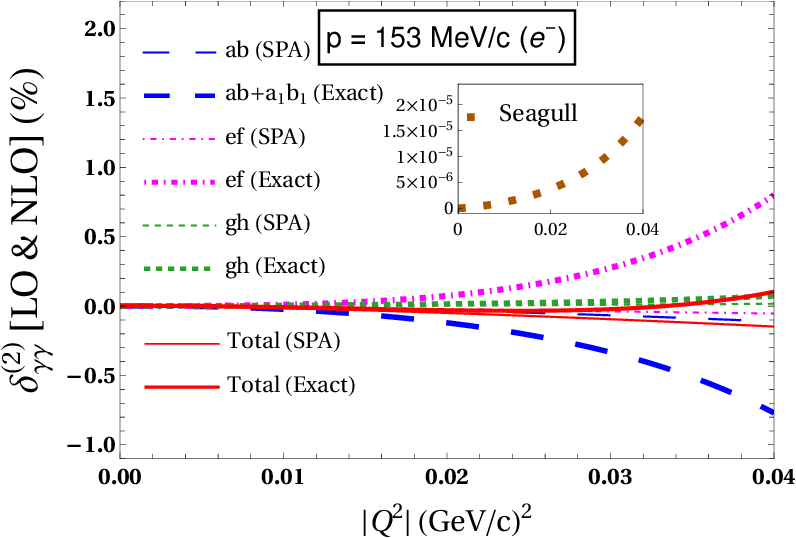}~\quad~\includegraphics[width=0.48\linewidth]{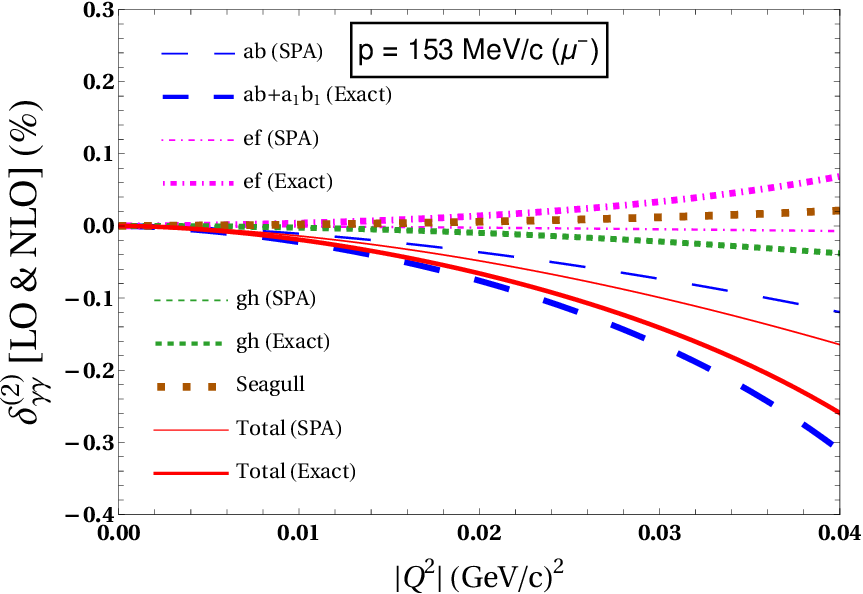} 

\vspace{0.5cm}

\includegraphics[width=0.48\linewidth]{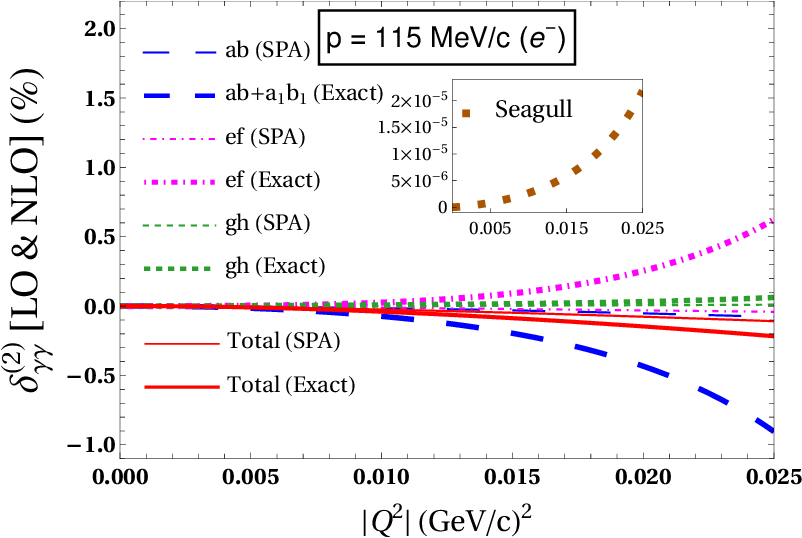}~\quad~\includegraphics[width=0.48\linewidth]{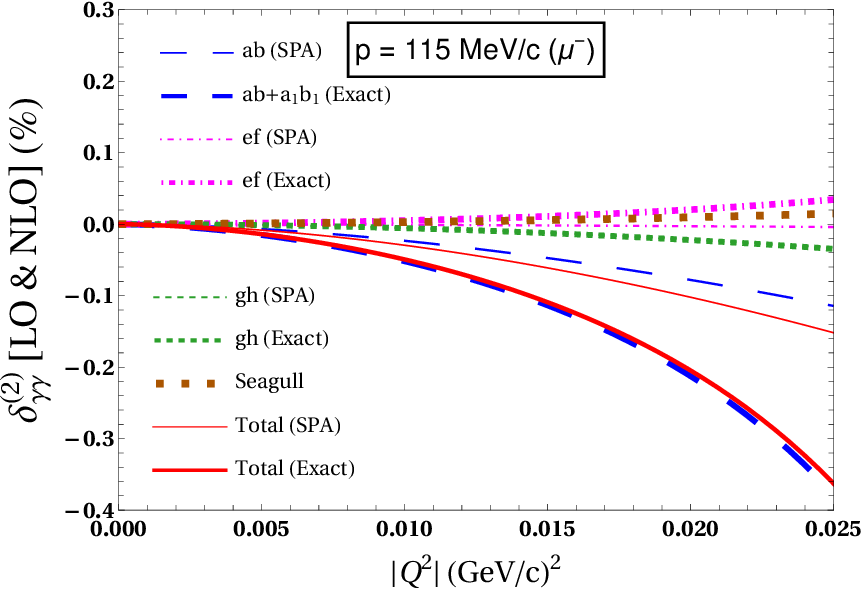} 

\caption{The box and crossed-box pairs of finite fractional \underline{kinematical NNLO} corrections 
arising from the LO and NLO TPE diagrams (a) - (h), are shown together with the contributions arising 
from the NLO seagull diagram (i) and the interference of the NLO Born (OPE) amplitude 
${\mathcal M}^{(1)}_\gamma$ with the LO TPE diagrams (a) and (b), entering the ${\mathcal O}(\alpha/M^2)$
component $\delta^{\rm (LO+NLO;2)}_{\gamma\gamma}$ [see Eq.~\eqref{eq:LO_NLO_TPE_delta2} ] of the 
corrections to the elastic lepton–proton differential cross section. Note that in the figure, we present
the combined {\cal O}($\alpha/M^2$) contribution from the LO pair of TPE diagrams (a) and (b), i.e.,
$\overline{\delta^{(ab;1/M^2)}_{\gamma\gamma}}+\overline{\delta^{(a_1b_1;1/M^2)}_{\gamma\gamma}}$, 
which is labeled as ``ab+${\rm a}_1{\rm b}_1$ (Exact)". For comparison, the corresponding SPA results of
Goswami {\it et al.}~\cite{Goswami:2025zoe} are also displayed. It is noteworthy that the tiny seagull 
contribution (inset panel) has no SPA counterpart. The results for e–p ($\mu$–p) elastic scattering are 
shown in the left (right) panel as functions of the squared four-momentum transfer $|Q^2|$, for three 
MUSE incident lepton beam momenta: 210 MeV/c, 153 MeV/c, and 115 MeV/c. All fractional corrections are 
given relative to the LO OPE differential cross section.}
\label{fig:LO_NLO_Exact_SPA} 
\end{center}
\end{figure*}

To conclude this section we display in Fig.~\ref{fig:LO_NLO_Exact_SPA} the relevant kinematically 
suppressed ${\mathcal O}(\alpha/M^2)$ components resulting from the LO and NLO diagrams entering their
total contribution to the elastic cross section, i.e.,
\begin{eqnarray}
\label{eq:LO_NLO_TPE_delta2}
\delta^{\rm (LO+NLO;2)}_{\gamma\gamma}(Q^2) &\equiv& \overline{\delta^{(ab;1/M^2)}_{\gamma\gamma}}(Q^2) 
+\overline{\delta^{(a_1b_1;1/M^2)}_{\gamma\gamma}}(Q^2) + \delta^{(ef;1/M^2)}_{\gamma\gamma}(Q^2) 
\nonumber\\ && +\, \delta^{(gh;1/M^2)}_{\gamma\gamma}(Q^2) + \delta^{({\rm seagull};1/M^2)}_{\gamma \gamma}(Q^2)\,.
\end{eqnarray}

In the following section we turn to the genuine dynamical TPE contributions that incorporate the 
proton–photon NNLO interaction at $\nu=2$, leading to the finite size modifications of the elastic 
differential cross section. 
%

\section{Dynamical NNLO corrections by including TPE diagrams up to NNLO} 
\label{sec:four}
\begin{figure*}[tbp] 
\begin{center} 
\includegraphics[scale=0.5]{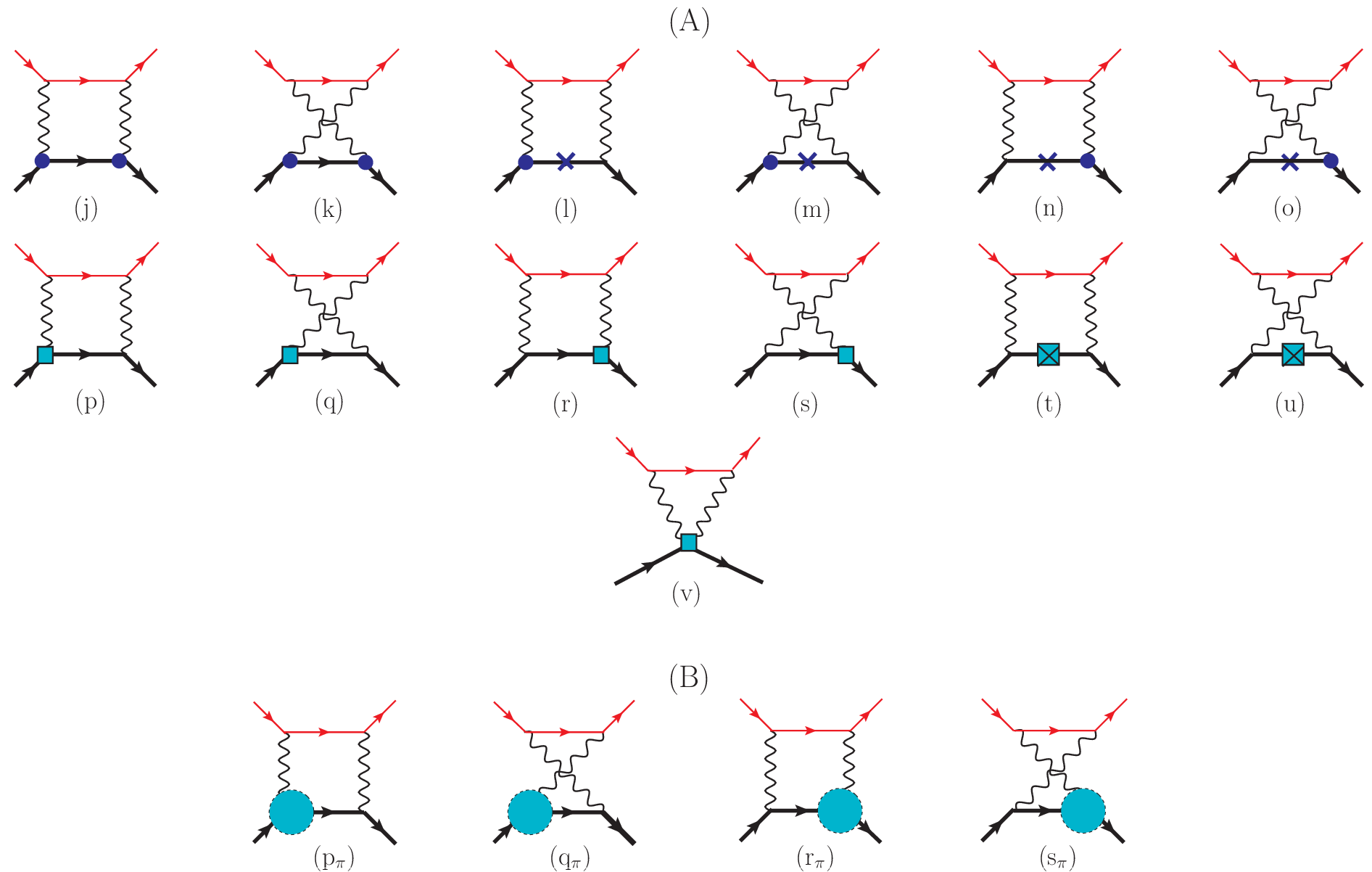}
	\caption{The one-loop NNLO [i.e., $\mathcal{O}(\alpha^2/M^2)$] TPE diagrams contributing to the 
            $\mathcal{O}(\alpha^3/M^2)$ lepton-proton elastic differential cross section, are shown. The 
            thick, thin, and wiggly lines denote the proton, lepton and photon propagators. The small 
            solid circles and square green boxes denote the insertions of NLO ($\nu=1$) and NNLO ($\nu=2$) 
            proton-photon interaction vertices, respectively. The crosses and green-square boxes represent
            the insertions of the ${\mathcal O}(1/M)$ and ${\mathcal O}(1/M^2)$ proton propagator 
            components. Panel (A) contains the true TPE one-loop diagrams, while panel (B) shows the 
            reducible two-loop TPE diagrams involving pion-loops and counter terms (not explicitly shown 
            above, instead see Fig.~1 of Ref.~\cite{Das:2025jfh}) at the proton-photon vertices, as 
            depicted by the large green blobs. These pion-loop insertions renormalize the diagrams (p) 
            – (s) with NNLO proton–photon vertex corrections, effectively reducing them to one-loop graphs
            {\it dressed} with proton form factors. } 
\label{fig:NNLO_TPE} 
\end{center} 
\end{figure*}

We restrict our attention to the four dominant ``reducible" two-loop TPE diagrams at chiral-order 
$\nu = 2$, where one of the loops is a pion-loop at the proton-photon vertex. In this case the pion-loop 
contribution factorizes, leading to a renormalization of the photon–proton vertices and effectively 
reduce the topologies to one-loop TPE diagrams dressed with the proton's Dirac ($F^{p}_{1}$) and Pauli 
($F^{p}_{2}$) form factors~\cite{Bernard:1998gv}, parameterizing the finite-size of the proton (cf. 
Fig.~\ref{fig:NNLO_TPE}). In other words, we here treat only our ``effective one-loop" approximation. 
Below we then extend the NLO results presented in the previous section to incorporate all one-loop 
dynamical NNLO corrections to the TPE arising from the interference between the TPE and the OPE Feynman
diagrams, which explicitly scale as ${\mathcal O}(\alpha/M^2)$. The extension also include insertions 
of the proton's propagator up to ${\mathcal O}(1/M^2)$, as given in Eq.\eqref{eq:p_prop}. To this end, 
the NNLO extension of Eq.~\eqref{eq:delta_TPE_LO} is given as
%
%
\begin{eqnarray}
\left[\frac{{\rm d}\sigma_{el}(Q^2)}{{\rm d}\Omega^\prime_l}\right]^{(\ell^\mp)}_{\rm NNLO} 
&=&\pm\left[\frac{{\rm d}\sigma_{el}(Q^2)}{{\rm d}\Omega^\prime_l}\right]_0
\delta^{\rm (NNLO)}_{\gamma\gamma}(Q^2)\,, 
\label{eq:delta_TPE_NNLO_1}
\end{eqnarray}
where
\begin{eqnarray}
\delta^{\rm (NNLO)}_{\gamma\gamma}(Q^2) 
&=& \frac{2{\mathcal R}e\!\!\sum\limits_{spins}\left[{\mathcal M}_{\gamma}^{(0)*}{\mathcal M}^{\rm (NNLO)}_{\gamma\gamma}
+ {\mathcal M}_{\gamma}^{(1)*}{\mathcal M}^{\rm (NLO)}_{\gamma\gamma} 
+ {\mathcal M}_{\gamma}^{(2)*}{\mathcal M}^{\rm (LO)}_{\gamma\gamma}\right]}{\sum\limits_{spins} 
\left|{\mathcal M}_{\gamma}^{(0)}\right|^2}
\nonumber\\
&=& \frac{2{\mathcal R}e\!\!\sum\limits_{spins}\left[{\mathcal M}_{\gamma}^{(0)*}{\mathcal M}^{\rm (NNLO)}_{\gamma\gamma}
+ {\mathcal M}_{\gamma}^{(2)*}{\mathcal M}^{\rm (LO)}_{\gamma\gamma}\right]}{\sum\limits_{spins} 
\left|{\mathcal M}_{\gamma}^{(0)}\right|^2}+ \mathcal{O}\left(\frac{\alpha}{M^3}\right)\, . 
\label{eq:delta_TPE_NNLO_2}
\end{eqnarray}
Note that the interference of ${\mathcal M}_{\gamma}^{(1)*}$ and ${\mathcal M}^{\rm (NLO)}_{\gamma\gamma}$ 
in the equation above leads to ${\mathcal O}(\alpha/M^3)$ terms, which are beyond our NNLO order 
calculation. They are therefore explicitly omitted in this analysis. As previously mentioned, if
$+\delta^{\rm (NNLO)}_{\gamma\gamma}$ represents to the TPE corrections for lepton-proton scattering, 
$-\delta^{\rm (NNLO)}_{\gamma\gamma}$ will corresponds to antilepton-proton scattering. In other words, 
the overall sign reflects the charge-odd nature of the radiative corrections. The last term 
${\mathcal M}^{(2)}_{\gamma}$ in the above expression denotes the NNLO OPE 
amplitude~\cite{Talukdar:2020aui,Goswami:2025zoe,Das:2025jfh}:
\begin{eqnarray} 
\label{eq:M2}
\mathcal{M}^{(2)}_{\gamma} &=& -\, \frac{e^2}{8M^2Q^2}[{\bar u}_l(p^\prime)\gamma^\mu u_l(p)]
\bigg[\chi^\dagger(p_p^\prime)\, \Big\{\left(2(v\cdot Q)^2-Q^2\right)v_\mu - (v\cdot Q)Q_\mu\Big\} \,\chi(p_p)\bigg]
\nonumber\\
&& -\, \frac{e^2}{ Q^2}[{\bar u}_l(p^\prime)\gamma^\mu u_l(p)]\,
\left[\chi^\dagger(p_p^\prime)\,{\cal V}^{(2)}_\mu \,\chi(p_p)\right]\,, 
\end{eqnarray}
where
\begin{eqnarray}
{\cal V}^{(2)}_\mu &=&   (F^p_1-1)v_\mu+\frac{1}{2M}\Bigg\{(F^p_1-1)\left(Q_\mu+\frac{Q^2}{2M}v_\mu \right)  
+2(F^p_1+F^p_2-1-\kappa_p)\left[S_\mu,S\cdot Q\right]\,\Bigg\}
\nonumber\\
&&\hspace{1.65cm} -\,\frac{Q^2}{8M^2}(F^p_1-2F^p_2-1)v_\mu+{\mathcal O}\left(\frac{1}{M^3}\right)\, , 
\label{eq:V2}
\end{eqnarray}
where the vertex factor ${\cal V}^{(2)}_\mu$  in Eq.~(\ref{eq:V2}) encodes the proton’s hadronic structure
{\it via} its form factors. It renormalizes the NNLO interactions originating from the 
${\mathcal L}^{(2)}_{\pi N}$ chiral Lagrangian.

The term ${\mathcal M}^{\rm (NNLO)}_{\gamma\gamma}$ denotes the sum of the box, the crossed-box and the seagull 
NNLO TPE amplitudes, labeled (j) – (v) in the upper panel (A) of Fig.~\ref{fig:NNLO_TPE}. 
In particular, the diagrams (p) - (s), which involve NNLO insertions at the proton-photon vertices, in addition
receive contributions from the the so-called “form-factor–type” graphs, $({\rm p}_\pi)$ - $({\rm s}_\pi)$, as 
depicted in the lower panel (B) of the same figure. Thus we have  
\begin{eqnarray}
{\mathcal M}^{\rm (NNLO)}_{\gamma\gamma} = {\mathcal M}^{(j)}_{\rm box} + \cdots + {\mathcal M}^{(u)}_{\rm xbox} 
+ {\mathcal M}^{(v)}_{\rm seagull} + \left[{\mathcal M}^{(p_\pi)}_{\rm box} + \cdots 
+ {\mathcal M}^{(s_\pi)}_{\rm xbox} \right]_{\rm form\,factors}\,.   
\end{eqnarray} 
The form factors displayed in Eq.~\eqref{eq:V2} admit the following low-energy Taylor expansion which to
$\mathcal{O}(1/M^2)$ can be expressed in terms of proton's Dirac and Pauli mean square radii, 
$\langle r^2_{1,2}\rangle\sim \mathcal{O}(1/M^{2})$, and the anomalous magnetic moment 
$\kappa_p\sim \mathcal{O}(M^{0})$ (see Ref.~\cite{Bernard:1998gv} for details):  
\begin{eqnarray}
F_1^p(Q^2) = 1+ \frac{Q^2}{6}\langle r_1^2\rangle + {\mathcal O}\left(\frac{1}{M^3}\right)\,, \quad \text{and} \quad
F_2^p(Q^2) = \kappa_p+ \frac{Q^2}{6}\langle r^2_{2}\rangle + {\mathcal O}\left(\frac{1}{M^3}\right)\,.
\end{eqnarray} 
However, for the radiatively corrected cross section evaluated up to $\mathcal{O}(\alpha/M^2)$, only the 
mean-square Dirac radius $\langle r^2_{1}\rangle$ contributes, while the mean-square Pauli radius
$\langle r^2_{2}\rangle$ enters beyond $\mathcal{O}(1/M^4)$. One can relate $\langle r^2_{1}\rangle$ to the 
proton’s root-mean-square (rms) electric or charge radius $r_p\equiv {\langle r^2_{E}\rangle}^{1/2}$ 
{\it via}
\begin{eqnarray}
\langle r^2_1 \rangle=\langle r^2_E \rangle-\frac{3\kappa_p}{2M^2}  + {\mathcal O}\left(M^{-3}\right)\,.
\label{eq:rp-r1}
\end{eqnarray}

The general frame-independent integral representations of the NNLO amplitudes, namely, 
${\mathcal M}^{(j)}_{\rm box}, \cdots, {\mathcal M}^{(s_\pi)}_{\rm xbox}$, along with their mutual partial 
cancellations when specialized to the lab frame, were presented without approximation in our recent SPA-based
analysis~\cite{Goswami:2025zoe}. In particular, the tilde notation over some of these amplitudes denotes their
residual parts after several terms proportional to $v_\mu v_\nu$ cancel among them in the lab frame, as 
detailed in that reference. Here, instead, we directly present the \underline{exact} expressions for the 
lepton-proton fractional TPE corrections to the lab frame cross section, as defined in 
Eqs.~\ref{eq:delta_TPE_NNLO_1} and \ref{eq:delta_TPE_NNLO_2}. The lab frame amplitudes (t) and (u) cancel exactly,
i.e., $\mathcal{M}^{(t)}_{\rm xbox}+\mathcal{M}^{(u)}_{\rm xbox}=0$, as was also found in our SPA 
analysis~\cite{Goswami:2025zoe}. Furthermore, it was shown therein that the NNLO seagull diagram (v) contributes
at $\mathcal{O}(\alpha/M^3)$ and thus omitted in that analysis. Consequently, the expressions of the amplitudes 
(t), (u), and (v) are omitted in the following analysis. The complete expression for the fractional 
contributions arising from the NNLO TPE diagrams, together with the interference between the NNLO OPE amplitude 
${\mathcal M}^{(2)}_\gamma$ and LO TPE diagrams (a) and (b), accurate to ${\mathcal O}(\alpha/M^2)$, is 
expressed as 
{
\begin{eqnarray}
\delta^{\rm (NNLO)}_{\gamma\gamma}(Q^2) &=& \delta^{(j)}_{\rm box}(Q^2) + \delta^{(k)}_{\rm xbox}(Q^2) 
+ \delta^{(l)}_{\rm box}(Q^2) + \delta^{(m)}_{\rm xbox}(Q^2) + \delta^{(n)}_{\rm box}(Q^2) 
+ \delta^{\rm (o)}_{\rm xbox}(Q^2) + \delta^{(p+p_\pi)}_{\rm box}(Q^2) 
\nonumber\\
&& + \delta^{(q+q_\pi)}_{\rm xbox}(Q^2) + \delta^{(r+r_\pi)}_{\rm box}(Q^2) + \delta^{(s+s_\pi)}_{\rm xbox}(Q^2) 
+ \delta^{(a_2)}_{\rm box}(Q^2) + \delta^{(b_2)}_{\rm xbox}(Q^2) 
+ {\mathcal O}\left(\frac{\alpha}{M^3}\right)\,,
\end{eqnarray} 
where
\begin{eqnarray}
\label{j-TPE}
\delta_{\rm box}^{(j)} (Q^2) &=& 2{\mathbb R}e\sum\limits_{spins}
\bigg[\mathcal{M}^{(0)*}_{\gamma} \widetilde{{\mathcal M}}^{(j)}_{\rm box}\bigg]\bigg/
\sum\limits_{spins}\left|\mathcal{M}^{(0)}_\gamma\right|^2
\nonumber\\
&=& -\,\frac{e^2}{8 M^2}\left[\frac{Q^2}{Q^2+4 E E^\prime}\right] {\mathbb R}e 
\bigg\{\int\frac{{\rm d}^4k}{(2\pi)^4 i} \frac{{\rm Tr}[(\slashed{p}^\prime+m_l) \gamma^{\mu}(\slashed{p}-\slashed{k}+m_l)\gamma^{\nu}
(\slashed{p}+m_l)\slashed{v}]}{(k^2+i0)\left[(Q-k)^2+i0\right] (k^2-2k \cdot p+i0) (v\cdot k+i0)} 
\nonumber\\
&&\hspace{4.5cm} \times\,\Big[ (k+Q)_{\mu} k_{\nu}- v_{\mu} k_{\nu} v\cdot (Q+k) -(Q+k)_{\mu} v_{\nu} (v\cdot k)\Big]\bigg\}
\\
&=& -\,\frac{e^2}{2 M^2}\left[\frac{Q^2}{Q^2+4 E E^\prime}\right] {\mathbb R}e 
\bigg\{4( E^\prime+E)  p_{\mu} p_{\nu} I^{-\mu \nu}_1(p,0|1,1,1,1)- 4Q^2 E v \cdot I_1^{-}(p,0|1,1,1,0)
\nonumber\\
&&\hspace{4cm} -\,(Q^2-4E^2-4E E^\prime)I^-(p,0|1,1,0,0)- 4E^2 I^-(p,0|1,0,1,0)
\nonumber\\
&&\hspace{4cm} +\,4Q^2 E^2 I^-(p,0|1,1,1,0)- E I^-(p,0|1,0,0,1) 
\nonumber\\
&&\hspace{4cm} +\,(2E+E^\prime) I^-(p,0|0,1,0,1) -( E+E^\prime) I_2^-(p,0|0,1,1,1) 
\nonumber\\
&&\hspace{4cm} +\,\left(2Q^2 E - Q^2E^\prime + 4E^2E^\prime - 4EE^{\prime 2}\right)I^-(p,0|1,1,0,1)
\nonumber\\
&&\hspace{4cm} +\,2(Q^2+2E E^\prime)I^-(p,0|0,1,1,0) \bigg\}\,,
\\
\nonumber\\
\nonumber\\
\label{k-TPE}
\delta_{\rm xbox}^{(k)} (Q^2) &=& 2{\mathbb R}e\sum\limits_{spins}
\bigg[\mathcal{M}^{(0)*}_{\gamma} \widetilde{{\mathcal M}}^{(k)}_{\rm xbox}\bigg]\bigg/
\sum\limits_{spins}\left|\mathcal{M}^{(0)}_\gamma\right|^2
\nonumber\\
&=& -\,\frac{e^2}{8 M^2}\left[\frac{Q^2}{Q^2+4 E E^\prime}\right] {\mathbb R}e 
\bigg\{\int\frac{{\rm d}^4k}{(2\pi)^4 i} 
\frac{{\rm Tr}[(\slashed{p}^\prime + m_l) \gamma^{\mu}(\slashed{p}+\slashed{k}-\slashed{Q}+m_l)\gamma^{\nu} 
(\slashed{p}+m_l)\slashed{v}]}{(k^2+i0)\left[(Q-k)^2+i0\right] (k^2+2k \cdot p^\prime+i0) (v\cdot k+i0)}
\nonumber\\
&&\hspace{4.5cm}  \times\,\Big[ (k+Q)_{\nu} k_{\mu}- v_{\nu} k_{\mu} v\cdot (Q+k) -(Q+k)_{\nu} v_{\mu} (v\cdot k)\Big]\bigg\}
\nonumber\\
&=&\! -\,\frac{e^2}{2 M^2}\left[\frac{Q^2}{Q^2+4 E E^\prime}\right]\! {\mathbb R}e 
\bigg\{4( E^\prime+E) p_{\mu}^\prime p_{\nu}^\prime I^{+\mu \nu}_1(p^\prime,0|1,1,1,1)- 4Q^2 E^\prime v\cdot I_1^{+}(p^\prime,0|1,1,1,0) 
\nonumber\\
&&\hspace{4cm} +\,(Q^2-4{E^\prime}^2-4E E^\prime)I^+(p^\prime,0|1,1,0,0)+ 4{E^\prime}^2 I^+(p^\prime,0|1,0,1,0) 
\nonumber\\
&&\hspace{4cm} -\,4 Q^2 E^{\prime 2} I^+(p^\prime,0|1,1,1,0) - E^\prime I^+(p^\prime,0|1,0,0,1) 
\nonumber\\
&&\hspace{4cm} +\,( 2E^\prime+E) I^+(p^\prime,0|0,1,0,1) - (E+E^\prime) I_2^+(p^\prime,0|0,1,1,1)
\nonumber\\
&&\hspace{4cm} +\,\left(2Q^2E^\prime - Q^2E + 4EE^{\prime 2} - 4E^2E^\prime\right)I^+(p^\prime,0|1,1,0,1) 
\nonumber\\
&&\hspace{4cm} -\,2(Q^2+2E E^\prime)I^+(p^\prime,0|0,1,1,0) \bigg\}\,,
\\
\nonumber\\
\nonumber\\
\label{l-TPE}
\delta_{\rm box}^{(l)} (Q^2) &=& 2{\mathbb R}e\sum\limits_{spins}
\bigg[\mathcal{M}^{(0)*}_{\gamma} \widetilde{{\mathcal M}}^{(l)}_{\rm box}\bigg]\bigg/
\sum\limits_{spins}\left|\mathcal{M}^{(0)}_\gamma\right|^2
\nonumber\\
&=& \frac{-e^2}{8 M^2}\left[\frac{Q^2}{Q^2+4 E E^\prime}\right] {\mathbb R}e 
\bigg\{\int\frac{{\rm d}^4k}{(2\pi)^4 i} 
\frac{{\rm Tr}[(\slashed{p}^\prime+m_l) \slashed{v}(\slashed{p}-\slashed{k}+m_l)\slashed{k} 
(\slashed{p}+m_l)\slashed{v}]}{(k^2+i0)\left[(Q-k)^2+i0\right] (k^2-2k \cdot p+i0) }
\left(1-\frac{k^2}{(v\cdot k)^2}\right) \bigg\}
\nonumber\\
&=& \frac{e^2 Q^2}{2 M^2}{\mathbb R}e \bigg\{I^-(p,0|1,1,0,0)-I^-(p,0|0,1,0,2)\bigg\}\,,
\\
\nonumber\\
\nonumber\\
\label{m-TPE}
\delta_{\rm xbox}^{(m)} (Q^2) &=& 2{\mathbb R}e\sum\limits_{spins}
\bigg[\mathcal{M}^{(0)*}_{\gamma}  \widetilde{{\mathcal M}}^{(m)}_{\rm xbox}\bigg]\bigg/
\sum\limits_{spins}\left|\mathcal{M}^{(0)}_\gamma\right|^2
\nonumber\\
&=& \frac{-e^2}{8 M^2}\left[\frac{Q^2}{Q^2+4 E E^\prime}\right] {\mathbb R}e 
\bigg\{\int\frac{{\rm d}^4k}{(2\pi)^4 i} 
\frac{{\rm Tr}[(\slashed{p}^\prime+m_l) \slashed{k}(\slashed{p}+\slashed{k}-\slashed{Q}+m_l)\slashed{v} 
(\slashed{p}+m_l)\slashed{v}]}{(k^2+i0)\left[(Q-k)^2+i0\right] (k^2+2k \cdot p^\prime+i0) }
\left(1-\frac{k^2}{(v\cdot k)^2}\right) \bigg\}
\nonumber\\
&=& -\,\frac{e^2 Q^2}{2 M^2}{\mathbb R}e \bigg\{I^+(p^\prime,0|1,1,0,0)-I^+(p^\prime,0|0,1,0,2)\bigg\}\,,
\end{eqnarray}
\begin{eqnarray}
\label{n-TPE}
\delta_{\rm box}^{(n)} (Q^2) &=& 2{\mathbb R}e\sum\limits_{spins}
\bigg[\mathcal{M}^{(0)*}_{\gamma} \widetilde{{\mathcal M}}^{(n)}_{\rm box}\bigg]\bigg/
\sum\limits_{spins}\left|\mathcal{M}^{(0)}_\gamma\right|^2 
\nonumber\\
&=& \frac{-e^2}{8 M^2}\left[\frac{Q^2}{Q^2+4 E E^\prime}\right] {\mathbb R}e 
\bigg\{\!\int\!\frac{{\rm d}^4k}{(2\pi)^4 i} 
\frac{{\rm Tr}[(\slashed{p}^\prime+m_l) (\slashed{Q}+\slashed{k})(\slashed{p}-\slashed{k}+m_l)\slashed{v} 
(\slashed{p}+m_l)\slashed{v}]}{(k^2+i0)\left[(Q-k)^2+i0\right] (k^2-2k \cdot p+i0)} 
\nonumber\\
&&\hspace{4.5cm} \times\, \left(1-\frac{k^2}{(v\cdot k)^2}\right) 
\nonumber\\
&&\hspace{3.7cm}+\, \int\frac{{\rm d}^4k}{(2\pi)^4 i} 
\frac{{\rm Tr}[(\slashed{p}^\prime+m_l) \slashed{v}(\slashed{p}-\slashed{k}+m_l)\slashed{v} 
(\slashed{p}+m_l)\slashed{v}]}{\left[(Q-k)^2+i0\right] (k^2-2k \cdot p+i0) (v \cdot k+i0) } \bigg\}
\nonumber\\
&=& \frac{e^2}{2 M^2}\left[\frac{Q^2}{Q^2+4 E E^\prime}\right] {\mathbb R}e 
\bigg\{ 3 (Q^2+4E E^\prime) I^-(p,0|0,1,1,0) + 4Q^2 E^2 I^-(p,0|1,1,1,0)
\nonumber\\
&&\hspace{3.7cm} +\,E(3 Q^2-8E E^\prime) I^-(p,0|0,1,1,1) + 4 E^2 I^-(p,0|0,0,1,2)
\nonumber\\
&&\hspace{3.7cm} -\,4 Q^2 E v\cdot I_1^-(p,0|1,1,1,0) - E^\prime I_2^-(p,0|0,1,1,1) 
\nonumber\\
&&\hspace{3.7cm} +\,(Q^2-4E^2) I^-(p,0|0,1,0,2) - E I^-(p,0|0,0,1,1) 
\nonumber\\
&&\hspace{3.7cm}  -\,4Q^2E^2 I^-(p,0|0,1,1,2) - 4E^2 I^-(p,0|1,0,1,0) 
\nonumber\\
&&\hspace{3.7cm} -\,2(Q^2+2E E^\prime)I_2^-(p,0|0,1,1,2) + (E+E^\prime) I^-(p,0|0,1,0,1)
\nonumber\\
&&\hspace{3.7cm} -\,(Q^2-4 E^2) I^-(p,0|1,1,0,0) \bigg\}\,,
\\
\nonumber\\
\nonumber\\
\label{o-TPE}
\delta_{\rm xbox}^{(o)} (Q^2) &=& 2{\mathbb R}e\sum\limits_{spins}
\bigg[\mathcal{M}^{(0)*}_{\gamma} \widetilde{{\mathcal M}}^{(o)}_{\rm xbox}\bigg]\bigg/
\sum\limits_{spins}\left|\mathcal{M}^{(0)}_\gamma\right|^2
\nonumber\\
&=& -\,\frac{e^2}{8 M^2}\left[\frac{Q^2}{Q^2+4 E E^\prime}\right] {\mathbb R}e \bigg\{\int\frac{{\rm d}^4k}{(2\pi)^4 i} 
\frac{{\rm Tr}[(\slashed{p}^\prime+m_l)\slashed{v} (\slashed{p}+\slashed{k}-\slashed{Q}+m_l)(\slashed{Q}+\slashed{k}) 
(\slashed{p}+m_l)\slashed{v}]}{(k^2+i0)\left[(Q-k)^2+i0\right] (k^2+2k \cdot p^\prime+i0) } 
\nonumber\\
&&\hspace{4.5cm} \times\,\left(1-\frac{k^2}{(v\cdot k)^2}\right) 
\nonumber\\
&&\hspace{4cm} +\,\int\frac{{\rm d}^4k}{(2\pi)^4 i} \frac{{\rm Tr}[(\slashed{p}^\prime+m_l) 
\slashed{v}(\slashed{p}+\slashed{k}-\slashed{Q}+m_l)\slashed{v} 
(\slashed{p}+m_l)\slashed{v}]}{\left[(Q-k)^2+i0\right] (k^2+2k \cdot p^\prime+i0) (v \cdot k+i0) } \bigg\}
\nonumber\\
&=& -\,\frac{e^2}{2 M^2}\left[\frac{Q^2}{Q^2+4 E E^\prime}\right] {\mathbb R}e 
\bigg\{ 3 (Q^2+4E E^\prime) I^+(p^\prime,0|0,1,1,0) + 4Q^2 E^{\prime 2} I^+(p^\prime,0|1,1,1,0)
\nonumber\\
&&\hspace{4cm} -\,E^\prime(3 Q^2-8E E^\prime) I^+(p^\prime,0|0,1,1,1) + 4E^{\prime 2} I^+(p^\prime,0|0,0,1,2)
\nonumber\\
&&\hspace{4cm} +\,4Q^2 E^\prime v\cdot I_1^
+(p^\prime,0|1,1,1,0)+ E I_2^+(p^\prime,0|0,1,1,1)
\nonumber\\
&&\hspace{4cm} +\,(Q^2-4{E^\prime}^2) I^+(p^\prime,0|0,1,0,2)+ E^\prime I^+(p^\prime,0|0,0,1,1) 
\nonumber\\
&&\hspace{4cm} -\,4{E^\prime}^2 Q^2 I^+(p^\prime,0|0,1,1,2) - 4{E^\prime}^2 I^+(p^\prime,0|1,0,1,0) 
\nonumber\\
&&\hspace{4cm} -\,2(Q^2+2E E^\prime) I_2^+(p^\prime,0|0,1,1,2)- (E^\prime+ E) I^+(p^\prime,0|0,1,0,1)
\nonumber\\
&&\hspace{4cm} -\,(Q^2-4 {E^\prime}^2) I^+(p^\prime,0|1,1,0,0) \bigg\}\,,
\\
\nonumber\\
\nonumber\\
\label{p-TPE}
\delta_{\rm box}^{(p+p_\pi)} (Q^2) &=& 2{\mathbb R}e \sum\limits_{spins}
\bigg[\mathcal{M}^{(0)*}_{\gamma} \left({\mathcal M}^{(p)}_{\rm box}+{\mathcal M}^{(p_\pi)}_{\rm box} \right)\bigg]\bigg/
\sum\limits_{spins}\left|\mathcal{M}^{(0)}_\gamma\right|^2
\nonumber\\
&=& -\,4 \pi \alpha\left[\frac{Q^4}{Q^2+4 E E^\prime}\right] \left(\frac{\langle r_1^2\rangle}{6} 
+\frac{\kappa_p}{4M^2}-\frac{1}{8 M^2}\right) 
\nonumber\\
&&\hspace{1cm} \times\, {\mathbb R}e \bigg\{\int\frac{{\rm d}^4k}{(2\pi)^4 i} 
\frac{{\rm Tr}[(\slashed{p}^\prime+m_l) \slashed{v}(\slashed{p}-\slashed{k}+m_l)\slashed{v} 
(\slashed{p}+m_l)\slashed{v}]}{(k^2+i0)\left[(Q-k)^2+i0\right] (k^2-2k \cdot p+i0) (v\cdot k+i0)} \bigg\}
\nonumber\\
&=& Q^2 \left(\frac{\langle r_1^2\rangle}{6} +\frac{\kappa_p}{4M^2}-\frac{1}{8 M^2}\right) \delta_{\rm box}^{(a)} (Q^2)\,,
\end{eqnarray}
\begin{eqnarray}
\label{q-TPE}
\delta_{\rm xbox}^{(q+q_\pi)} (Q^2) &=& 2{\mathbb R}e \sum\limits_{spins}
\bigg[\mathcal{M}^{(0)*}_{\gamma} \left({\mathcal M}^{(q)}_{\rm xbox}+{\mathcal M}^{(q_\pi)}_{\rm xbox}\right)\bigg]\bigg/
\sum\limits_{spins}\left|\mathcal{M}^{(0)}_\gamma\right|^2
\nonumber\\
&=& -\,4 \pi \alpha\left[\frac{Q^4}{Q^2+4 E E^\prime}\right] \left(\frac{\langle r_1^2\rangle}{6} 
+\frac{\kappa_p}{4M^2}-\frac{1}{8 M^2}\right)
\nonumber\\
&&\hspace{1cm} \times\, {\mathbb R}e \bigg\{\int\frac{{\rm d}^4k}{(2\pi)^4 i} 
\frac{{\rm Tr}[(\slashed{p}^\prime+m_l) \slashed{v}(\slashed{p}+\slashed{k}-\slashed{Q}+m_l)\slashed{v} 
(\slashed{p}+m_l)\slashed{v}]}{(k^2+i0)\left[(Q-k)^2+i0\right] (k^2+2k \cdot p^\prime+i0) (v\cdot k+i0)} \bigg\}
\nonumber\\
&=& Q^2 \left(\frac{\langle r_1^2\rangle}{6} +\frac{\kappa_p}{4M^2}-\frac{1}{8 M^2}\right) \delta_{\rm xbox}^{(b)} (Q^2)\,,
\\
\nonumber\\
\nonumber\\
\label{r-TPE}
\delta_{\rm box}^{(r+r_\pi)} (Q^2) &=& 2{\mathbb R}e \sum\limits_{spins}
\bigg[\mathcal{M}^{(0)*}_{\gamma} \left({\mathcal M}^{(r)}_{\rm box}+ {\mathcal M}^{(r_\pi)}_{\rm box}\right)\bigg]\bigg/
\sum\limits_{spins}\left|\mathcal{M}^{(0)}_\gamma\right|^2
\nonumber\\
&=& -\,4 \pi \alpha\left[\frac{Q^4}{Q^2+4 E E^\prime}\right] \left(\frac{\langle r_1^2\rangle}{6} 
+\frac{\kappa_p}{4M^2}-\frac{1}{8M^2}\right)
\nonumber\\
&&\hspace{1cm} \times\, {\mathbb R}e \bigg\{\int\frac{{\rm d}^4k}{(2\pi)^4 i} 
\frac{{\rm Tr}[(\slashed{p}^\prime+m_l) \slashed{v}(\slashed{p}-\slashed{k}+m_l)\slashed{v}
(\slashed{p}+m_l)\slashed{v}]}{(k^2+i0)\left[(Q-k)^2+i0\right] (k^2-2k \cdot p+i0) (v\cdot k+i0)} \bigg\}
\nonumber\\
&=& Q^2 \left(\frac{\langle r_1^2\rangle}{6} +\frac{\kappa_p}{4M^2}-\frac{1}{8M^2}\right) \delta_{\rm box}^{(a)} (Q^2)\, ,
\\
\nonumber\\
\nonumber\\
\label{s-TPE}
\delta_{\rm xbox}^{(s+s_\pi)} (Q^2) &=& 2{\mathbb R}e \sum\limits_{spins}
\bigg[\mathcal{M}^{(0)*}_{\gamma} \left({\mathcal M}^{(s)}_{\rm xbox}+{\mathcal M}^{(s_\pi)}_{\rm xbox}\right)\bigg]\bigg/
\sum\limits_{spins}\left|\mathcal{M}^{(0)}_\gamma\right|^2
\nonumber\\
&=& -\,4 \pi \alpha\left[\frac{Q^4}{Q^2+4 E E^\prime}\right] \left(\frac{\langle r_1^2\rangle}{6} 
+\frac{\kappa_p}{4M^2}-\frac{1}{8 M^2}\right)
\nonumber\\
&&\hspace{1cm} \times\, {\mathbb R}e \bigg\{\int\frac{{\rm d}^4k}{(2\pi)^4 i} 
\frac{{\rm Tr}[(\slashed{p}^\prime+m_l) \slashed{v}(\slashed{p}+\slashed{k}-\slashed{Q}+m_l)\slashed{v} 
(\slashed{p}+m_l)\slashed{v}]}{(k^2+i0)\left[(Q-k)^2+i0\right] (k^2+2k \cdot p^\prime+i0) (v\cdot k+i0)} \bigg\}
\nonumber\\
&=& Q^2 \left(\frac{\langle r_1^2\rangle}{6} +\frac{\kappa_p}{4M^2}-\frac{1}{8 M^2}\right) \delta_{\rm xbox}^{(b)} (Q^2)\,.
\end{eqnarray}
It is noteworthy that the last four NNLO box and crossed-box contributions, Eqs.~\eqref{p-TPE}– \eqref{s-TPE},
are proportional to contributions $\delta_{\rm box}^{(a)}$ and $\delta_{\rm xbox}^{(b)}$ arising from the LO 
TPE diagrams (a) and (b), respectively. Since the latter exactly vanish at LO [i.e., ${\mathcal O}(\alpha M^0)$]
in the SPA analysis, it follows that the ${\mathcal O}(\alpha/M^2)$ contributions from the (p) – (s) diagrams 
also vanish in SPA, as shown in Ref.~\cite{Goswami:2025zoe}. In contrast our \underline{exact} TPE analysis 
exhibits non-zero proton structure effects at NNLO as manifested through the LEC such as $\kappa_p$
and $\langle r_1^2\rangle$ (equivalently the proton radius $r_p$).\footnote{ It is worth re-emphasizing that in 
standard HB$\chi$PT power counting arguments, the proton’s $\langle r_1^2\rangle \sim {\mathcal O}(1/M^2)$, 
whereas $\kappa_p$ scales as ${\mathcal O}(M^0)$. Consequently, the NNLO results in Eqs.~\eqref{p-TPE} – 
\eqref{s-TPE} are consistent with the expected ${\mathcal O}(1/M^2)$ scaling. } The additional interference 
between the NNLO OPE amplitude ${\mathcal M}^{(2)}_\gamma$ and the LO TPE diagrams generate the following two
contributions:
\begin{eqnarray}
\delta^{(a_2)}_{\rm box}(Q^2) &=& 2{\mathbb R}e\sum\limits_{spins}
\bigg[\mathcal{M}^{(2)*}_{\gamma} {\mathcal M}^{(a)}_{\rm box}\bigg]\bigg/
\sum\limits_{spins}\left|\mathcal{M}^{(0)}_\gamma\right|^2
\nonumber\\
&=& -\,4 \pi\alpha\left[\frac{Q^4}{Q^2+4E E^\prime}\right] \left(\frac{\langle r_1^2\rangle}{6} 
+\frac{\kappa_p}{4M^2}-\frac{1}{8 M^2}\right)
\nonumber\\
&&\hspace{1cm} \times\, {\mathbb R}e \Bigg\{\int \frac{{\rm d}^4 k}{{(2 \pi)}^4 i}
\frac{{\rm Tr}\left[(\slashed{p}+m_l)\,\slashed{v}\,(\slashed{p}^{\prime}+m_l)\,\slashed{v}\, 
(\slashed{p}-\slashed{k}+m_l)\,\slashed{v}\right]}{(k^2+i0)\,
[(Q-k)^2+i0]\,(k^2-2 k\cdot p+i0)\,(v\cdot k+i0)}\Bigg\} + \mathcal{O}\left(\frac{\alpha}{M^3}\right)
\nonumber\\
\label{a2-TPE}
&=&  Q^2 \left(\frac{\langle r_1^2\rangle}{6} +\frac{\kappa_p}{4M^2}-\frac{1}{8 M^2}\right) \delta^{(a)}_{\rm box}(Q^2) 
+ \mathcal{O}\left(\frac{\alpha}{M^3}\right)\,, 
\end{eqnarray}
and
\begin{eqnarray}
\label{b2-TPE}
\delta^{(b_2)}_{\rm xbox}(Q^2) &=& 2{\mathbb R}e\sum\limits_{spins}
\bigg[\mathcal{M}^{(2)*}_{\gamma} {\mathcal M}^{(b)}_{\rm xbox}\bigg]\bigg/
\sum\limits_{spins}\left|\mathcal{M}^{(0)}_\gamma\right|^2
\nonumber\\
&=& -\,4 \pi\alpha\left[\frac{Q^4}{Q^2+4E E^\prime}\right] \left(\frac{\langle r_1^2\rangle}{6} 
+\frac{\kappa_p}{4M^2}-\frac{1}{8 M^2}\right)
\nonumber\\
&&\hspace{1cm} \times\, {\mathbb R}e \Bigg\{\int \frac{{\rm d}^4 k}{{(2 \pi)}^4 i}
\frac{{\rm Tr}\left[(\slashed{p}+m_l)\,\slashed{v}\,(\slashed{p}^{\, \prime}+m_l)\,\slashed{v}\, 
(\slashed{p}+\slashed{k}-\slashed{Q}+m_l)\,\slashed{v}\right]}{(k^2+i0)\,
[(Q-k)^2+i0]\,(k^2 + 2k\cdot p^\prime+i0)\,(v\cdot k+i0)}\Bigg\} + \mathcal{O}\left(\frac{\alpha}{M^3}\right)\,,
\nonumber\\
&=&   Q^2 \left(\frac{\langle r_1^2\rangle}{6} +\frac{\kappa_p}{4M^2}-\frac{1}{8 M^2}\right) \delta^{(b)}_{\rm xbox}(Q^2) 
+ \mathcal{O}\left(\frac{\alpha}{M^3}\right)\,.
\end{eqnarray} 
As discussed in the previous section, the IBP decomposition and partial fractioning reduce each of these 
NNLO fractional contributions to a sequence of 2-, 3-, and 4-point scalar and tensor loop-integrals. This 
reduction involves not only the loop-integrals already encountered in the preceding works of 
Refs.~\cite{Choudhary:2023rsz} and \cite{Goswami:2025zoe}, but also an additional set of fourteen 2- and 
3-point master-integrals, as well as five reducible 3- and 4-point tensor loop-integrals. These 
loop-integrals are analytically evaluated and their expressions are given in Appendix~B. All such 2-point
master-integrals appearing above exhibit UV divergence, with the exception of 
$I^{-}(p,0|1,0,0,1)=I^{+}(p^\prime,0|1,0,0,1)$, which correspond to scaleless integrals that vanish in DR.

Our analytical expressions for the finite parts of the TPE contributions arise from the pairs of box and 
crossed-box NNLO amplitudes (j) - (s$_\pi$) [cf. Fig.~\ref{fig:NNLO_TPE}] and incorporate all corrections up 
to ${\mathcal O}(\alpha/M^2)$ in each case. As in the previous section, we replace $E^\prime\to E$ and 
$\beta^\prime \to \beta$ in all ${\mathcal O}(\alpha/M^2)$ terms. As noted among the LO and NLO diagrams 
(a) - (d), a substantial cancellation occur between the NNLO box and cross–box pairs (the exception is the
(n) and (o) pair of diagrams), leading to markedly simplified results. In fact the total contribution from
the box (l) and crossed-box (m) diagrams vanishes identically, as obtained in our SPA analysis in 
Ref.~\cite{Goswami:2025zoe}, 
i.e., $\delta^{(lm)}_{\gamma\gamma}(Q^2) = \delta_{\rm box}^{(l)}(Q^2) + \delta_{\rm xbox}^{(m)}(Q^2)=0\,$.
It is also noteworthy that the additional IR-divergences appearing at NNLO, namely, contributions arising 
from diagrams $(p),\, (p_\pi),\cdots (s)$ and $(s_\pi)$, as well as the interference of the NNLO OPE 
amplitude ${\mathcal M}^{(2)}_\gamma$ with the LO TPE diagrams (a) and (b), cancel at 
${\mathcal O}(\alpha/M^2)$. Their remnants constitute only residual divergences at ${\mathcal O}(\alpha/M^3)$,
which is beyond our accuracy, and therefore not needed in this analysis. The remaining non-vanishing finite 
results are discussed below:\\ 

\noindent$\bullet$ First, the total contribution from the box (j) and crossed-box (k) diagrams is expressed as
\begin{eqnarray}
\label{eq:jk-TPE}
\delta^{(jk)}_{\gamma\gamma}(Q^2) &=& \delta^{(jk;1/M^2)}_{\gamma\gamma}(Q^2) 
+ {\mathcal O}\left(\frac{\alpha}{M^3}\right)\,, \quad \text{where}
\\
\delta^{(jk;1/M^2)}_{\gamma\gamma}(Q^2) &=& \, \frac{Q^2}{8M^2}\delta^{(0)}_{\gamma\gamma}(Q^2) 
=  -\frac{\pi\alpha (-Q^2)^{3/2}}{16M^2 E}\left[\frac{1}{1+\frac{Q^2}{4 E^2}}\right]\, . 
\end{eqnarray} 
where $\delta^{(0)}_{\gamma\gamma}$ is the finite $\mathcal O(\alpha M^0)$ contribution of the LO TPE 
diagrams (a) and (b); see Eq.~\eqref{eq:delta-ab_LO}.

\noindent$\bullet$ Second, the total contribution arising from the box (n) and crossed-box (o) diagrams is 
given by
\begin{eqnarray}
\label{eq:no-TPE}
\delta^{(no)}_{\gamma\gamma}(Q^2) &=&  \delta_{\rm box}^{(n)}(Q^2) +  \delta_{\rm xbox}^{(o)}(Q^2) 
\nonumber\\
&\equiv&  \delta^{(no;1/M)}_{\gamma\gamma}(Q^2) + \delta^{(no;1/M^2)}_{\gamma\gamma}(Q^2) 
+ {\mathcal O}\left(\frac{\alpha}{M^3}\right)\,,  
\end{eqnarray}
where 
\begin{eqnarray}
\label{eq:delta-no_NLO}
\delta^{(no;1/M)}_{\gamma\gamma}(Q^2)
&=& -\,\frac{8\pi\alpha Q^4}{M}\,  
{\mathbb R}e \left[ I^{(0)}(p,0|0,1,1,2) \right], \quad\,\,\,
\end{eqnarray}
and
\begin{eqnarray}
\label{eq:delta-no_NNLO}
\delta^{(no;1/M^2)}_{\gamma\gamma}(Q^2) 
\!&=&\!  \frac{2\pi\alpha Q^2}{M^2E^2 \beta^2(Q^2+4E^2)}\, {\mathbb R}e \Bigg\{
2E^2Q^2 \beta^2(Q^2+4E^2)\delta^{(M^0)}I^+(p^\prime,0|0,1,1,2)
\nonumber\\
&&\hspace{3.8cm} +\,(Q^2+2E^2)(Q^2+2E^2\beta^2)\!\bigg(\!I^{(0)}(p,0|0,0,1,2)-I(Q|0,1,0,2)\!\bigg)\! \Bigg\}\,.
\end{eqnarray}
Note the surprising appearance of the kinematically enhanced term 
$\delta^{(no;1/M)}_{\gamma\gamma}\sim {\mathcal O}(\alpha/M)$ which arise in the NNLO diagrammatic 
contributions. This originates from the 3-point reducible loop-integrals $I^-_2(p,0|0,1,1,2)$ and 
$I^+_2(p^\prime,0|0,1,1,2)$, as demonstrated in Appendix~B, and are expressed in terms of the 
master-integrals $I^-(p,0|0,1,1,2)$ and $I^+(p^\prime,0|0,1,1,2)$, respectively. The latter 
integrals contain contributions proportional to $MI^{(0)}(p,0|0,1,1,2)\sim{\mathcal O}(M)$, as 
discussed in the last section [see Eqs.~\eqref{eq:I-0112} and \eqref{eq:I+0112}].\\

\noindent$\bullet$ Third, the total finite contribution arising from diagrams (p) - (s) and their 
pionic counterparts (${\rm p}_{\pi}$) - (${\rm s}_{\pi}$) (after subtracting the residual 
${\mathcal O}(\alpha/M^3)$ IR divergence) is given by 
\begin{eqnarray}
\label{eq:ps-TPE}
\overline{\delta^{(p+p_{\pi},\cdots,s+s_{\pi})}_{\gamma\gamma}}(Q^2) &=&  \delta_{\rm box}^{(p+p_{\pi})}(Q^2) 
+ \delta_{\rm xbox}^{(q+q_{\pi})}(Q^2) + \delta_{\rm box}^{(r+r_{\pi})}(Q^2) +  \delta_{\rm xbox}^{(s+s_{\pi})}(Q^2) 
\nonumber\\
&& -\, 2Q^2 \left(\frac{ r_p^2}{6} -\frac{1}{8M^2}\right) \delta^{\rm (box)}_{\rm IR}(Q^2)
\nonumber\\
&=& \overline{\delta^{(p+p_{\pi},\cdots,s+s_{\pi};1/M^2)}_{\gamma\gamma}}(Q^2) 
+ {\mathcal O}\left(\frac{\alpha}{M^3}\right)\,, 
\end{eqnarray}
where
\begin{eqnarray}
\overline{\delta^{(p+p_{\pi},\cdots,s+s_{\pi};1/M^2)}_{\gamma\gamma}}(Q^2) &=& 
-\,\frac{\pi \alpha (-Q^2)^{3/2}}{E} \left(\frac{ r_p^2}{6} -\frac{1}{8M^2}\right)\left[\frac{1}{1+\frac{Q^2}{4 E^2}}\right]  \,.
\end{eqnarray}
 
\noindent$\bullet$ Fourth, the total contribution arising from the interference of the NNLO OPE amplitude 
${\mathcal M}^{(2)}_\gamma$ with the LO TPE diagrams (a) and (b) (after subtracting the residual 
${\mathcal O}(\alpha/M^3)$ IR divergence) is given by
\begin{eqnarray}
\label{eq:ab2-TPE}
\overline{\delta^{(a_2b_2)}_{\gamma\gamma}}(Q^2) &=&  \delta_{\rm box}^{(a_2)}(Q^2) +  \delta_{\rm xbox}^{(b_2)}(Q^2) 
-  Q^2 \left(\frac{ r_p^2}{6} -\frac{1}{8M^2}\right) \delta^{\rm (box)}_{\rm IR}(Q^2) 
\nonumber\\
&=& \overline{\delta^{(a_2b_2;1/M^2)}_{\gamma\gamma}}(Q^2) +  {\mathcal O}\left(\frac{\alpha}{M^3}\right)\,,
\end{eqnarray}
where
\begin{eqnarray}
\overline{\delta^{(a_2b_2;1/M^2)}_{\gamma\gamma}}(Q^2) &=&  -\, \frac{\pi \alpha (-Q^2)^{3/2}}{2E} 
\left(\frac{ r_p^2}{6} -\frac{1}{8M^2}\right)\left[\frac{1}{1+\frac{Q^2}{4 E^2}}\right]  \,.  
\end{eqnarray}
Note that, in presenting the above results, we have eliminated the proton’s iso-vector mean-square radius
$\langle r^2_1\rangle$ and anomalous magnetic moment $\kappa_p$ in favor of the proton mean square 
charge radius, $r^2_p\equiv \langle r^2_{E}\rangle$ [see Eq.~\eqref{eq:rp-r1}], which is treated as a 
phenomenologically fixed input parameter in our analysis.

We finally present the consolidated analytical expression for the finite fractional part of the dynamical 
NNLO TPE contributions to the elastic differential cross section, retaining all terms up to 
${\mathcal O}(\alpha/M^2)$, as obtained from Eqs.~\eqref{eq:jk-TPE} - \eqref{eq:ab2-TPE}: 
\begin{eqnarray}
\overline{\delta^{\rm (NNLO)}_{\gamma\gamma}} (Q^2) &=& \delta^{\rm (NNLO)}_{\gamma\gamma} (Q^2) 
- 3Q^2 \left(\frac{ r_p^2}{6} -\frac{1}{8M^2}\right) \delta^{\rm (box)}_{\rm IR}(Q^2)  
\nonumber\\
&=& \delta^{\rm (NNLO;1)}_{\gamma\gamma} (Q^2) + \delta^{\rm (NNLO;2)}_{\gamma\gamma} (Q^2) 
+ {\mathcal O}\left(\frac{\alpha}{M^3}\right)\,,
\end{eqnarray}
where the ${\mathcal O}(\alpha/M)$ enhanced components arise solely from the NNLO TPE box (n) and
crossed-box (o) pair of diagrams and are collected in 
\begin{eqnarray}
\delta^{\rm (NNLO;1)}_{\gamma\gamma} (Q^2) \equiv  \delta^{(no;1/M)}_{\gamma\gamma} (Q^2) 
= \frac{2 \alpha Q^2}{\pi ME \beta  } \ln{\sqrt{\frac{1+\beta}{1-\beta}}}\, . 
\end{eqnarray}
We emphasize that, in presenting our final consolidated numerical results in the next section (see later in 
Fig.~\ref{fig:TPE_compare}), the above terms involving the NNLO TPE diagrams will be grouped along with the
NLO modifications to the elastic corrections rather than being treated as NNLO contributions. By contrast, 
the genuine NNLO components [i.e., ${\mathcal O}(\alpha/M^2)$] components involving the NNLO TPE diagrams 
are collected in the term
\begin{eqnarray}
\delta^{\rm (NNLO;2)}_{\gamma\gamma} (Q^2) 
&\equiv& \delta^{(jk)}_{\gamma\gamma}(Q^2) + \delta^{(no;1/M^2)}_{\gamma\gamma}(Q^2) 
+ \overline{\delta^{(p+p_{\pi},\cdots,s+s_{\pi})}_{\gamma\gamma}}(Q^2) + \overline{\delta^{(a_2b_2)}_{\gamma\gamma}}(Q^2)
\nonumber\\
&=& -\,\frac{\alpha Q^4}{2\pi M^2 E^2 (Q^2+4E^2)^2} \Bigg\{ \bigg(8E^2(Q^2+2E^2)\beta^2 - Q^4(1-\beta^2)\bigg) \ln{\sqrt{\frac{1+\beta}{1-\beta}}}
\nonumber\\
&&\hspace{4.3cm} -\,\beta^3(Q^2+4E^2)^2\Bigg\} - \frac{\pi \alpha E (-Q^2)^{3/2}}{4 M^2 (Q^2+4E^2)}
\nonumber\\
&&  -\, \frac{3 \pi \alpha (-Q^2)^{3/2}}{2E} \left(\frac{ r_p^2}{6} -\frac{1}{8M^2}\right)\left[\frac{1}{1+\frac{Q^2}{4 E^2}}\right]  \,. 
\label{eq:NNLO_TPE_delta2}
\end{eqnarray}
In these \underline{exact} NNLO TPE results we find non-vanishing, structure-dependent proton corrections at
${\mathcal O}(\alpha/M^2)$ from the form factor-type diagrams (p) – (${\rm s}_\pi$), whereas in the 
corresponding SPA-based evaluation of Ref.~\cite{Goswami:2025zoe}, they are kinematically suppressed.   
Fig.~\ref{fig:NNLO_Exact_SPA} displays the exactly evaluated numerical results for each of these box and 
crossed-box paired contributions, along with their total fractional contribution 
$\delta^{\rm (NNLO;2)}_{\gamma\gamma}$. We also compare these results with the corresponding SPA-based 
results reported in Ref.~\cite{Goswami:2025zoe}. 
\begin{figure*}[tbp]
\begin{center}
\includegraphics[width=0.48\linewidth]{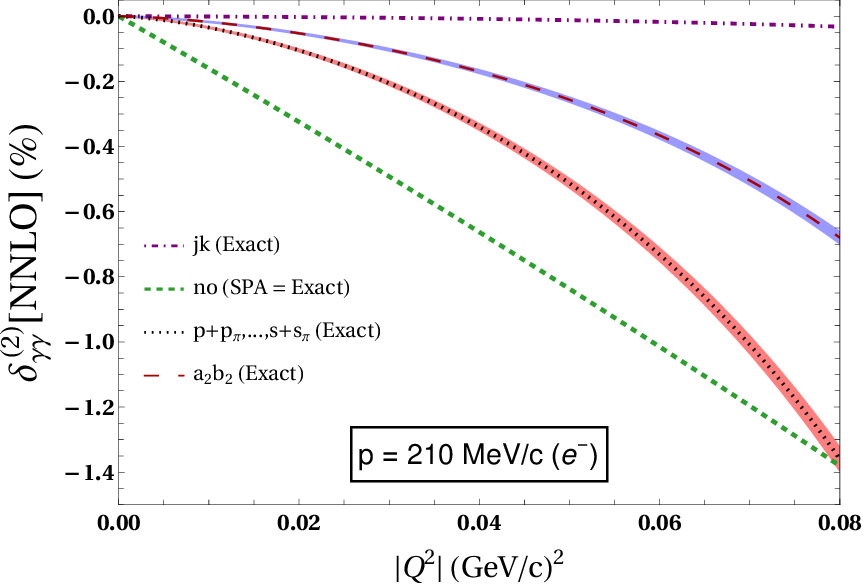}~\quad~\includegraphics[width=0.48\linewidth]{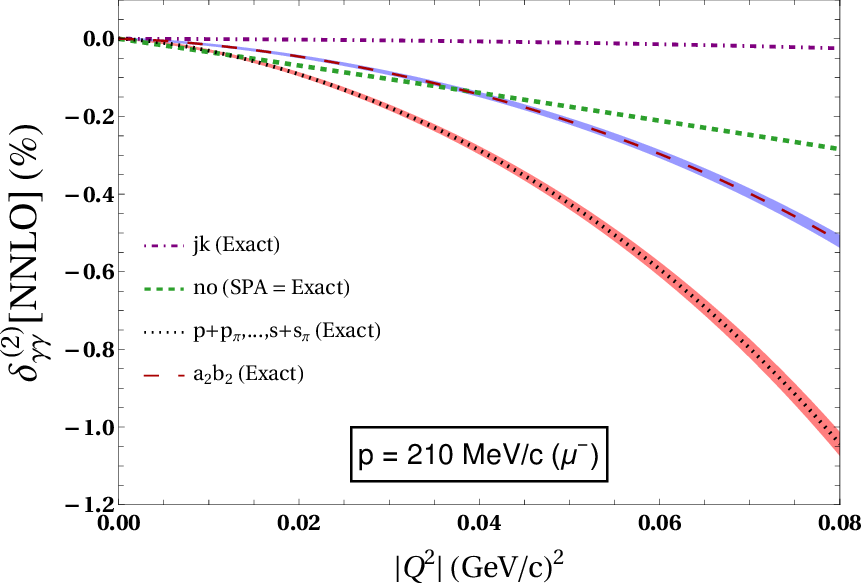}

\vspace{0.5cm}
    
\includegraphics[width=0.48\linewidth]{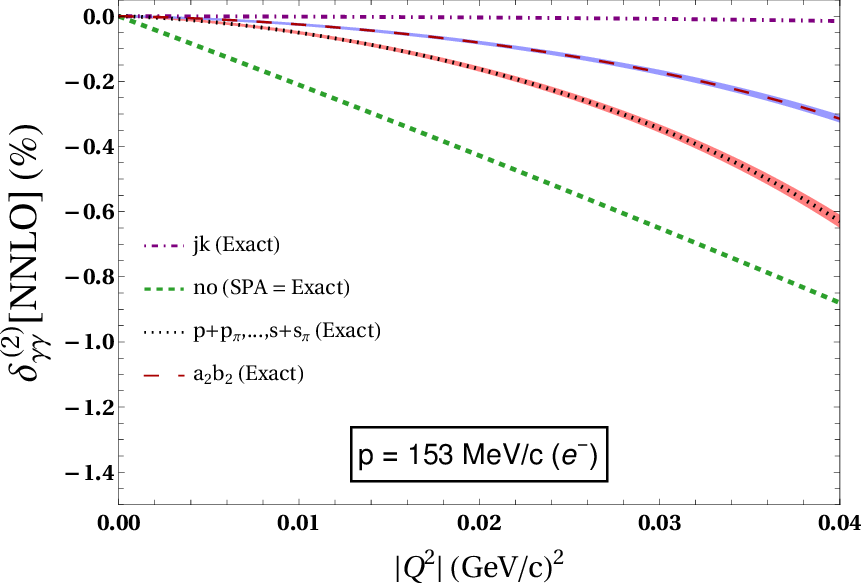}~\quad~\includegraphics[width=0.48\linewidth]{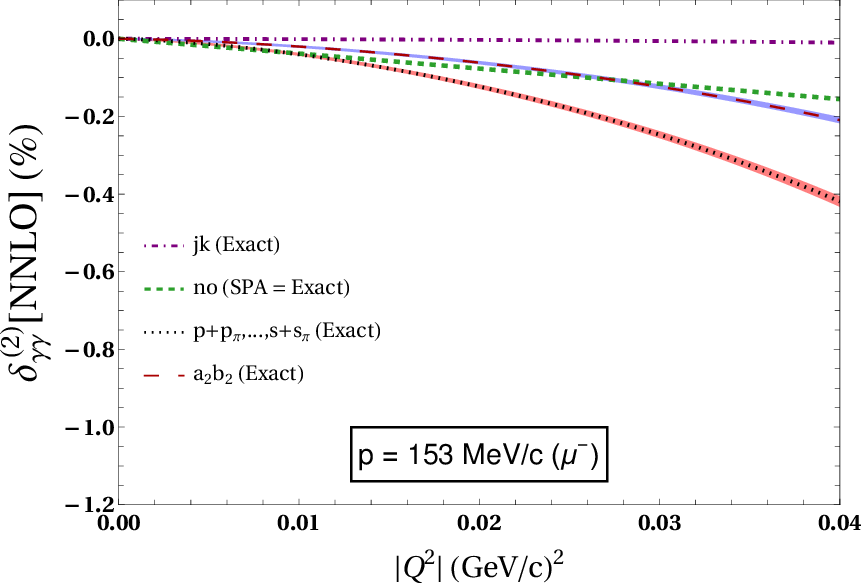} 

\vspace{0.5cm}

\includegraphics[width=0.48\linewidth]{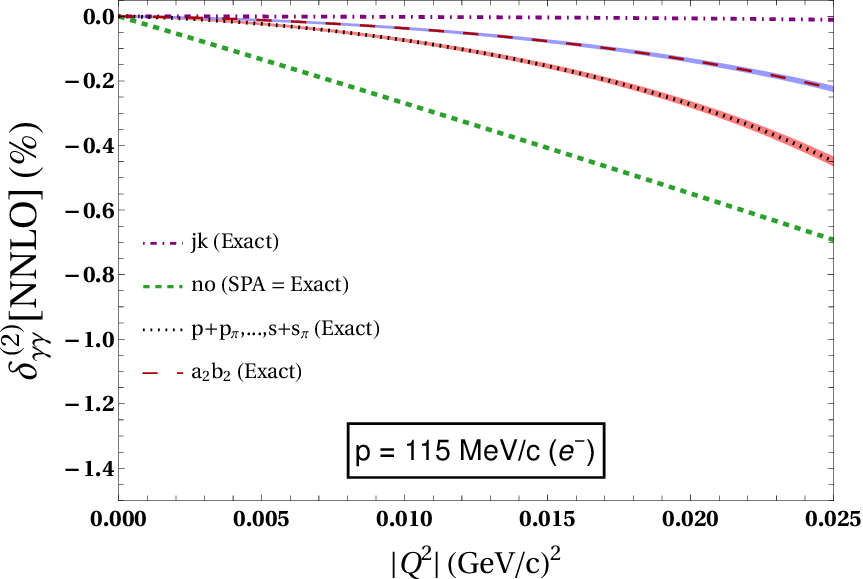}~\quad~\includegraphics[width=0.48\linewidth]{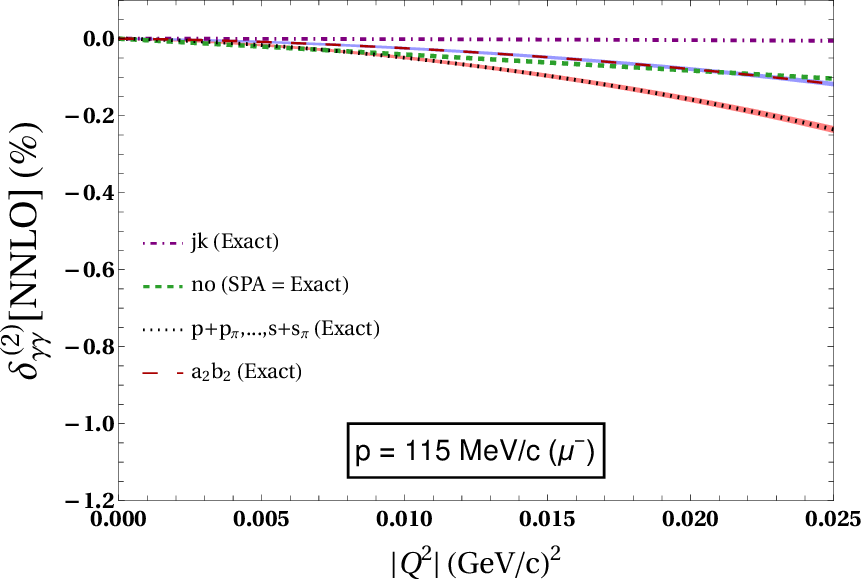}
\caption{The box and crossed-box pairs of finite fractional \underline{dynamical NNLO} corrections arising 
from the NNLO TPE diagrams, (j) – (${\rm s}_\pi$), are shown together with the interference of the NNLO OPE 
amplitude ${\mathcal M}^{(2)}_\gamma$ with the LO TPE diagrams (a) and (b), entering the 
${\mathcal O}(\alpha/M^2)$ component $\delta^{\rm (NNLO;2)}_{\gamma\gamma}$ [Eq.~\eqref{eq:NNLO_TPE_delta2}] 
of the electron-proton (left panel) and muon-proton (right panel) elastic differential cross sections. For 
comparison the corresponding SPA results of Goswami {\it et al.}~\cite{Goswami:2025zoe} are also shown. 
Within the SPA evaluation the only non-vanishing NNLO TPE contribution arises from diagrams (n) and (o). 
The result is identical to the one obtained in our exact calculation, as indicated by the label 
“no (SPA = Exact)” for the curves. The purple and red shaded bands reflect the uncertainties associated 
with varying the proton rms radius across the range spanned by the CREMA determination 
[$r_p=0.84087(39)$~fm]~\cite{Antognini:2013txn,Pohl:2013yb} and the MAMI-ISR measurement 
[$r_p=0.87\pm (0.014){\rm stat.}\pm (0.024){\rm syst.}\pm (0.003)_{\rm mod.}$~fm]~\cite{Mihovilovic:2019jiz}.
The central curve in each band corresponds to result obtained using the mean value, 
$\overline{r_p}\sim 0.855$~fm. All fractional corrections are given relative to the LO OPE differential 
cross section relevant to MUSE kinematical range.}
\label{fig:NNLO_Exact_SPA} 
\end{center}
\end{figure*}


\section{Results and Discussion} 
Here we present our numerical evaluations of the \underline{exact} TPE and the SPA versions of the 
corrections to the lepton-proton elastic differential cross section. We incorporate all radiative effects
modified by the dominant proton chiral structure-dependent corrections up to ${\mathcal O}(\alpha/M^2)$
(NNLO accuracy) in HB$\chi$PT, which is of relevance to the low energy MUSE kinematical 
regime~\cite{Gilman:2013eiv}.  This analysis supersedes the earlier $\mathcal{O}(1/M)$ NLO analysis of 
Choudhary {\it et al.}~\cite{Choudhary:2023rsz} where, for example, no finite-size proton effects appeared,   
effectively treating the proton as point-like. As detailed in the two preceding sections, the NNLO TPE 
contributions to the elastic cross section, i.e., $\mathcal{O}(\alpha/M^{2})$, arise from the following 
two principal sources: 
\begin{enumerate}
\item The \underline{kinematical NNLO} recoil contributions generated by the interference of the TPE and 
OPE amplitudes at LO and NLO when the outgoing lepton kinematical variables $(E^\prime,\beta^\prime)$ are 
re-expressed in terms of the incoming ones $(E,\,\beta)$ by using Eq.~\eqref{eq:E_beta_expansions}. 
\item The \underline{dynamical NNLO} contributions originating from the interference of TPE and OPE 
amplitudes that involve NNLO chiral order Feynman diagrams. 
\end{enumerate}
While point 1 involves either two NLO $\gamma$pp vertex insertions or a single NLO $\gamma$pp vertex 
combined with a $\mathcal{O}(1/M)$ correction to the proton propagator, point 2 originates from 
contributions involving either a single NNLO $\gamma$pp vertex insertion or a $\mathcal{O}(1/M^2)$ 
correction to the proton propagator. The numerical results correspond to each of the two types of 
contributions.   

However, before turning to the ${\mathcal O}(\alpha/M^2)$ discussion of the TPE contributions, for 
completeness, we first revisit the ${\mathcal O}(\alpha/M)$ corrections to the cross section. As 
discussed in Sec.~\ref{sec:three}, these terms arise from the LO and NLO TPE diagrams (a) – (i), 
evaluated in the \underline{exact} TPE analysis of Choudhary {\it et al.}~\cite{Choudhary:2023rsz}. 
However, as demonstrated in Sec.~\ref{sec:four}, there exists an additional and significant recoil 
contribution arising from the kinematically enhanced NNLO component $\delta^{(no;1/M)}_{\gamma\gamma}$, 
Eq.~\eqref{eq:delta-no_NLO}, which are associated with the box (n) and crossed-box (o) diagrams. This 
term modifies the result of Ref.~\cite{Choudhary:2023rsz}. Figure~\ref{fig:NLO_Exact_SPA} displays the 
numerical values of all the relevant NLO [i.e, ${\mathcal O}(\alpha/M)$] finite fractional contributions 
to the elastic differential cross section arising from the exactly evaluated TPE diagrams in this work.
The corresponding total NLO contribution is defined as: 
\begin{eqnarray}
\delta^{(1)}_{\gamma\gamma} (Q^2) \equiv  \overline{\delta^{(ab;1/M)}_{\gamma\gamma}} (Q^2) 
+ \delta^{(ef;1/M)}_{\gamma\gamma} (Q^2) + \delta^{(gh;1/M)}_{\gamma\gamma} (Q^2) 
+ \delta^{({\rm seagull};1/M)}_{\gamma\gamma} (Q^2) +\delta^{(no;1/M)}_{\gamma\gamma} (Q^2)\,.  
\label{eq:TPE_delta1}
\end{eqnarray}
The first four NLO radiative corrections were first resolved in Ref.~\cite{Choudhary:2023rsz}, whereas 
the last term \eqref{eq:delta-no_NLO}] contains a term first evaluated in this paper. In other words, 
Fig.~\ref{fig:NLO_Exact_SPA} provides an updated version the our previous NLO results published in 
Choudhary {\it et al.}~\cite{Choudhary:2023rsz}. As noted before, the \underline{exact} TPE loop-integrals
contributing to the corrections $\delta^{(ef)}_{\gamma\gamma}$ and $\delta^{(gh)}_{\gamma\gamma}$, due to
the four diagrams (e) - (h), add a positive contribution, constructively -- unlike the other box and 
crossed-box pairs. Therefore, we find that the (e) - (h) diagrams add sizable contributions to the e-p 
cross section. The large numerical size of these evaluated contributions are partly driven by the sizable
3-point scalar loop-integrals, such as $Z^-(\Delta,i\sqrt{-Q^2}/2,m_l,E)$, 
$Z^+(\Delta^\prime,i\sqrt{-Q^2}/2,m_l,-E^\prime)$, $I^-(p,0|0,1,1,2)$ and $I^+(p^\prime,0|0,1,1,2)$ (cf.
Appendices for their analytic expressions) that enter them. For the same reason the individual 
contributions from the LO TPE diagrams $\delta^{(a)}_{\rm box}$ and $\delta^{(b)}_{\rm xbox}$ are
also quite sizable; however, their mutual cancellation substantially reduces the net magnitude of their 
IR-finite combination, $\overline{\delta^{(ab)}_{\gamma\gamma}}$. It is also worth noting that the absence
of such sizable loop-integrals, most notably the $Z^\pm$ functions, in our SPA-based 
treatments~\cite{Talukdar:2019dko,Goswami:2025zoe}, is the primary reason why the SPA results are 
substantially smaller than those obtained from the \underline{exact} treatment. In 
Fig.~\ref{fig:NLO_Exact_SPA}, we compare our recently obtained SPA results of Ref.~\cite{Goswami:2025zoe}
with the present \underline{exact} numerical evaluation of the ${\mathcal O}(\alpha/M)$ corrections. 
Furthermore, in Fig.~\ref{fig:NLO_Exact_SPA} we remark that in our HB$\chi$PT framework, the contributions
arising from the (n) and (o) diagrams turn our equal in both the SPA and \underline{exact} evaluations. 
The other qualitative features visible in Fig.~\ref{fig:NLO_Exact_SPA} are enumerated as follows. 
\begin{enumerate}
\item The \underline{exact} evaluations reveal substantial cancellations among the various contributions.
This is in sharp contrast to the SPA results where such cancellations are less pronounced. 
\item Our results for muon scattering are about a magnitude smaller than for the electron case.
\item For electron scattering prominent contributions originate from the exact (e) and (f) diagrams, 
which exhibit a sign change as $Q^2$ increases across the MUSE kinematic range. The magnitude of the 
remaining corrections increase monotonically with increasing momentum transfer $Q^2$. These 
contributions also shown the largest discrepancy between the \underline{exact} and the SPA results at the
highest $Q^2$ values. 
\item The muon scattering results are in stark contrast to the electron results. The only prominent 
contribution in our \underline{exact} evaluation are the positive contributions from the (g) and (h) 
diagrams. However, for our SPA evaluation from the (e) and (f) diagrams, the contributions are negative,
which become more negative with increasing $Q^2$. We find similar behavior (but smaller in magnitude) 
for the contributions from diagrams (n) and (o) which become increasing more negative with increasing 
$|Q^2|$. The other contributions are quite small. 
\end{enumerate}

\begin{figure*}[tbp]
\begin{center}
\includegraphics[width=0.48\linewidth]{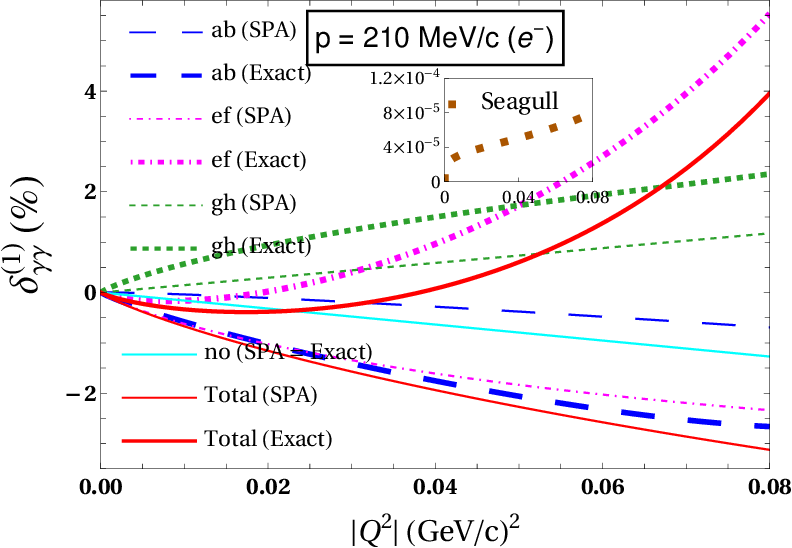}~\quad~\includegraphics[width=0.48\linewidth]{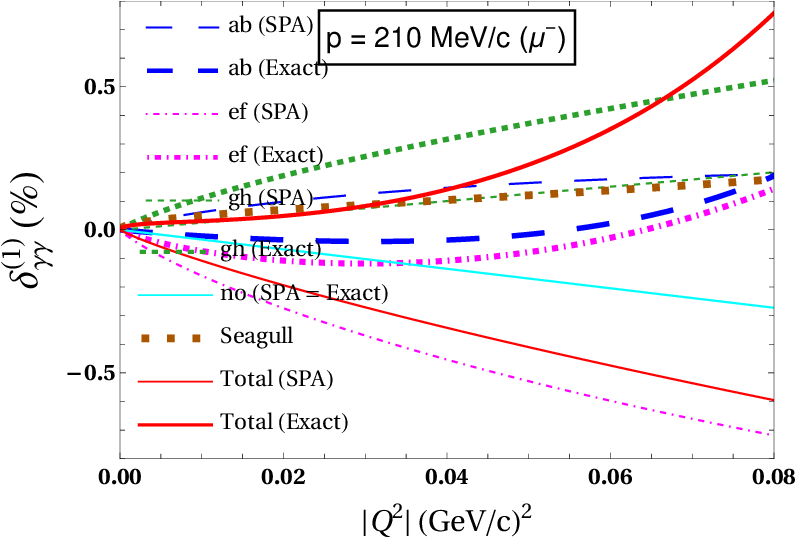}

\vspace{0.5cm}
    
\includegraphics[width=0.48\linewidth]{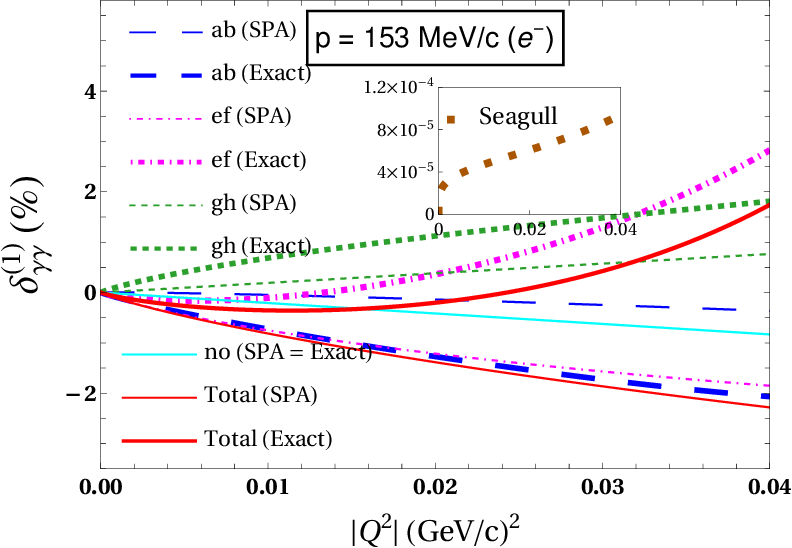}~\quad~\includegraphics[width=0.48\linewidth]{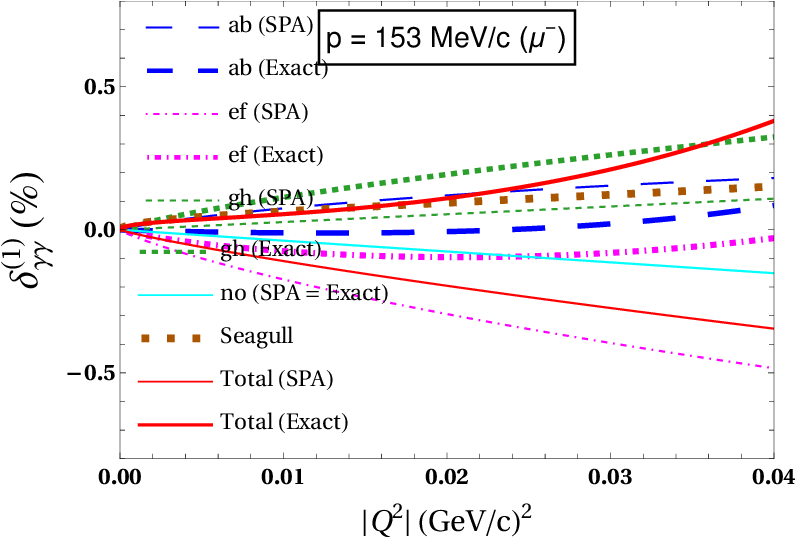} 

\vspace{0.5cm}

\includegraphics[width=0.48\linewidth]{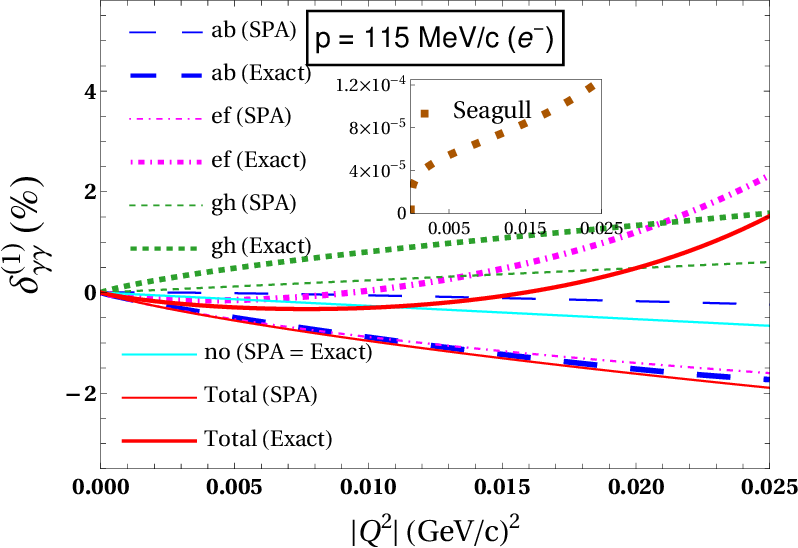}~\quad~\includegraphics[width=0.48\linewidth]{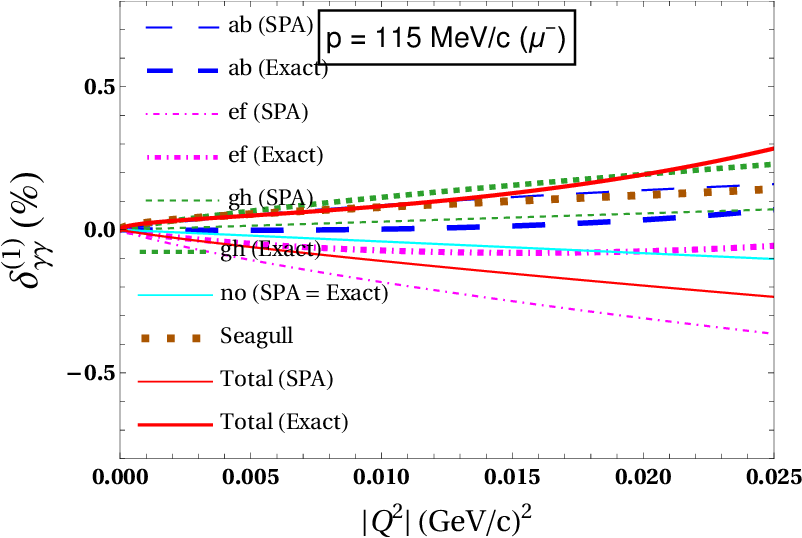}
\caption{The finite fractional NLO [i.e., ${\mathcal O}(\alpha/M)$] corrections 
$\delta^{\rm (1)}_{\gamma\gamma}$ [see Eq.~\eqref{eq:TPE_delta1}] arising from relevant box and crossed-box
pairs of TPE diagrams, contributing to the elastic lepton-proton differential cross section 
up-to-and-including NLO [i.e., ${\mathcal O}(\alpha^3/M)$] accuracy, is presented. For comparison the NLO
SPA results of Goswami {\it et al.}~\cite{Goswami:2025zoe} are also displayed. The results for e–p ($\mu$–p)
elastic scattering are shown in the left (right) panel as functions of the squared four-momentum transfer 
$|Q^2|$, for three MUSE incident lepton beam momenta: 210 MeV/c, 153 MeV/c, and 115 MeV/c. All corrections 
are given relative to the LO OPE differential cross section.  }
\label{fig:NLO_Exact_SPA} 
\end{center}
\end{figure*} 

We now turn to the $\mathcal{O}(\alpha/M^{2})$ \underline{kinematical} TPE results, which are presented in 
Fig.~\ref{fig:LO_NLO_Exact_SPA}. Here we display the kinematically suppressed NLO contributions to the cross
section arising from the LO and NLO diagrams (a) – (i), and compare them with our NLO part of the previously
obtained SPA results of Goswami \textit{et al.}~\cite{Goswami:2025zoe}. The salient features of the these 
NNLO corrections, as evident from Fig.~\ref{fig:LO_NLO_Exact_SPA} are enumerated as below:
\begin{enumerate}
\item The \underline{exact} evaluation reveals sizable cancellations among the individual positive and 
negative contributions, rendering the net results numerically smaller than the corresponding 
$\mathcal{O}(\alpha/M)$ corrections shown in Fig.~\ref{fig:NLO_Exact_SPA}. Such cancellations are not 
manifest in the corresponding SPA results.  
\item The $\mathcal{O}(\alpha/M^{2})$ contribution from the box (e) and crossed-box (f) diagrams 
constitutes the dominant part of the correction. Moreover the \underline{exact} results are significantly 
larger in magnitude and opposite in sign compared to the corresponding SPA results. 
\item Regarding the contributions from the box (a) and crossed-box (b) diagrams, the SPA results are smaller
than those from the \underline{exact} evaluation. In the \underline{exact} case the amplitudes receive 
additional contributions from the interference between the NLO OPE amplitude and the LO TPE diagrams,
$$\mathcal{M}_{\gamma}^{(1)*}\,\big(\mathcal{M}^{(a)}_{\text{box}} + \mathcal{M}^{(b)}_{\text{xbox}}\big)\,,$$  
[denoted as ${\rm a}_{1}{\rm b}_{1}$ in the figure]. These contributions are absent in the SPA treatment.
Notably the above interference terms are numerically tiny compared to the dominant contributions arising
from interference of the LO OPE amplitude $\mathcal M^{(0)}_{\gamma}$ with the (a) and (b) TPE diagrams.
\item The result from the box (g) and crossed-box (h) diagrams is also tiny relative to the other TPE 
terms. In this case, the SPA and \underline{exact} results are numerically very close, while exhibiting
opposite signs for electron and muon scattering across all beam momenta.
\item As discussed earlier, the seagull contributions exhibit a strong dependence on the lepton mass. 
Thus, the seagull term is extremely small for electrons but becomes noticeably larger for muons.
\end{enumerate}

Next, we consider our exactly evaluated $\mathcal{O}(\alpha/M^{2})$ \underline{dynamical} TPE  
contributions from the NNLO diagrams (j) – (${\rm s}_{\pi}$), which include non-vanishing proton structure
effects arising from the pion-loop diagrams (${\rm p}_{\pi}$) – (${\rm s}_{\pi}$). These results stand in 
contrast to the SPA results of Ref.~\cite{Goswami:2025zoe}, where all such $\mathcal{O}(\alpha/M^{2})$ 
contributions from the diagrams above canceled once the residual $\mathcal{O}(\alpha/M^{3})$ terms were
dropped. We adopt the strategy of Ref.~\cite{Goswami:2025zoe} to estimate the proton structure-dependent 
corrections parameterized by the rms charge radius $r_p$,  and we select a representative range of 
recently extracted experimental values. This allows us to quantify the sensitivity of our NNLO results to 
the existing phenomenological discrepancy associated with the charge-radius 
puzzle~\cite{Pohl:2010zza,Pohl:2013,Mohr:2012tt,Antognini:1900ns,Bernauer:2014,Carlson:2015,Bernauer:2020ont,Gao:2021sml}.
We vary our results over a reasonably wide range of proton charge-radius values, specifically between
$r_p=0.87\pm (0.014){\rm stat.}\pm (0.024){\rm syst.}\pm (0.003)_{\rm mod.}$~fm and $0.84087(39)$~fm. The 
former corresponds to the recent extraction from electron proton scattering ISR measurements by the A1 
Collaboration (MAMI)~\cite{Mihovilovic:2019jiz}, while the latter reflects the high precision muonic 
hydrogen spectroscopy determination reported about a decade ago by the CREMA Collaboration 
(PSI)~\cite{Antognini:2013txn,Pohl:2013yb}. The corresponding results for the dynamical NNLO results are
displayed in Fig.~\ref{fig:NNLO_Exact_SPA}, which exhibit the following features enumerated as below:
\begin{enumerate}
\item Unlike the kinematical NNLO corrections shown in Fig.~\ref{fig:LO_NLO_Exact_SPA}, a notable feature
of the dynamical NNLO corrections is that they are all negative and add constructively, thereby yielding
net contributions that are larger in magnitude than the former. Moreover, they exhibit only a weak 
dependence on the lepton mass, with the sole exception of the contribution from diagrams (n) and (o), 
which displays a pronounced sensitivity to the lepton mass.
\item In SPA, the only non-vanishing dynamical contribution at this order arises from the (n) and (o) 
diagrams, and this contribution is identical to that obtained in our \underline{exact} evaluation.
\item The dynamical contributions at $\mathcal{O}(\alpha/M^{2})$ are numerically comparable in size to the
corresponding kinematical corrections with sole exception of vanishingly small contribution from diagrams 
(j) and (k).
\item A distinguishing feature between the SPA and the \underline{exact} TPE evaluations is that only in 
the latter approach there are additional dynamical contributions from the interference between the NNLO 
OPE and LO TPE amplitudes, 
$$\mathcal{M}_{\gamma}^{(2)*}\,\big(\mathcal{M}^{(a)}_{\text{box}} + \mathcal{M}^{(b)}_{\text{xbox}}\big)\,,$$ 
[denoted as ${\rm a}_2{\rm b_2}$ in the figure]. In contrast to the tiny $\mathcal{O}(\alpha/M^{2})$ 
kinematical suppressed interference contribution involving the NLO OPE amplitude 
$\mathcal{M}_{\gamma}^{(1)*}$ and the LO TPE diagrams, apparent from Fig.~\ref{fig:LO_NLO_Exact_SPA} 
(labeled as ${\rm a}_1{\rm b_1}$), the ${\rm a}_2{\rm b_2}$ interference contribution is numerically 
sizable. 
\end{enumerate}

Finally, in Fig.~\ref{fig:TPE_compare} we consolidate all our \underline{exact} TPE results obtained 
from the analysis of the LO, NLO and NNLO TPE diagrams contributing to the charge-odd component of the 
virtual radiative corrections to the elastic $\ell^\mp$-p differential cross section up to NNLO [i.e, 
${\mathcal O}(\alpha/M^2)$], i.e.,  
\begin{eqnarray}
\left[\frac{{\rm d}\sigma_{el}(Q^2)}{{\rm d}\Omega^\prime_l}\right]^{(\ell^\mp)}_{\gamma\gamma;{\rm odd}}
&=& \left[\frac{{\rm d}\sigma_{el}(Q^2)}{{\rm d}\Omega^\prime_l}\right]^{(\ell^\mp)}_{\rm LO+NLO} 
+ \left[\frac{{\rm d}\sigma_{el}(Q^2)}{{\rm d}\Omega^\prime_l}\right]^{(\ell^\mp)}_{\rm NNLO} 
\nonumber\\
&=& \left[\frac{{\rm d}\sigma_{el}(Q^2)}{{\rm d}\Omega^\prime_l}\right]_0 \delta^{(\ell^\mp)}_{\rm TPE}(Q^2)\,,
\end{eqnarray}
where the total fractional TPE radiative corrections is given by 
\begin{eqnarray}
\delta^{(\ell^-)}_{\rm TPE}(Q^2) = -  \delta^{(\ell^+)}_{\rm TPE}(Q^2)
&=& \delta^{\rm (LO+NLO)}_{\gamma\gamma} (Q^2) + \delta^{\rm (NNLO)}_{\gamma\gamma} (Q^2)
\nonumber\\
&\equiv& \delta^{(0)}_{\gamma\gamma} (Q^2) + \delta^{(1)}_{\gamma\gamma} (Q^2) 
+ \delta^{(2)}_{\gamma\gamma} (Q^2) + o\left(\frac{\alpha}{M^2}\right) \,.
\end{eqnarray}
The last term, denoted by ``o”,  represents additional ${\mathcal O}(\alpha/M^2)$ correction terms that 
can arise from  the inclusion of irreducible two-loop pionic TPE diagrams, which were omitted in the 
present analysis, as discussed in Sec.~\ref{sec:intro}.  

Fig.~\ref{fig:TPE_compare} displays the numerical values of the ``true" LO, NLO and NNLO TPE corrections 
$\delta^{(0,1,2)}_{\gamma\gamma}$ to the elastic differential cross section, namely, the different 
$1/M$-recoil order radiative corrections, where $\delta^{(0)}_{\gamma\gamma}\sim {\mathcal O}(\alpha M^0)$ 
and $\delta^{(1)}_{\gamma\gamma}\sim {\mathcal O}(\alpha/M)$ are given in Eqs.~\eqref{eq:delta-ab_LO} and 
\eqref{eq:TPE_delta1}, respectively, and $\delta^{(2)}_{\gamma\gamma}\sim {\mathcal O}(\alpha/M^2)$ is 
given by the sum of Eqs.~\eqref{eq:LO_NLO_TPE_delta2} and \eqref{eq:NNLO_TPE_delta2}:
\begin{equation} 
\delta^{(2)}_{\gamma\gamma} (Q^2) = \delta^{{\rm (LO+NLO};2)}_{\gamma\gamma} (Q^2) 
+  \delta^{({\rm NNLO};2)}_{\gamma\gamma} (Q^2)\, .
\end{equation} 
The following features are extracted from Fig.~\ref{fig:TPE_compare}. 
\begin{figure*}[tbp]
\begin{center}
\includegraphics[width=0.48\linewidth]{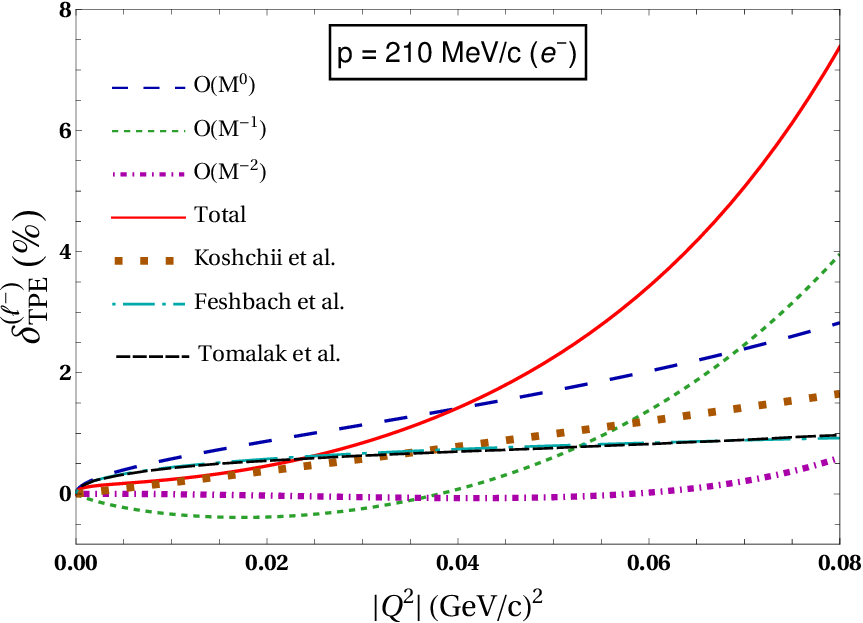}~\quad~\includegraphics[width=0.48\linewidth]{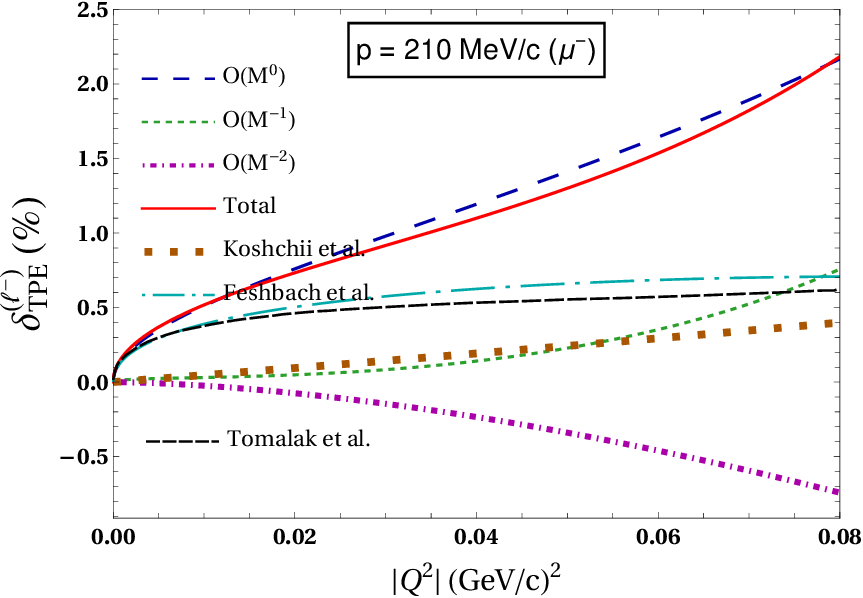}

\vspace{0.5cm}
    
\includegraphics[width=0.48\linewidth]{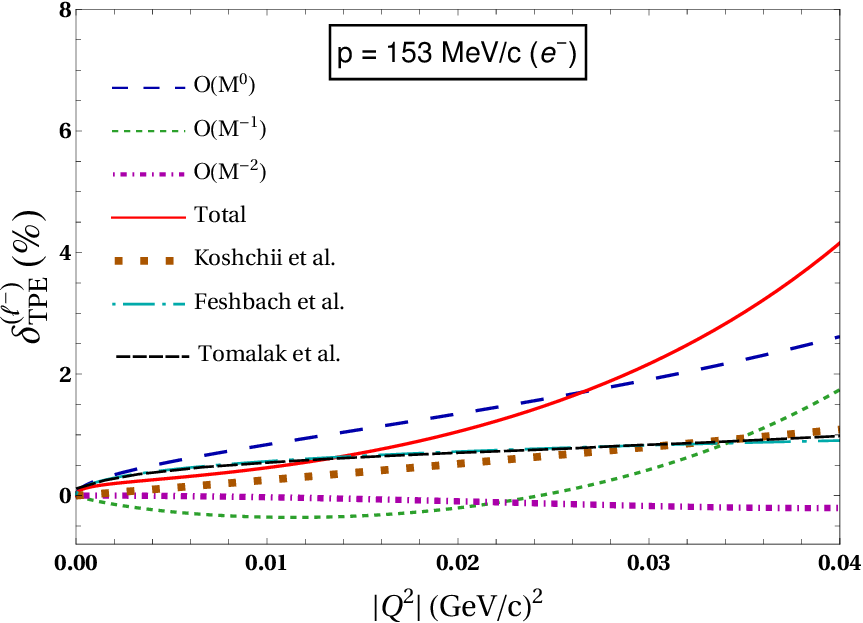}~\quad~\includegraphics[width=0.48\linewidth]{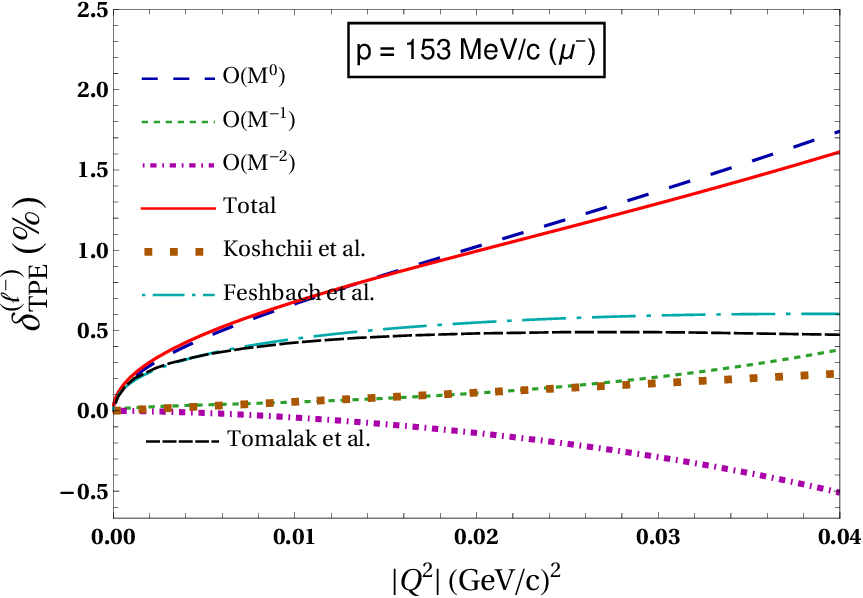} 

\vspace{0.5cm}

\includegraphics[width=0.48\linewidth]{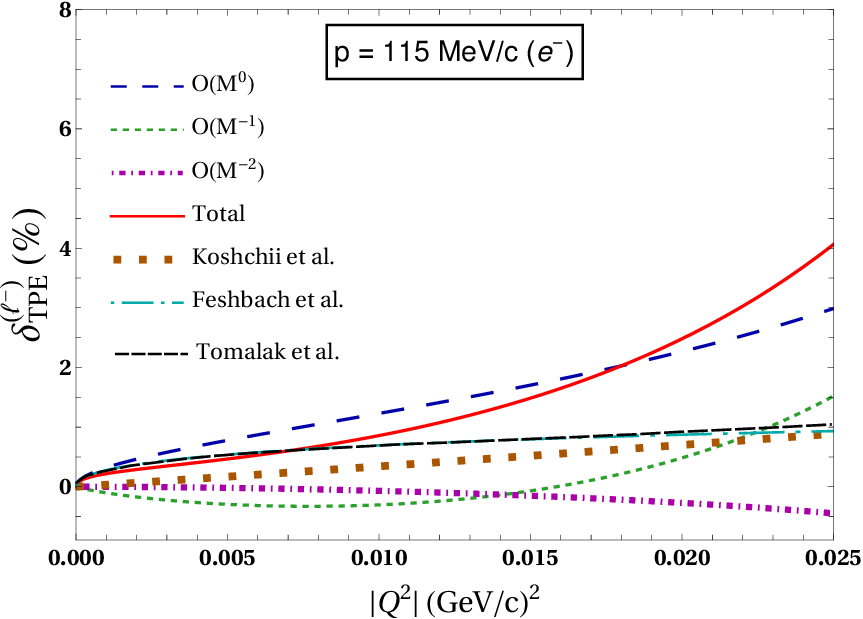}~\quad~\includegraphics[width=0.48\linewidth]{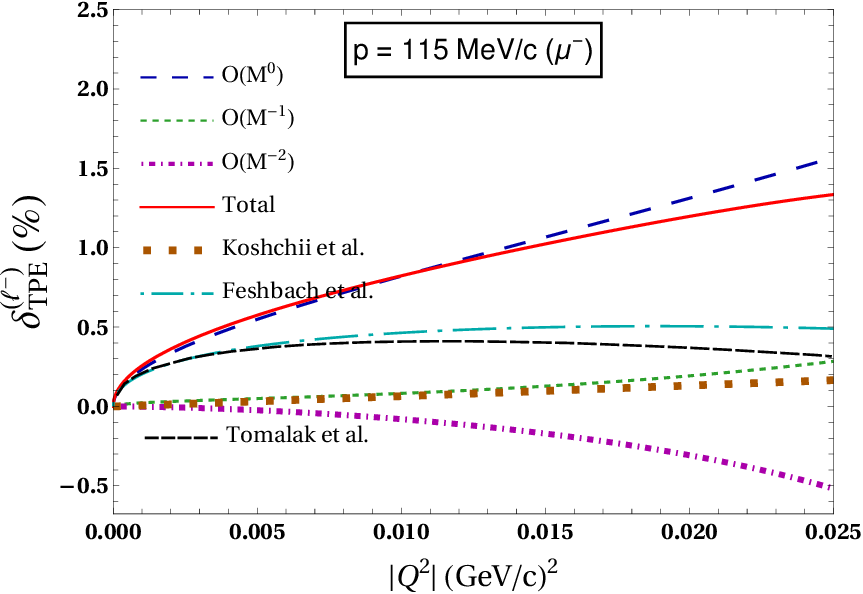}
\caption{Our HB$\chi$PT results for the LO [i.e., $\mathcal{O}(\alpha M^0)$], NLO [i.e., 
$\mathcal{O}(\alpha/M)$] and NNLO [i.e., $\mathcal{O}(\alpha/ M^2)$] finite fractional TPE corrections
to the electron-proton (left panel) and muon-proton (right panel) elastic differential cross sections up
to NNLO [i.e., $\mathcal{O}(\alpha^3/M^2)$] accuracy, is presented. The figure incorporates all LO, NLO 
and NNLO TPE diagrams considered in this work. In particular, the hadronic structure-dependent 
$\mathcal{O}(\alpha/M^2)$ corrections, relying on the proton rms charge radius ($r_p$) are evaluated by 
fixing its value to $r_p=\overline{r_p}\sim 0.855$~fm. This choice corresponds to the mean of the CREMA
determination [$r_p=0.84087(39)$~fm]~\cite{Antognini:2013txn,Pohl:2013yb} and the MAMI–ISR extraction
[$r_p=0.87\pm (0.014){\rm stat.}\pm (0.024){\rm syst.}\pm (0.003)_{\rm mod.}$~fm]~\cite{Mihovilovic:2019jiz}.
Our ``Total" result (solid red line) is compared with the well known potential scattering result of 
McKinley and Feshbach~\cite{McKinley:1948zz}, the SPA-based hadronic model result of Koshchii and 
Afanasev~\cite{Koshchii:2017dzr} and the dispersion technique based result of Tomalak and 
Vanderhaeghen~\cite{Tomalak:2015aoa,Tomalak:2015hva}. The fractional contributions are considered relative
to the LO Born differential cross section relevant to MUSE kinematical range. }
\label{fig:TPE_compare} 
\end{center}
\end{figure*}
\begin{enumerate}
\item The LO corrections $\delta^{\rm (0)}_{\gamma\gamma}$ exhibit a roughly linear dependence on the $|Q^2|$
but depend only weakly on the incident lepton beam momentum and the lepton mass.
\item For electron scattering the NLO corrections $\delta^{\rm (1)}_{\gamma\gamma}$ are negative for very 
low-$Q^2$ values and turn positive at higher momentum transfers. In contrast for muon scattering the NLO 
corrections remain positive throughout and grow monotonically with $|Q^2|$ across the entire MUSE kinematic
range. 
\item The net NNLO (kinematical and dynamical) corrections $\delta^{\rm (2)}_{\gamma\gamma}$ are negative and
much smaller in magnitude compared to the lower orders. Thus, we conclude that the TPE results exhibit a 
reasonable perturbative convergence. 
\item It is noteworthy that the total TPE corrections $\delta^{(e^-\!\!\!,\,\mu^-)}_{\rm TPE}$ for 
electrons and muons turn out to be roughly of the same magnitude. For electron scattering, we observe 
delicate cancellations among several large contributions, whereas for muon scattering the dominant terms 
add constructively, resulting in a net correction of comparable magnitude in both cases. The total 
corrections increase with both $Q^2$ and beam momentum, reaching values of approximately $7.5\%$ for electron
scattering and $4.4\%$ for muon scattering. 
\item A comparison with other recent TPE predictions, such as 
Refs.~\cite{Tomalak:2015aoa,Tomalak:2015hva,Koshchii:2017dzr}, which are largely based on the assumption of 
elastic proton–intermediate–state dominance, indicates that our analytical HB$\chi$PT-based \underline{exact}
TPE calculation yields corrections of substantially larger magnitudes. In particular, for electron scattering, 
the predicted total correction increases markedly at the highest beam momentum, in clear contrast to the 
behavior observed in these alternative approaches. By comparison, the corresponding correction for muon 
scattering exhibits only a mild dependence on the incident beam momentum.
\end{enumerate}

\section{Summary and Conclusion}
In this work, we have presented a comprehensive and fully analytic evaluation of TPE corrections to elastic 
lepton–proton scattering within the framework of SU(2) HB$\chi$PT. Our analysis extended the previous studies
of the \underline{exact} analytical evaluation of the fractional TPE corrections to the elastic differential
cross section up-to-and-including NLO by Choudhary {\it et al.}~\cite{Choudhary:2023rsz}. Here we have 
additionally incorporated the kinematically suppressed NNLO corrections from the LO and NLO TPE diagrams, as
well as the dynamical NNLO corrections from the NNLO TPE diagrams. The latter includes the proton's hadronic
structure-dependent corrections arising from the dominant reducible two-loop form factor TPE diagrams with 
factorizable pion-loops. Our previous \underline{exact} TPE analysis in Ref.~\cite{Choudhary:2023rsz} 
demonstrated that the inclusion of the NLO diagrams (c) – (i) leads to sizable corrections compared with 
those that arise from the LO diagrams (a) and (b). The large NLO effect raised concerns regarding the 
efficacy of low-energy perturbative convergence of the HB$\chi$PT expansion scheme. This concern motivated a
more robust investigation into the behavior of the NNLO perturbative corrections of TPE, as pursued in this
work. 

Firstly, we demonstrated that the \underline{exact} evaluation yields NLO results that differ substantially 
from those obtained in our earlier HB$\chi$PT SPA analyses of Refs.~\cite{Talukdar:2019dko} and
\cite{Goswami:2025zoe}. Besides resolving the inherent shortcoming associated with the SPA formulation within
HB$\chi$PT, namely, that the Feshbach-type correction~\cite{McKinley:1948zz} is absent at LO, our results 
clearly demonstrate that the net NNLO (kinematical and dynamical) contributions are numerically small 
compared to the LO and NLO contributions, thereby exhibit good convergence in the $1/M$-recoil expansion, 
while the genuine NNLO dynamical corrections are uniformly negative and yield non-negligible contributions
to the elastic differential cross section. In particular, the net NNLO corrections were found to be 
approximately $0.5\%$ ($0.8\%$) of the LO Born contribution for electron-proton (muon-proton) scattering in 
the MUSE kinematic range. In other words, we find a rather nominal sensitivity to lepton mass effects 
arising at NNLO. In addition, our exactly evaluated results identifies several contrasting features between 
the kinematical and dynamical NNLO corrections which are qualitatively very different for the corresponding 
SPA results reported in Ref.~\cite{Goswami:2025zoe}. Especially, the \underline{exact} treatment yields 
non-vanishing proton structure-dependent ${\mathcal O}(\alpha/M^2)$ correction terms, which were found to 
cancel at this order in the SPA analysis, where they are instead kinematically suppressed and 
effectively pushed beyond ${\mathcal O}(\alpha/M^3)$. Finally, we demonstrate that no additional 
IR-divergent contributions to the cross section arise from the NNLO TPE diagrams beyond those already 
present at LO. Specifically, the IR divergences originating from the box (a) and crossed-box (b) diagrams 
are fully accounted for and are systematically canceled by the inclusion of the corresponding soft-photon 
bremsstrahlung contributions, as explicitly shown in our recent work in Ref.~\cite{Das:2025jfh}.

Despite these advances, the present calculation up to NNLO accuracy does not yet constitute a fully 
systematic treatment. In particular, we have neglected contributions from irreducible pionic two-loop TPE 
diagrams, whose exact analytical evaluation is substantially more involved. Furthermore, according to the 
HB$\chi$PT power counting, additional NNLO contributions arise from TPE diagrams involving inelastic 
intermediate excited nucleon states, most notably the $\Delta(1232)$ resonance. Such effects are known to 
become increasingly important at higher energies and momentum transfers and may yield non-negligible 
corrections, especially in muon–proton scattering at kinematics relevant to the MUSE experiment. A 
comprehensive investigation of these missing contributions, and their implications for future 
high-precision extractions of lepton–proton TPE corrections, is therefore essential and will be addressed
in a forthcoming work.

In conclusion, we find that an SPA-based analysis of the TPE suffers from serious shortcomings due to an 
improper treatment of the kinematics associated with the loop-integrations in the box and crossed-box 
diagrams. In particular, we demonstrate that significant contributions arise from kinematical regions of 
the loop-momentum where both exchanged photons are hard, i.e., far off their mass-shell, and are therefore 
not reliably captured within the SPA framework.

\acknowledgments
RG and UR acknowledges financial support from the Science and Engineering Research Board, Republic of India, 
(Grant No.\,CRG/2022/000027). RG also acknowledges the organizers of the HADRON2025 International Conference 
at Osaka University for their financial support and hospitality during the conference.

\appendix 
\section{Loop-integrals entering the kinematical NNLO recoil corrections }
In Sec.~3, we analyzed the NNLO [i.e., ${\mathcal O}(\alpha/M^2)$] kinematical TPE corrections to the elastic
lepton–proton differential cross section. This required an exact analytic evaluation of the LO and NLO TPE 
box and crossed-box diagrams. The resulting expressions can be written in terms of a set of master integrals, 
which are discussed in the main text. Several of these integrals were previously evaluated using dimensional
regularization (DR) up to ${\mathcal O}(1/M^2)$ in our SPA-based TPE analysis of Ref.~\cite{Goswami:2025zoe}. 
The remaining loop integrals, which did not appear in that work, are listed below to ${\mathcal O}(1/M^2)$ 
accuracy. In particular, a subset of latter already appeared in our earlier \underline{exact} TPE study in 
Ref.~\cite{Choudhary:2023rsz}, where they were computed to ${\mathcal O}(1/M)$ accuracy. Their 
${\mathcal O}(1/M^2)$ extended results are also included here.

The kinematics of the elastic lepton-proton scattering is defined in terms of the incoming and outgoing lepton
four-momenta $p_\mu=(E,{\bf p})$ and $p^\prime_\mu=(E^\prime, {\bf p^\prime})$, respectively, with 
$|{\bf p}|=\beta E$ and $|{\bf p^\prime}|=\beta^\prime E^\prime$, such that the four-momentum transfer is 
$Q_\mu=(p-p^\prime)_\mu$, while $k_\mu$ represents the TPE loop four-momentum. First we present the result of 
the 3-point master-integral $Z^-\left(\Delta,i \sqrt{-Q^2}/2,m_l,E\right)$, computed up to $\mathcal{O}(1/M^2)$
as follows:
\begin{eqnarray}
Z^-(\Delta,i\sqrt{-Q^2}/2,m_l,E) 
&=& Z^{(0)}(\Delta,i\sqrt{-Q^2}/2,m_l,E) + \delta^{(1/M)}Z^{-}(\Delta,i\sqrt{-Q^2}/2,m_l,E)
\nonumber\\
&+&\delta^{(1/M^2)}Z^{-}(\Delta,i\sqrt{-Q^2}/2,m_l,E)+ {\mathcal O}\left(\frac{1}{M^3}\right)\,
\label{eq:Z-}
\end{eqnarray}
where the four-vector $\Delta_\mu=(p-Q/2)_\mu$, and the ${\mathcal O}(M^0)$, ${\mathcal O}(1/M)$, and 
${\mathcal O}(1/M^2)$ components are respectively given by
\begin{eqnarray}
Z^{(0)}(\Delta,i\sqrt{-Q^2}/2,m_l,E) &=& 
-\,\frac{1}{(4\pi)^2\sqrt{E^2-\Delta^2}}\Bigg[\frac{1}{2}\text{Li}_2\left(\frac{\Delta^2}{\Delta^2-m^2_l}\right)
- \frac{1}{2}\text{Li}_2\left(\frac{\Delta^2-m^2_l}{\eta^2}\right) 
\nonumber\\
&&\hspace{3cm} -\, \frac{1}{2}\text{Li}_2\left(\frac{\eta^2}{\Delta^2-m^2_l}\right) 
+ \text{Li}_2\left(1+\frac{E(1+\beta)}{\eta}\right) 
\nonumber\\
&&\hspace{3cm}
+\, \text{Li}_2\left(1+\frac{E(1-\beta)}{\eta}\right) \Bigg]\,,
\label{eq:Z0}
\end{eqnarray}
\begin{eqnarray}
\delta^{(1/M)}Z^{-}(\Delta,i\sqrt{-Q^2}/2,m_l,E) &=& 
\frac{-Q^2}{4 (4 \pi)^2 M (E^2-\Delta^2)}\! \Bigg[(4 \pi)^2 E Z^{(0)} 
\!-\! \sqrt{\frac{\Delta^2}{\Delta^2-m^2_l}} \ln \!\left(\frac{\sqrt{\Delta^2} 
+ \sqrt{\Delta^2-m^2_l}}{\sqrt{\Delta^2}-\sqrt{\Delta^2-m^2_l}}\right) 
\nonumber \\
&& \hspace{3.2cm} -\, \ln \left(\frac{m_l^2}{\Delta^2-m^2_l}\right) 
\!-\! i\pi\left(1+\frac{\sqrt{\Delta^2}+E}{\sqrt{\Delta^2-m^2_l}} \right)\Bigg]\,,\quad\,
\label{eq:Z-delta1}
\end{eqnarray}
and
\begin{eqnarray}
\delta^{(1/M^2)}\! Z^-\left(\Delta,i \sqrt{-Q^2}/2,m_l,E\right) 
&=& \frac{Q^4 \Delta^2 Z^{(0)}}{32 M^2 (E^2 - \Delta^2)^2}  \!-\! \frac{E Q^2}{4 M (E^2 - \Delta^2)} \delta^{(1/M)}\! 
Z^-\left(\Delta,i \sqrt{-Q^2}/2,m_l,E\right) 
\nonumber\\
&&  +\,\frac{Q^4}{32 (4\pi)^2 M^2 (E^2 - \Delta^2)^2 } \Bigg[-\frac{2E (E^2 - \Delta^2)}{\Delta^2 -m_l^2}
\nonumber\\
&& -\, \left(2E-\frac{2E^2}{\eta} -\frac{m_l^2 \sqrt{E^2-\Delta^2}}{\Delta^2- m_l^2}\right) 
\ln{\left(\frac{m_l^2}{\Delta^2 -m_l^2}\right)}
\nonumber\\
&& +\,\frac{E\left(E+\sqrt{E^2 - \Delta^2}\right)\left(\sqrt{E^2 - \Delta^2} -\beta^2 E\right)}{\Delta^2-m_l^2} 
\ln{\left(\frac{\Delta^2-m_l^2}{\eta^2}\right)}
\nonumber\\
&& -\,\frac{2E \sqrt{\Delta^2}}{\sqrt{\Delta^2-m_l^2} }
\ln{\left(\frac{\sqrt{\Delta^2}+\sqrt{\Delta^2-m_l^2}}{\sqrt{\Delta^2}-\sqrt{\Delta^2-m_l^2}}\right)} 
+ E\left(E+\sqrt{E^2-\Delta^2}\right)
\nonumber\\
&& \times\,\bigg\{\frac{1+\beta}{\sqrt{E^2-\Delta^2}+\beta E} \ln{\bigg(-\frac{E(1+\beta)}{\eta}\bigg)} 
-\frac{1-\beta}{\sqrt{E^2-\Delta^2} -\beta E }
\nonumber\\
&&\hspace{0.6cm} \times\,\ln{\left(-\frac{E(1-\beta)}{\eta}\right)}\bigg\} \Bigg]\,, 
\label{eq:Z-delta2}
\end{eqnarray}   
where $\eta=-E+\sqrt{E^2-\Delta^2}$, $\Delta^2= (p-Q/2)^2 = m^2_l-Q^2/4$, and 
\begin{eqnarray}
{\rm Li}_2(z) =- \int_0^z {\rm d}t\, \frac{\ln(1-t)}{t}\,,\quad  \forall z\in {\mathbb C}\,,
\end{eqnarray}
is the standard dilogarithm (or Spence) function.

Next we present the result of the 3-point master-integral 
$Z^+\left(\Delta^\prime,i \sqrt{-Q^2}/2,m_l,-E^\prime\right)$ computed up to $\mathcal{O}(1/M^2)$: 
\begin{eqnarray}
Z^+(\Delta^\prime,i\sqrt{-Q^2}/2,m_l,-E^\prime) 
&=& -\,Z^{(0)}(\Delta,i\sqrt{-Q^2}/2,m_l,E) + \delta^{(1/M)}Z^+(\Delta^\prime,i\sqrt{-Q^2}/2,m_l,-E^\prime)
\nonumber\\
&& +\,\delta^{(1/M^2)}Z^+(\Delta^\prime,i\sqrt{-Q^2}/2,m_l,-E^\prime)+{\mathcal O}\left(\frac{1}{M^3}\right)\,,
\label{eq:Z+}
\end{eqnarray}
where the four-vector $\Delta^\prime_\mu=-\Delta_\mu=-(p-Q/2)_\mu$, and the ${\mathcal O}(1/M)$, and ${\mathcal O}(1/M^2)$ 
components are respectively given by
\begin{eqnarray}
\delta^{(1/M)} Z^+\left(\Delta^\prime,i \sqrt{-Q^2}/2,m_l,-E^\prime\right) 
&=& \frac{Q^2}{4 (4 \pi)^2 M (E^2-\Delta^2)}\Bigg[(4\pi)^2 E Z^{(0)} +\sqrt{\frac{\Delta^2}{\Delta^2-m_l^2}}
\nonumber\\
&& \times\,\ln\bigg({\frac{\sqrt{\Delta^2}+\sqrt{\Delta^2-m_l^2}}{\sqrt{\Delta^2}-\sqrt{\Delta^2-m_l^2}}}\bigg) 
-\frac{4}{\beta}\ln{\sqrt{\frac{1+\beta}{1-\beta}}} -\ln{\left(\frac{m_l^2}{\Delta^2-m_l^2}\right)}
\nonumber\\
&& +\,i \pi \bigg(\frac{\sqrt{\Delta^2}-E}{\sqrt{\Delta^2-m_l^2}}-3\bigg)\Bigg] \,, 
\end{eqnarray}
and
\begin{eqnarray}
\delta^{(1/M^2)} Z^+\left(\Delta^\prime,i \sqrt{-Q^2}/2,m_l,-E^\prime\right) 
&=& \frac{Q^4}{32 M^2 (4\pi)^2 (E^2- \Delta^2)^{5/2}} \Bigg[ (4 \pi)^2 Z^{(0)}\sqrt{E^2- \Delta^2} (2E^2 +\Delta^2) 
\nonumber\\
&& -\,\frac{8(4 \pi)^2 ME}{Q^2} (E^2- \Delta^2)^{3/2} \delta^{(1/M)} Z^+\left(\Delta^\prime,i \sqrt{-Q^2}/2,m_l,-E^\prime\right) 
\nonumber\\
&& -\,\frac{2E\sqrt{\Delta^2\left(E^2- \Delta^2\right)}}{\sqrt{\Delta^2 -m_l^2}}\ln{\left(\frac{\sqrt{\Delta^2}+\sqrt{\Delta^2-m_l^2}}{\sqrt{\Delta^2}-\sqrt{\Delta^2-m_l^2}}\right)} 
\nonumber\\
&& -\,\frac{2(E^2- \Delta^2)^{3/2}}{E \beta^2} \left(4-\frac{E^2 \beta^2}{\Delta^2 -m_l^2}\right) +\frac{16 (E^2- \Delta^2)}{\beta} \ln{\sqrt{\frac{1+\beta}{1-\beta}}} 
\nonumber\\
&& +\, \left(\frac{4(E^2- \Delta^2)(1+\beta^2)}{\beta^2} -\frac{m_l^2 (E^2-\Delta^2)}{\Delta^2-m_l^2} -2E \sqrt{E^2- \Delta^2}\right)  
\nonumber\\
&& \times\,\ln\left(\frac{m_l^2}{\eta^2}\right) +\frac{\sqrt{E^2-\Delta^2}}{\Delta^2 -m_l^2} \ln\left(\frac{\Delta^2-m_l^2}{\eta^2}\right)
\bigg\{\left(\sqrt{E^2 -\Delta^2} - 3E\right) 
\nonumber\\
&& \times\,\left(\Delta^2 -m_l^2\right) - \Delta^2 \sqrt{E^2-\Delta^2} \bigg\} - \frac{E(1-\beta)}{\sqrt{E^2-\Delta^2}-\beta E} 
\bigg\{ 5 (E^2-\Delta^2) 
\nonumber\\
&& \times\,\ln{\left( -\frac{E(1-\beta)}{\eta}\right)} - 3E \sqrt{E^2-\Delta^2} -\frac{8(E^2- \Delta^2)^{3/2}}{\beta E} 
\nonumber\\ 
&& +\,\frac{4 (1-\beta)(E^2 -\Delta^2)^2}{E^2 \beta^3} \bigg\} - \frac{E(1+\beta)}{\sqrt{E^2-\Delta^2}+\beta E} \ln{\left( -\frac{E(1+\beta)}{\eta}\right)}  
\nonumber\\
&& \times\,\bigg\{ 5 (E^2-\Delta^2) - 3E\sqrt{E^2-\Delta^2} + \frac{8 (E^2- \Delta^2)^{3/2}}{\beta E} 
\nonumber\\
&& -\,\frac{4 (1-\beta) (E^2 -\Delta^2)^2}{E^2 \beta^3} \bigg\}\Bigg] \,. 
\end{eqnarray}

Finally we require the 3-point master-integral $I (Q|1,1,0,1)=I^- (p,0|1,1,0,1)=I^+ (p^\prime,0|1,1,0,1)$ 
for the kinematical $\mathcal O(\alpha/M^2)$ corrections. The integral was previously evaluated only up to
$\mathcal O(1/M)$ accuracy in Ref.~\cite{Talukdar:2019dko}. Its $\mathcal O(1/M^2)$ extension is given as 
below
\begin{eqnarray}
\label{eq:1101}
I^- (p,0|1,1,0,1) &=& \frac{1}{i} \int \frac{{\rm d}^4k}{(2\pi)^4}\frac{1}{(k^2+i0) [(k-Q)^2+i0] (v\cdot k +i0)}
\nonumber\\
&=& I^{(0)}(Q|1,1,0,1) +\delta^{(1/M)}I(Q|1,1,0,1) + \delta^{(1/M^2)}I(Q|1,1,0,1) 
+ \mathcal{O}\left(\frac{1}{M^3}\right)\,,
\end{eqnarray}
where
\begin{eqnarray}
I^{(0)}(Q|1,1,0,1) &=& -\,\frac{1}{16}\sqrt{\frac{1}{-Q^2}} \,,
\\
\delta^{(1/M)}I(Q|1,1,0,1) &=& -\,\frac{1}{8 M \pi^2}\left[1-\ln{\left(-\frac{\sqrt{-Q^2}}{M}+\frac{Q^2}{M^2}\right)}\right]\,, \quad \text{and} 
\\
\delta^{(1/M^2)}I(Q|1,1,0,1) &=& \frac{\sqrt{-Q^2}}{M^2}\left(\frac{1}{128}-\frac{1}{8\pi^2}\right)\,.
\end{eqnarray}

\section{Loop-integrals entering the dynamical NNLO corrections}
In Sec.~4 the evaluation of the genuine ${\mathcal O}(\alpha/M^2)$ dynamical TPE corrections 
requires incorporating a new set 2-, 3-, and 4-point scalar and tensor loop-integrals, that were not present
in the earlier TPE analyses of Choudhary {\it et al.}~\cite{Choudhary:2023rsz} and Goswami 
{\it et al.}~\cite{Goswami:2025zoe}. We first consider the two 4-point tensor loop-integrals appearing
in the fractional contributions $\delta^{(j)}_{\gamma\gamma}$ and $\delta^{(k)}_{\gamma\gamma}$ from the box
(j) and crossed-box (k) diagrams [see Eqs.~\eqref{j-TPE} and \eqref{k-TPE}], namely,  
\begin{eqnarray}
I_1^{-\mu \nu} (p,0|1,1,1,1)  &=& \frac{1}{i} \int \frac{{\rm d}^4k}{(2\pi)^4}
\frac{k^\mu k^\mu}{(k^2+i0) [(k-Q)^2+i0] (k^2-2k\cdot p+i0) (v\cdot k +i0)}\,,\quad \text{and}\qquad\,\,
\\
I_1^{+\mu \nu} (p,0|1,1,1,1)  &=& \frac{1}{i} \int \frac{{\rm d}^4k}{(2\pi)^4}
\frac{k^\mu k^\nu}{(k^2+i0) [(k-Q)^2+i0] (k^2+2k\cdot p^\prime+i0) (v\cdot k +i0)}\,,
\end{eqnarray}
respectively. In order to compute the sum of the box-crossed-box pair of fractional contributions to the 
cross section, $\delta_{\gamma \gamma}^{(jk)}\sim \mathcal{O}(\alpha/M^2)$, we are required to evaluate 
only to LO [i.e., $\mathcal{O}(M^0)$] accuracy in the recoil expansion: 
\begin{eqnarray}
p_{\mu} p_{\nu}I_1^{-\mu \nu} (p,0|1,1,1,1) &+& p_{\mu}^\prime p_{\nu}^\prime I_1^{+\mu \nu} (p,0|1,1,1,1) 
\nonumber\\
&=&  \frac{1}{2} \bigg[p\cdot T_1^-(p,0|0,1,1,1) - p^\prime\cdot T_1^+(p^\prime,0|0,1,1,1) - Q \cdot I_1(Q|1,1,0,1)
\nonumber\\
&&\hspace{0.4cm} +\,m_l^2 \Big(I^- (p,0|0,1,1,1)+ I^+ (p^\prime,0|0,1,1,1) \Big)\bigg]
\nonumber\\
&=& -\,\frac{Q^2}{4} I^{(0)}(Q|1,1,0,1)+\mathcal{O}\left(\frac{1}{M}\right)\, .
\end{eqnarray}
The above scalar product $Q \cdot I_1(Q|1,1,0,1)$ can be further reduced {\it via} partial-fraction 
decomposition into an appropriate combination of 2- and 3-point scalar integrals, namely,
\begin{eqnarray}
Q\cdot I_1(Q|1,1,0,1) 
&=& \frac{1}{i} \int \frac{{\rm d}^4k}{(2\pi)^4}\frac{k\cdot Q}{(k^2+i0) [(k-Q)^2+i0] (v\cdot k+i0)}
\nonumber\\
&=& \frac{1}{2} \bigg[I^-(p,0|0,1,0,1)+Q^2 I^-(p,0|1,1,0,1)-I^-(p,0|1,0,0,1)\bigg]\, .
\end{eqnarray}
The following 2-point functions are evaluated in dimensional regularization {\it via} analytic continuation
to $D$-dimensional space-time: 
\begin{eqnarray}
I(Q|1,0,0,1) &\equiv &  I^-(p,0|1,0,0,1) \equiv  I^+(p^\prime ,0|1,0,0,1) = 0 \,, \quad \text{and}
\\ 
I(Q|0,1,0,1) &\equiv& I^- (p,0|0,1,0,1) \equiv I^+ (p^\prime ,0|0,1,0,1)   
\nonumber\\
&=& \frac{Q^2}{16 \pi^2 M} \left[\frac{1}{\epsilon}+\gamma_E 
- \ln\left({\frac{4\pi \mu^2}{m_l^2}}\right)-\ln{\left(\frac{m_l^2 M^2}{Q^4}\right)}-2\right]\, ,    
\end{eqnarray}
where $\gamma_E=0.577216...$ denotes the Euler-Mascheroni constant  and $\epsilon = (4-D)/2<0$. The former
scaleless loop-integrals vanish identically in dimensional regularization, while the latter UV–divergent 
integral $I(Q|0,1,0,1)\sim {\mathcal O}(1/M)$ generates only ${\mathcal O}(\alpha/M^3)$ contributions to 
$\delta_{\gamma \gamma}^{(jk)}$. These lie beyond our working accuracy, therefore they may be safely 
discarded. The two 3-point tensor integrals $T_1^{\pm\mu}$ are reduced into a combination of 2- and 3-point
function using the well-known {\it Passarino-Veltman decomposition}~\cite{Passarino:1979}:
\begin{eqnarray}
T_1^{-\mu} (p,0|0,1,1,1)  &=& \frac{1}{i} \int \frac{{\rm d}^4k}{(2\pi)^4}
\frac{(k-p)^{\mu}}{ [(k-Q)^2+i0] (k^2-2k\cdot p+i0)(v \cdot k+i0)} 
= v^{\mu} B^-_1 + p^{\prime \mu} B^-_2\,, \quad \text{and}
\\
T_1^{+\mu} (p^\prime,0|0,1,1,1)  &=& \frac{1}{i} \int \frac{{\rm d}^4k}{(2\pi)^4} 
\frac{(k+p^\prime)^{\mu}}{ [(k-Q)^2+i0] (k^2+2k\cdot p^\prime +i0)(v \cdot k+i0)} 
= v^{\mu} B^+_1 + p^{ \mu} B^+_2\,,   
\end{eqnarray}
where
\begin{eqnarray}
B^-_1  &=& \left[1-\frac{1}{{\beta^\prime}^2}\right] I^-(p,0|0,1,1,0) 
- \left[E-\frac{E}{{\beta^\prime}^2}+\frac{m_l^2}{E^\prime {\beta^\prime}^2}\right] I^-(p,0|0,1,1,1)
\nonumber\\
&& +\,\frac{1}{2 E^\prime {\beta^\prime}^2}\left[I^-(p,0|0,0,1,1)-I(Q|0,1,0,1)\right]\,,
\nonumber\\
B^-_2 &=& \frac{1}{E^\prime {\beta^\prime}^2} I^-(p,0|0,1,1,0) 
- \left[\frac{E}{E^\prime {\beta^\prime}^2}-\frac{m_l^2}{{E^\prime }^2{\beta^\prime}^2}\right] I^-(p,0|0,1,1,1)
\nonumber\\
&& -\,\frac{1}{2 {E^\prime }^2{\beta^\prime}^2}\left[I^-(p,0|0,0,1,1)-I(Q|0,1,0,1)\right]\,,
\nonumber\\
B^+_1  &=& \left[1-\frac{1}{\beta^2}\right] I^+(p^\prime,0|0,1,1,0) 
+ \left[E^\prime-\frac{E^\prime}{\beta^2}+\frac{m_l^2}{E\beta^2}\right] I^+(p^\prime,0|0,1,1,1)
\nonumber\\
&& -\,\frac{1}{2 E \beta^2}\left[I^+(p^\prime,0|0,0,1,1)-I(Q|0,1,0,1)\right]\,,
\nonumber\\
B^+_2 &=& \frac{1}{E \beta^2} I^+(p^\prime,0|0,1,1,0) +\left[\frac{E^\prime}{E\beta^2} 
-\frac{m_l^2}{E^2\beta^2}\right] I^+(p^\prime,0|0,1,1,1)
\nonumber\\
&& +\,\frac{1}{2 E^2\beta^2}\left[I^+(p^\prime,0|0,0,1,1)-I(Q|0,1,0,1)\right]\,.
\end{eqnarray} 
While the 3-point scalar integrals $I^-(p,0|0,1,1,1)$ and $I^+(p^\prime,0|0,1,1,0)$ were already evaluated
in our earlier works, Refs.~\cite{Choudhary:2023rsz} and \cite{Goswami:2025zoe}, the 2-point scalar 
master-integrals entering the $B^\pm$ functions are evaluated in DR, yielding the following UV-divergent 
expressions:
\begin{eqnarray}
I^-(p,0|0,1,1,0) &\equiv& I^+(p^\prime,0|0,1,1,0) \equiv  I^-(p,0|1,0,1,0) \equiv I^+(p^\prime,0|1,0,1,0) 
\nonumber\\
&=& -\,\frac{1}{(4\pi)^2} \left[\frac{1}{\epsilon} +\gamma_E -\ln{\left(\frac{4\pi \mu^2}{m_l^2}\right)}-2 \right]\,,
\\
I^-(p,0|0,0,1,1)  &=& -\,\frac{2 E}{(4 \pi)^2}\left[\left\{\frac{1}{\epsilon}+\gamma 
-\ln{\left(\frac{4\pi \mu^2}{m_l^2}\right)}\right\}-2 +2 \beta \ln\sqrt{{\frac{1+\beta}{1-\beta}}}\,\,\right]\,,
\\
I^+(p^\prime,0|0,0,1,1)  &=&  \frac{2 E^\prime}{(4 \pi)^2}\left[\left\{\frac{1}{\epsilon}+\gamma 
-\ln{\left(\frac{4\pi \mu^2}{m_l^2}\right)}\right\}-2 + 2\beta^\prime \ln\sqrt{{\frac{1+\beta^\prime}{1-\beta^\prime}}} \, \,\right]\,.
\end{eqnarray}
There is one additional 2-point master-integral $I(Q|1,1,0,0)$, appearing in the TPE diagrams (j), (k), 
(n) and (o), which is likewise UV-divergent and evaluated to yield
\begin{eqnarray}
I(Q|1,1,0,0) \equiv  I^-(p,0|1,1,0,0) \equiv I^+(p^\prime,0|1,1,0,0) = -\,\frac{1}{(4\pi)^2} \left[\frac{1}{\epsilon} +\gamma_E 
-\ln{\left(-\frac{4\pi \mu^2}{Q^2}\right)}-2 \right]\,.
\end{eqnarray}
Finally we evaluate several new 3-point loop-integrals, which also contribute to the cross section at 
NNLO. These reducible functions are decomposed into combinations of UV-divergent 2- and 3-point scalar 
master-integrals as follows:
\begin{eqnarray}    
I_1^{+\mu} (p^\prime,0|1,1,1,0)  &=& \frac{1}{i} \int \frac{{\rm d}^4k}{(2\pi)^4}
\frac{k^{\mu}}{(k^2+i0) [(k-Q)^2+i0] (k^2+2k\cdot p^\prime+i0)}
\nonumber\\
&=& \frac{1}{8 \pi^2 Q^2 \nu^2}\left[\left(p^{\prime{\mu}}+\frac{1}{2}Q^{\mu}\right) 
\ln{\left(-\frac{Q^2}{m_l^2}\right)}-8\pi^2(Q^2 p^{\prime{\mu}}+2m_l^2 Q^{\mu}) I(Q|1,1,1,0)\right]\,,
\end{eqnarray}
\begin{eqnarray}
 I_2^- (p,0|0,1,1,1)  &=& \frac{1}{i} \int \frac{{\rm d}^4k}{(2\pi)^4}
\frac{k^2}{ [(k-Q)^2+i0] (k^2-2k\cdot p+i0) (v\cdot k +i0) }
\nonumber\\
&=& I^- (p,0|0,1,0,1) + 2 p \cdot T_1^- (p,0|0,1,1,1) + 2 m_l^2 I^- (p,0|0,1,1,1)\,,
\\
 I_2^+ (p^\prime,0|0,1,1,1)  &=& \frac{1}{i} \int \frac{{\rm d}^4k}{(2\pi)^4}
\frac{k^2}{ [(k-Q)^2+i0] (k^2+2k\cdot p^\prime +i0) (v\cdot k +i0) }
\nonumber\\
&=& I^- (p,0|0,1,0,1) - 2 p^\prime \cdot T_1^+ (p^\prime,0|0,1,1,1) + 2 m_l^2 I^+ (p^\prime,0|0,1,1,1)\,,
\\
 I_2^- (p,0|0,1,1,2)  &=& \frac{1}{i} \int \frac{{\rm d}^4k}{(2\pi)^4}
\frac{k^2}{ [(k-Q)^2+i0] (k^2-2k\cdot p+i0) (v\cdot k +i0)^2 }
\nonumber\\
&=& I^- (Q|0,1,0,2) + 2 p \cdot T_1^- (p,0|0,1,1,2) + 2 m_l^2 I^- (p,0|0,1,1,2)\,,
\\
I_2^+ (p^\prime,0|0,1,1,2)  &=& \frac{1}{i} \int \frac{{\rm d}^4k}{(2\pi)^4}
\frac{k^2}{ [(k-Q)^2+i0] (k^2+2k\cdot p^\prime+i0) (v\cdot k +i0)^2 }
\nonumber\\
&=&  I^- (Q|0,1,0,2) - 2 p^\prime \cdot T_1^+ (p^\prime,0|0,1,1,2) + 2 m_l^2 I^+ (p^\prime,0|0,1,1,2)\, .
\end{eqnarray}
Here $I^-(p,0|0,1,1,2)$ and $I^+(p^\prime,0|0,1,12)$ are given in Eqs.~\eqref{eq:I-0112} and 
\eqref{eq:I+0112}, respectively. The expression for $I(Q|0,1,0,2)$ can be found in 
Ref.~\cite{Choudhary:2023rsz}. We emphasize that although each of these scalar integrals contains
UV divergences, all such divergences cancel exactly once the box and crossed-box TPE diagram pairs are 
combined. The net contribution from these diagrams to the cross section is therefore finite.



\end{document}